\documentclass[3p, nopreprintline, compress]{elsarticle}

\usepackage{float}
\usepackage{hyperref}
\usepackage{color}
\usepackage{array}
\usepackage{epstopdf}
\usepackage[cmex10]{amsmath}
\usepackage{amsthm,amsfonts,amssymb,amscd, amsbsy}
\usepackage{amssymb}
\usepackage{mathtools}
\usepackage{enumitem}
\usepackage{dsfont}
\usepackage{stackrel}
\usepackage{mathtools}
\usepackage{bbm}
\usepackage[dvipsnames,table,xcdraw]{xcolor}
\usepackage{tikz}
\usepackage{caption}
\usepackage{subcaption}
\usetikzlibrary{arrows,automata}
\newcommand{\norm}[1]{\left\lVert#1\right\rVert}

\usepackage{multirow}
\usepackage{nicefrac}

\hypersetup{pdfauthor=Mert Kalfa}

\DeclareMathSymbol{\mlq}{\mathord}{operators}{``}
\DeclareMathSymbol{\mrq}{\mathord}{operators}{`'}

\begin{document}

\begin{frontmatter}

\title{Towards Goal-Oriented Semantic Signal Processing:\\Applications and Future Challenges}

\author[bilkent_addr]{Mert Kalfa\corref{mycorrespondingauthor}}
\cortext[mycorrespondingauthor]{Corresponding author}
\ead{kalfa@ee.bilkent.edu.tr}

\author[bilkent_addr]{Mehmetcan Gok}
\ead{mehmetcan@ee.bilkent.edu.tr}
\author[bilkent_addr]{Arda Atalik}
\ead{arda.atalik@bilkent.edu.tr}
\author[bilkent_addr]{Busra Tegin}
\ead{btegin@ee.bilkent.edu.tr}
\author[bilkent_addr]{Tolga M. Duman}
\ead{duman@ee.bilkent.edu.tr}
\author[bilkent_addr]{Orhan Arikan}
\ead{oarikan@ee.bilkent.edu.tr}

\address[bilkent_addr]{Department of Electrical and Electronics Engineering, Bilkent University, 06800, Ankara, Turkey}

\begin{abstract}

Advances in machine learning technology have enabled real-time extraction of semantic information in signals which can revolutionize signal processing techniques and improve their performance significantly for the next generation of applications. With the objective of a concrete representation and efficient processing of the semantic information, we propose and demonstrate a formal graph-based semantic language and a goal filtering method that enables goal-oriented signal processing.  The proposed semantic signal processing framework can easily be tailored for specific applications and goals in a diverse range of signal processing applications. To illustrate its wide range of applicability, we investigate several use cases and provide details on how the proposed goal-oriented semantic signal processing framework can be customized. We also investigate and propose techniques for communications where sensor data is semantically processed and semantic information is exchanged across a sensor network.

\end{abstract}


\end{frontmatter}

\section{Introduction}
\label{sec:Introduction}

Signal processing has played a major role in the era of the digital revolution. It has been essential in the design and development of complex systems that integrate the contributions of multiple scientific disciplines. Significant challenges in critical areas including remote sensing, healthcare, finance, transportation, entertainment, communications, and security have been successfully overcome with advances in signal processing techniques~\cite{kaveh2011}.

Semantic information theory (SIT)~\cite{Carnap1952} has been around almost as long as the classical information theory (CIT)~\cite{shannon1948mathematical}. As its name implies, SIT is focused on the \textit{semantic problem} which is the accurate generation, processing, storage, and transmission of the \textit{intended meaning} rather than individual bits and symbols, which is referred to as the \textit{technical problem}. Despite its long history, semantic information has been used sporadically and only at a basic level in various signal processing applications so far. An important example of semantic information utilization is the target tracking by an individual sensor or by a sensor network. For instance, in computer vision applications, following the detection of objects via real-time processing of images or videos, typically, only a small subset of detected object reports are transferred to the tracking system that initiates new tracks or maintains the existing ones. In such processes, objects that are moving closely can be identified and tracked as a group. In this application, detection and tracking of objects with their identified group properties can be viewed as a semantic extraction of information from image or video signals. Furthermore, if multiple sensors operate as part of a network, they can share the semantic information in the form of object tracks for improved surveillance and object handover operations~\cite{javed2003tracking, khan2003consistent, kumar2010multiple}.

Recent advances in machine learning have far-reaching implications in a diverse range of disciplines including but not limited to healthcare~\cite{zacharaki2009, wiens2017, erickson2017}, finance~\cite{krollner2010, Dingli2017FinancialTS, dixonFinance2020}, entertainment~\cite{zhou2008, vinyals2019grandmaster}, and communications~\cite{farsad2018, wang2019}. Among many achievements of machine learning technology, real-time extraction of rich semantic information in streaming data is of critical importance in future signal processing applications. For instance, the identification and classification of objects in images and video streams can now be performed in real time~\cite{Redmon_2016_CVPR,scaledyolov4, Tijtgat_2017_ICCV, Liu2019}. Availability of such rich semantic data can significantly change the signal processing techniques and improve their performance. 

The 6G and beyond communication systems have the potential to bring about an \textit{Internet of Intelligence} with connected people, connected things, and connected intelligence~\cite{letaief2019roadmap, Strinati2020, akyildiz20206g, dang2020should, chen2020connected}. It will possess the potential to build up a wireless network of \textit{everything sensing}, \textit{everything connected} and \textit{everything intelligent}. Besides the traditional radio technology innovations, 6G and beyond will explore many innovative architectures and technologies, among which the goal-oriented semantic signal processing and communications will be a core component. Semantic communications will enable effective utilization of the available network capacity. Furthermore, the users of the network such as autonomous driving cars will benefit significantly from the goal filtered semantic data exchanges among themselves and the smart city backbone. 

In this paper, for a structured and universal representation and efficient processing of the semantic information, we propose a formal semantic signal processing framework that includes a graph-based semantic language and a goal-oriented parsing method. The proposed framework can easily be tailored for specific signal processing applications and goals. In the proposed framework, a semantic information extractor identifies and classifies components of a signal into a set of application-specific classes. The semantic relations among the identified components are described by a set of application-specific predicates. Furthermore, along with the identification and classification of the components, each node in the graph is associated with a hierarchical set of attributes that provides additional information, ordered in increasing detail. Those components that are semantically related to each other form directed bipartite graphs where only edges between a component and a predicate are allowed to exist. Since typically there are very few nodes in these graphs, operations on these graphs can be implemented very efficiently. Furthermore, the proposed bipartite structure enables a complete yet relatively simple representation of signals and reduces the computational complexity of graph-based signal processing applications considerably.

In the proposed semantic signal processing framework, based on an internally or externally defined set of goals that can vary with time, graphs can be grouped as those which will be processed further and those that are currently not of interest. The further processing stages may include spatio-temporal tracking of graph parameters and conducting various operations on their attributes. At any point in the processing chain, the desired level of semantic information of those graphs which are of interest can be locally stored or shared with another processor through appropriate communication protocols. Since typical high bandwidth sensor data have a sparse occurrence of interesting events, the corresponding semantic signal processing algorithms result in remarkable compression rates. To illustrate the significant reduction in the communication rate compared to classical approaches, we investigate and propose techniques for communication over a sensor network where sensor data is semantically processed and semantic information is exchanged in the network. Furthermore, to demonstrate the wide range of applicability of the proposed goal-oriented semantic signal processing framework, we also investigate several use cases and provide details on how the proposed framework can be adapted for different signal modalities.

The rest of the paper is organized as follows. In the next section, we review existing approaches to semantic and goal-oriented information in signals. In Section~\ref{sec:SurveyExtractorsComms}, we present a detailed literature survey on the recently developed semantic information extraction and semantic communication techniques. The proposed goal-oriented semantic language and signal processing framework is introduced in Section~\ref{sec:ProposedLanguage}. In Section~\ref{sec:ApplicationsOfSPExtraction}, we demonstrate how the proposed semantic signal processing framework can be customized for a few use cases. In Section~\ref{sec:TXOfSPExtraction}, we investigate the transmission and storage of semantic information, as well as the data compression potential of the proposed framework. Concluding remarks and promising research directions are given in Section~\ref{sec:Conclusion}.

\section{Semantic and Goal-Oriented Information}
\label{sec:SemanticAndGoalOrientedInformation}
In his seminal paper~\cite{shannon1948mathematical}, Claude E. Shannon introduced CIT and paved the way for modern communications as we know it. However, as recognized and later formalized by Shannon and Weaver~\cite{shannon1949mathematical}, CIT only addresses the low-level \textit{technical problem}, i.e., only the accurate representation and transmission of bits and symbols are taken into account. At higher levels of abstraction, there are the \textit{semantic problem} and the \textit{effectiveness problem}, which focus on transmission of the \textit{intended meaning} and execution of the \textit{intended effect}, respectively. This work focuses on the mid-level \textit{semantic problem}. 

Consider the speech communication between two people. We use words as part of a shared language to convey certain information to one another. If person-A describes a visual scene to person-B, the exact image that forms in the mind of person-B is almost certainly not the same as that in person-A's mind. However, if the \textit{intended} information is correctly conveyed, no semantic information is lost during transmission. Moreover, from a throughput perspective, a complex scene with a few meaningful components can be conveyed with a comparable number of sentences, greatly reducing the number of \textit{symbols} that are transmitted compared to a pixel-by-pixel breakdown of the scene. Another important factor in human communications is the ability to ask questions, which restricts the domain of knowledge considerably when conveying information. In this work, we refer to questions that can be asked internally or externally as \textit{goals}. Hence, the term \textit{goal-oriented information} is used throughout the paper as semantic information that is filtered by pertinent questions.

With the introduction of the Internet of Things (IoT), the next-generation communication networks will be dominated by massive machine-type communications (mMTC)~\cite{Strinati2020}. Introducing a \textit{language} as a medium to describe signals and form queries about them enables a mode of communication between machines that is similar to humans and can greatly reduce the overall throughput and increase efficiency. Although this increase in efficiency is readily expected and has been known almost since the introduction of CIT~\cite{shannon1949mathematical,Carnap1952}, efficient extraction of semantic information was not possible until the recent advances in machine learning (ML) and artificial intelligence (AI) technologies.

ML and AI techniques enable the development of novel signal processing algorithms and systems that can extract and use the semantic information in their inputs in real-time. Modern ML methods including convolutional and recurrent deep neural network (DNN) architectures~\cite{schuster1997bidirectional,simard2003best,larochelle2009}, Long-Short Term Memory (LSTM) networks~\cite{hochreiter1997long}, and more recently, scene graph generation techniques~\cite{Li_2017_ICCV,Xu_2017_CVPR,Li_2018_ECCV,Yang_2018_ECCV,Gu_2019_CVPR,Wang_2019_CVPR} make it possible to efficiently extract semantic information in signals of widely different modality such as speech, image and video signals. The resulting semantic signals allow for the next generation of signal processing techniques to be implemented at a semantic level with a goal-oriented approach in real-time.

Unlike the classical signal processing approaches, in semantic signal processing, goals are expressed in a semantic language. To be able to define suitable performance metrics, and implement compression/coding schemes based on conveyed or inferred meaning of transmission, one needs to define a \textit{language} that maps these meanings to a predefined syntactic structure.  Therefore, it is critical to establish a semantic information and language model that is sufficiently general to be suitable in various signal processing applications, yet simple enough to be used by low-end agents with stringent power and computing limitations. In the rest of this section, a brief review of previously proposed semantic language modalities is presented.

\subsection{Natural Languages}
The most popular and intuitive candidates for a shared language in signal processing and mMTC applications are Natural Languages (NL). There is a vast amount of work on Natural Language Processing (NLP)~\cite{chowdhury2003natural} for the generation of NL sentences given an input signal, especially for human-machine interface applications. Examples include question answering~\cite{waltz1978english,hirschman2001natural,choi2018quac}, captioning of images and videos~\cite{you2016image,aneja2018convolutional, pan2017video, krishna2017dense}, and discourse parsing~\cite{marcu2000theory, soricut2003sentence, li2014recursive}. Relatively recently, there has been an increasing focus on semantic communications using NL as a basis as well~\cite{Kountouris2007,Juba2011,Bao2011,xie2020lite}.

NLs are attractive since they offer a universal language for all agents, regardless of their specific capabilities. However, this universality also comes with the requirement of processing a massive knowledge base (e.g., the English language) with inherent ambiguities and inconsistencies. This makes the NL approaches unnecessarily complex for simple IoT sensors and similar machine-type applications. For such cases, an unambiguous and simple language structure is required. 

\subsection{Propositional Logic}
The seminal work on semantic communications by Carnap and  Bar-Hillel~\cite{Carnap1952} introduces propositional logic as a semantic language. In a propositional logic language, the information is encoded in Boolean symbols corresponding to \textit{components} and \textit{primitive properties}. Common logical operations such as NOT~($\neg$), AND~($\land$), OR~($\lor$), IMPLIES~($\Rightarrow$) etc., are used to combine several symbols to form \textit{molecular sentences} describing a state (e.g., $\neg (A \Rightarrow B) \land C \,$). Goals in this language can be formed similarly using propositional logic, effectively parsing the existing truth table of symbols for the desired pattern.

The propositional logic as a semantic language is attractive, since it can be tailored for specific applications and as a result, may not suffer from the complexity and ambiguity of NLs. However, it is challenging to incorporate numerical attributes such as position and velocity into the language; therefore, the resulting descriptions may lack complete information about the signals of interest.

\subsection{Graph-Based Languages}
Graph-based languages offer significant advantages over NLs as they are mathematical constructs that can represent components in a signal, as well as their relationships and states. Most popular applications of graph-based languages include scene graph generations from images and videos ~\cite{Li_2017_ICCV,Xu_2017_CVPR,Li_2018_ECCV,Yang_2018_ECCV,Gu_2019_CVPR,Wang_2019_CVPR}, knowledge graphs and graph-based question answering~\cite{wang2014knowledge, wang2017knowledge,zou2014natural, zheng2019interactive}, and semantic web applications~\cite{carroll2004jena, bellur2007improved, segaran2009programming}. The question answering problems use sub-graphs called \textit{motifs} to search for a pattern within a graph representation, enabling a goal-oriented approach~\cite{fan2010, fan2013, serratosa2014, sun2005}. Moreover, graph nodes and edges can include additional attributes to convey a more complete description of the underlying scene~\cite{vcebiric2019summarizing}. 

With the aforementioned attractive qualities, we believe that graphs with attributes can provide a structured and complete description for a variety of signals of interest. However, the majority of graph-based language models still depend on NLs as their backbone; hence, they can still be unnecessarily complex for simple machine-type applications with stringent efficiency requirements. Therefore, we advocate the use of a graph-based language that can be tailored to specific signal domains and applications of interest with a unique and relatively small knowledge base that would be of great advantage for machine-type applications. 

\section{Survey of Semantic Transformations and Goal-Oriented Semantic Communications}
\label{sec:SurveyExtractorsComms}

Before rigorously defining our proposed language and signal processing framework in Section~\ref{sec:ProposedLanguage}, we present the state-of-the-art in existing semantic transformations that can effectively process signals of different modalities and goal-oriented semantic communications in this section.

\subsection{State-of-the-Art in Semantic Transformations}
Semantic information exists in many signal modalities such as a textual description of an image, a knowledge graph derived from a paragraph, and even in correlation functions of random processes.
We refer to \textit{semantic transformation} or \textit{semantic extraction} as the mapping from an input modality to a target semantic modality where the target semantic modality can be anything ranging from vectors, texts, and graphs as discussed in Section~\ref{sec:SemanticAndGoalOrientedInformation}.

In this part, we give a detailed survey of the state-of-the-art semantic transformations. It is important to point out that although there are many input-target modalities, we only discuss a small, albeit an intuitive subset of transformation modalities mostly focused on visual input types. 

\subsubsection{Object Detection and Segmentation}
\begin{figure}[!t]
    \centering
    \includegraphics[width=0.8\textwidth]{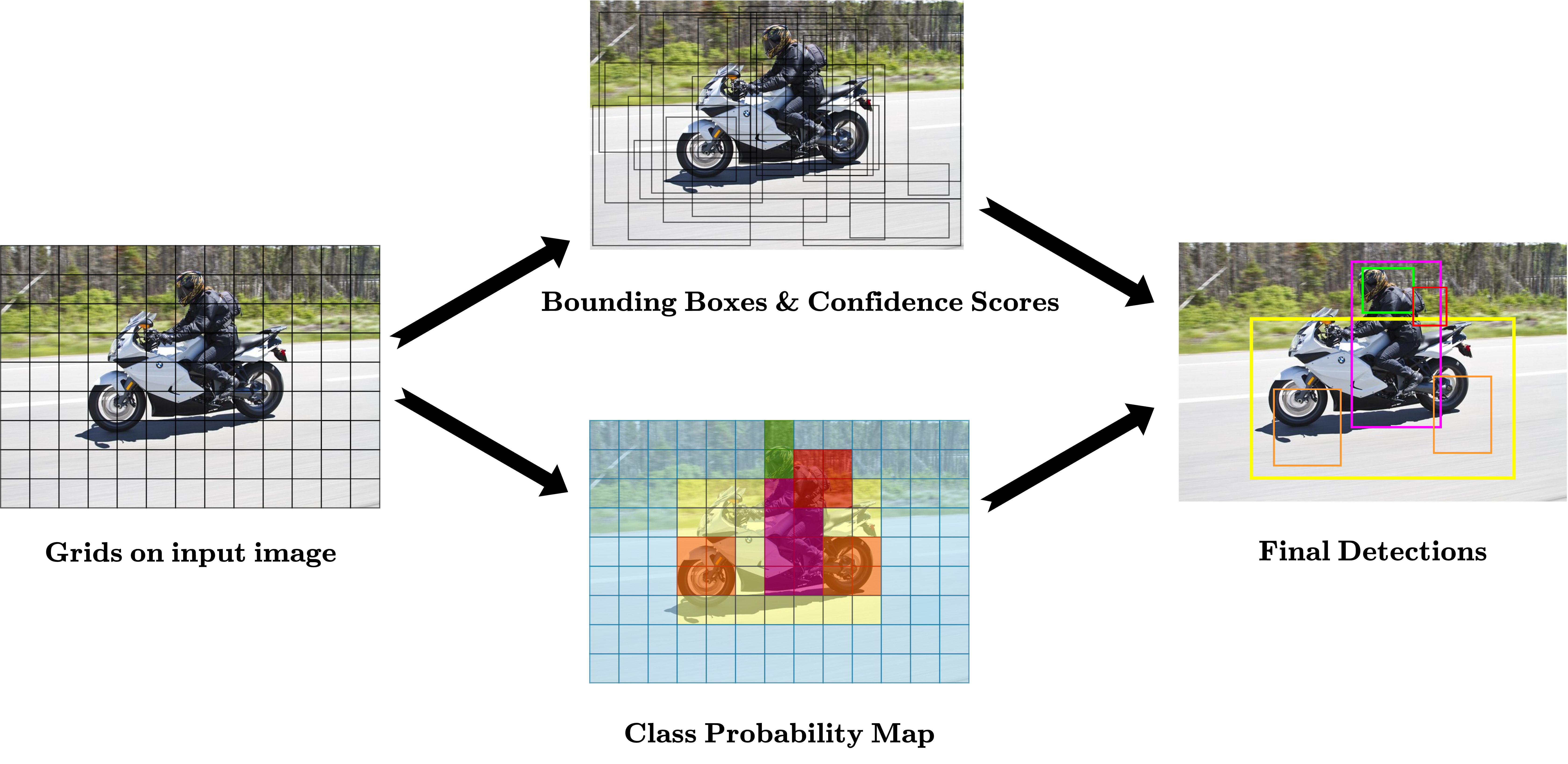}
    \caption{YOLO~\cite{Redmon_2016_CVPR} object detection model. Unlike Faster-RCNN~\cite{fasterrcnn}, YOLO does not use region proposals, instead, it splits the input image into grids and performs inference on each grid.}
    \label{fig:yolov1}
\end{figure}

Object detection can be interpreted as one of the core semantic transformations from visual domain to domain of object classes. Convolutional Neural Networks (CNNs)~\cite{imagenet} constitute fundamental processing modules of object detection methods. Recurrent CNN (RCNN) models are introduced in~\cite{rcnn}, where the selective search is utilized to propose candidate regions and each region is processed independently by a CNN and classified using support vector machines (SVMs). The use of RCNN models is further diversified in~\cite{fastrcnn, fasterrcnn}. In~\cite{fastrcnn} instead of extracting region features separately, a single forward pass through a CNN is performed and the resulting feature map is branched into regions using  Region-of-Interest (RoI) pooling. RCNNs with real-time processing capabilities are introduced in~\cite{fasterrcnn}, where object regions are proposed by Region Proposal Networks (RPNs) instead of selective search. YOLO is introduced in~\cite{Redmon_2016_CVPR} as the very first attempt on fast object detection task and it is further improved in~\cite{yolov2, yolov3}. Unlike Faster-RCNN ~\cite{fasterrcnn}, YOLO, and its later versions, do not utilize region proposals. Instead, they split the input image into cells and perform inference on a limited number of boxes in each cell~\cite{Redmon_2016_CVPR} as shown in Fig.~\ref{fig:yolov1}. Moreover, recent works~\cite{efficientdet, scaledyolov4} use network scaling approaches to achieve higher detection accuracy within shorter inference times, and obtain state-of-the-art results for real-time object detection. Particularly,~\cite{efficientdet} proposes a weighted bi-directional feature pyramid network to fuse features from multiple levels, and in~\cite{scaledyolov4} a network scaling method is applied to the YOLO architecture.  

Segmentation is another semantic transformation that is applicable for visual inputs or inputs that are transformed to 2-D domains such as time-frequency or time-scale. Semantic segmentation can be considered as a higher resolution version of object detection where a category label is assigned to each pixel. More formally, semantic segmentation aims to find a way to assign a label from categories to a set of random variables that correspond to each pixel in the image. Here, each label can represent an object class such as ``person'', ``plane'', ``car'', etc., or labels can be the distinct but unspecified clusters in an unsupervised setting~\cite{van2021unsupervised, xia2017w, ji2018invariant}. In~\cite{shotton2008semantic}, texton forests are proposed as efficient low-level features for image segmentation. Alternative approaches include random forest classifiers~\cite{shotton2011real} and combination of SVMs and Markov Random Fields (MRFs)~\cite{tighe2014scene}. The state-of-the-art in semantic segmentation typically employs convolutional architectures in supervised, semi-supervised, and weakly-supervised settings~\cite{hao2020brief}. It is important to note that, the features extracted by the deeper layers of a CNN are more concentrated on concise semantics with low spatial details whereas shallow layers of a CNN are more aware of spatial details such as edge orientations. Convolutional architectures for image segmentation include dilated convolutions~\cite{yu2015multi} to incorporate context information from multiple scales, Fully Convolutional Network (FCN) architectures~\cite{long2015fully} using skip-connections~\cite{ronneberger2015u}, a symmetrical encoder-decoder structure called DeconvNet~\cite{noh2015learning}, and its simpler version in~\cite{badrinarayanan2017segnet}.

Recurrent architectures are also employed in semantic segmentation. In~\cite{poudel2016recurrent}, Recurrent Fully Convolutional Networks are proposed for multi-slice Magnetic Resonance Imaging (MRI) segmentation where Gated Recurrent Unit (GRU) is incorporated into the bottleneck of the U-Net architecture given in~\cite{ronneberger2015u}. Moreover, several Generative Adversarial Network (GAN) architectures are proposed~\cite{luc2016semantic, xue2018segan, luo2018macro}, where adversarial training is introduced to semantic segmentation in~\cite{luc2016semantic}. 

\begin{figure}
    \centering
    \includegraphics[width=0.6\textwidth]{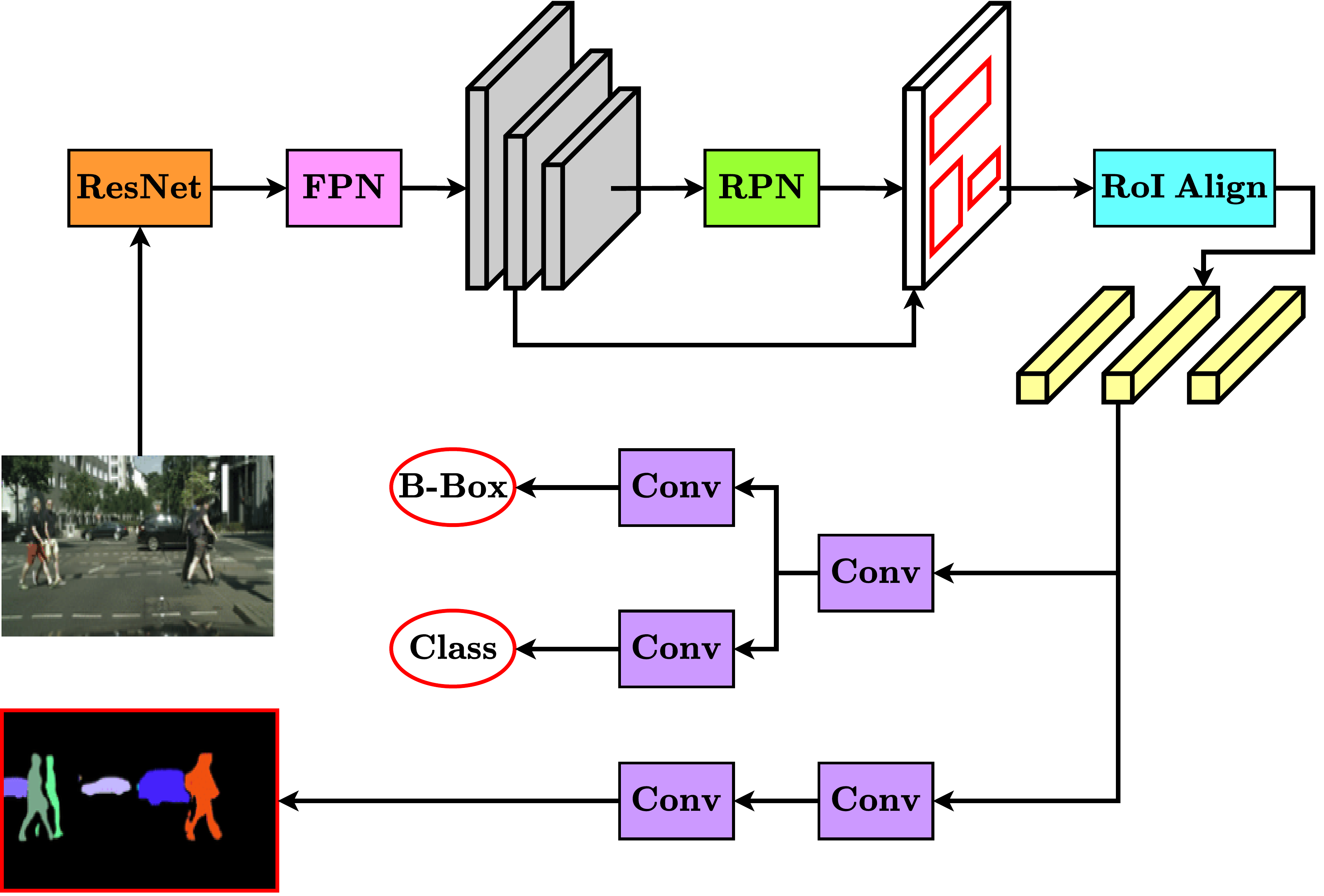}
    \caption{Mask-RCNN~\cite{he2017mask} model for instance segmentation. As its backbone, Mask-RCNN uses ResNet which is followed by a Feature Pyramid Network (FPN) and a Region Proposal Network (RPN). Features for proposed regions are extracted with RoI Alignment. Finally, bounding-box regression, instance classification, and segmentation mask inference are performed.}
    \label{fig:mask_rcnn}
\end{figure}

\begin{figure}[ht]
    \centering
    \begin{subfigure}{0.40\textwidth}
        \centering
        \includegraphics[width=\linewidth]{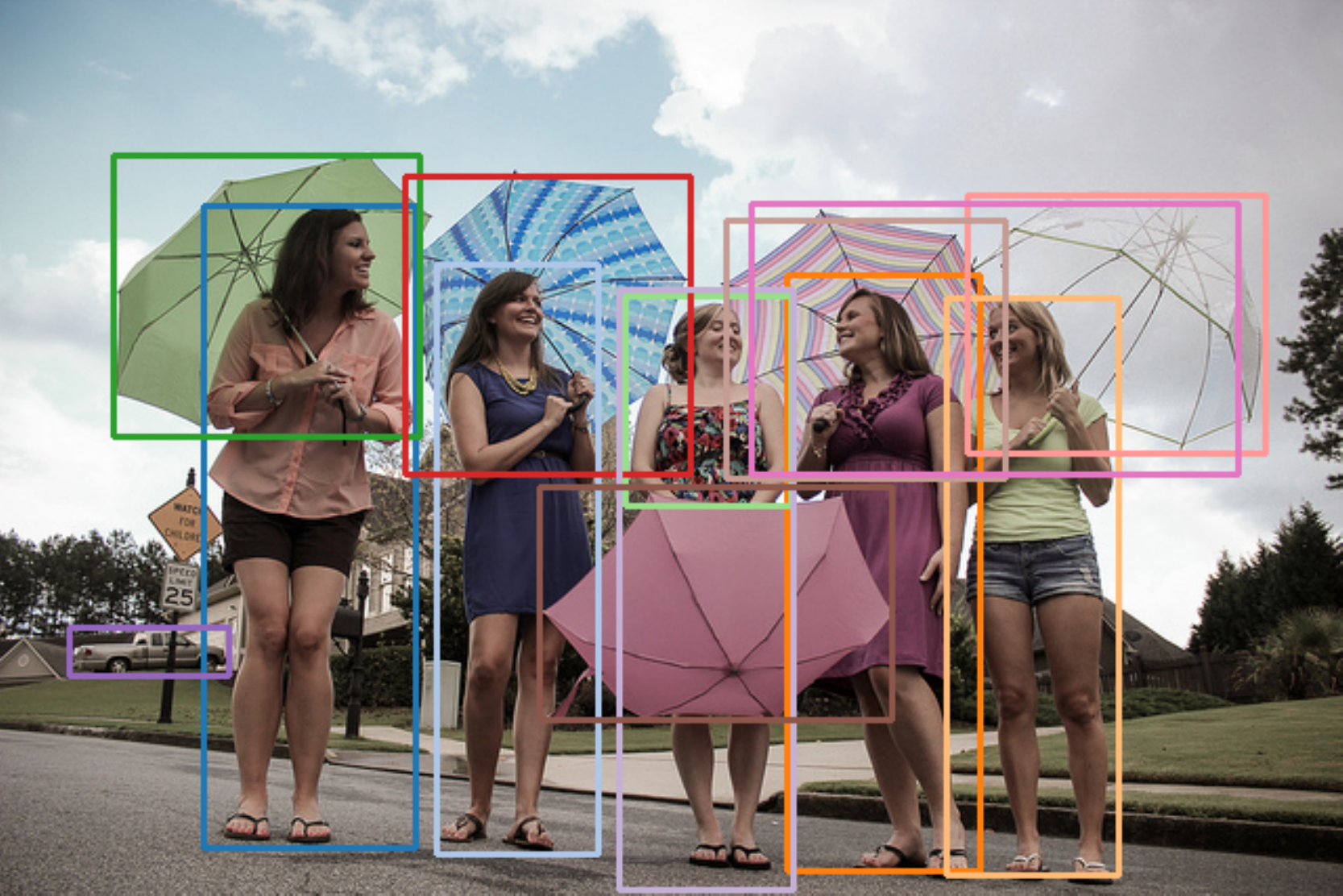}
        \caption{Object Detection} 
        \label{fig:img_objdet}
    \end{subfigure}
    \hspace*{\fill}
    \begin{subfigure}{0.40\textwidth}
        \centering
        \includegraphics[width=\linewidth]{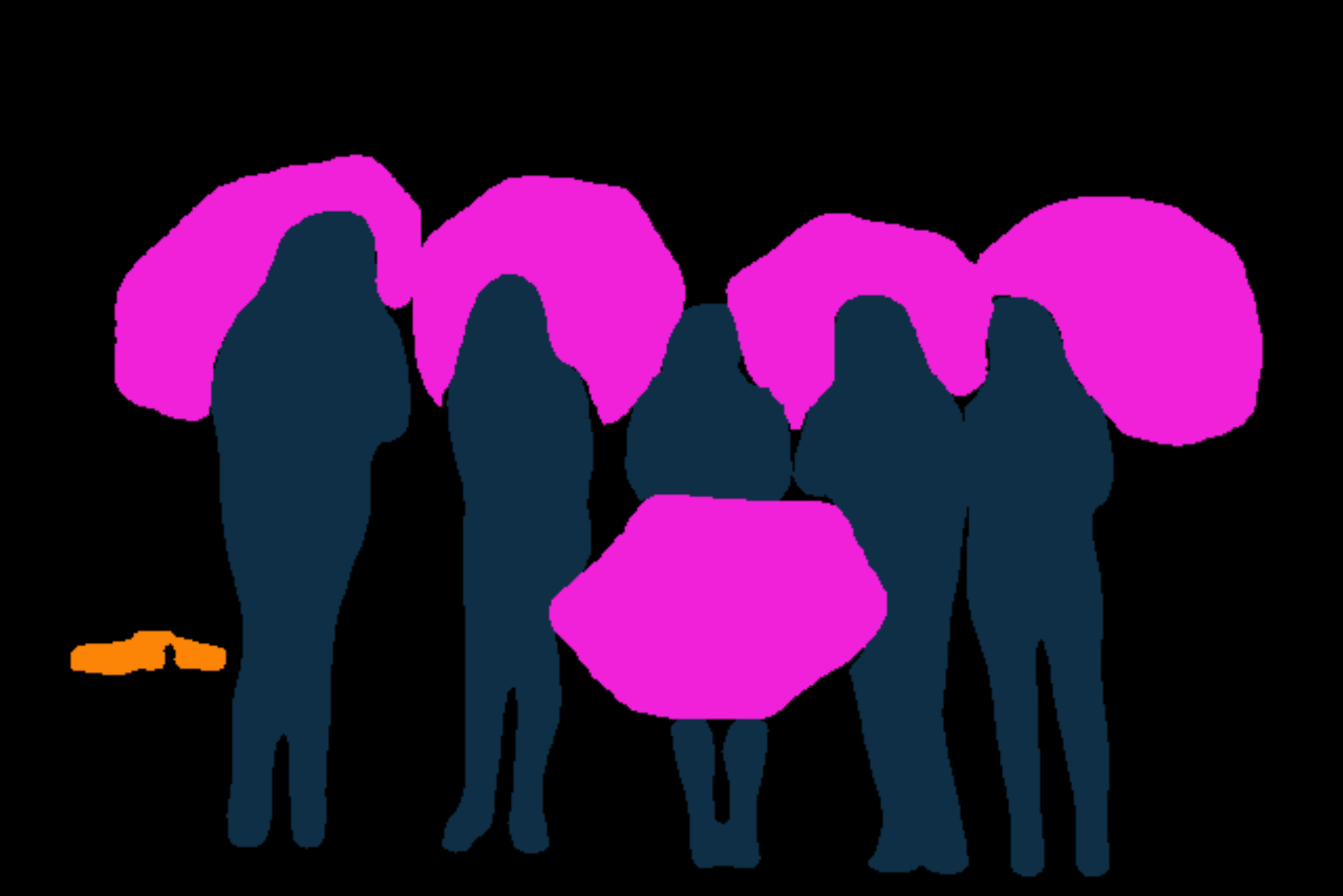}
        \caption{Semantic Segmentation} 
        \label{fig:img_semseg}
    \end{subfigure}
    \medskip
    
    \begin{subfigure}{0.40\textwidth}
        \centering
        \includegraphics[width=\linewidth]{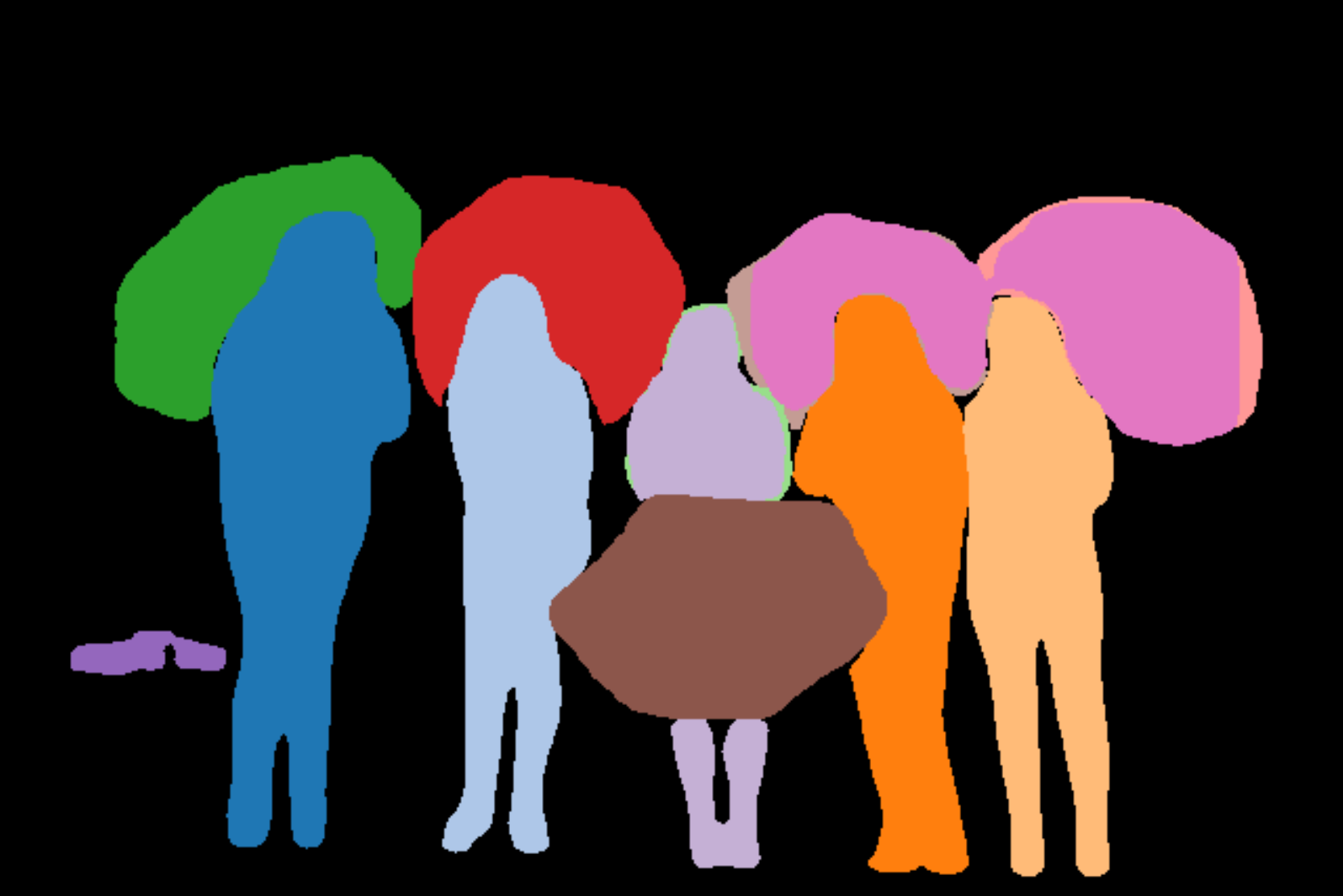}
        \caption{Instance Segmentation} 
        \label{fig:img_instseg}
    \end{subfigure}
    \hspace*{\fill}
    \begin{subfigure}{0.40\textwidth}
        \centering
        \includegraphics[width=\linewidth]{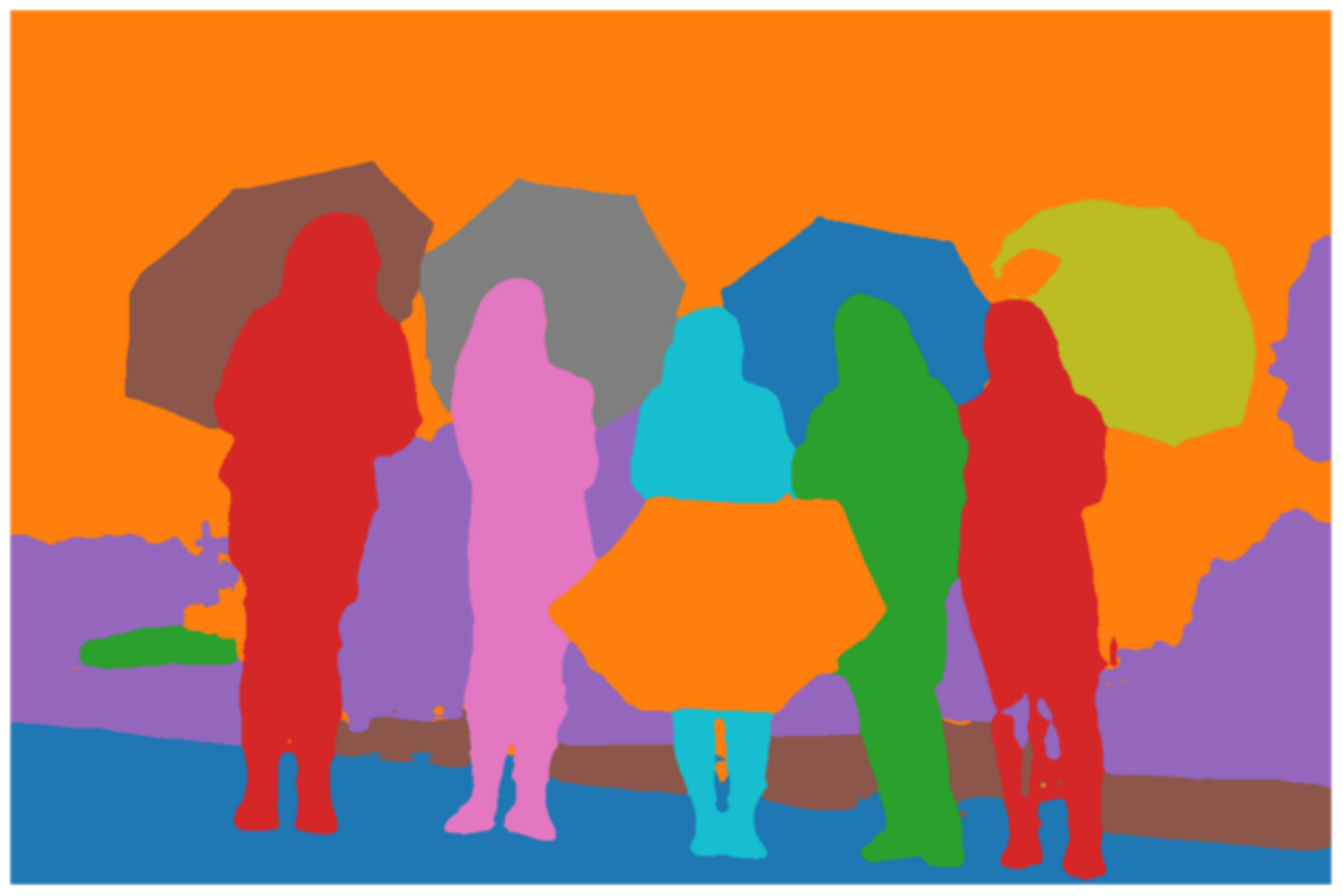}
        \caption{Panoptic Segmentation} 
        \label{fig:img_panseg}
    \end{subfigure}

    \caption{Different object detection and segmentation techniques applied to an image. (a) In object detection, detected/classified objects are localized with bounding boxes. (b) Semantic segmentation aims to assign class labels at the pixel level. (c) Instance segmentation is applied to the object detection for finer localization. (d) In panoptic segmentation, both background and foreground are segmented such that each pixel is associated with an instance and a class simultaneously.}
    \label{fig:segmentations}
\end{figure}

Semantic segmentation architectures that offer real-time operation are proposed in~\cite{poudel2019fast, park2019extremec3net, li2019dfanet}. In~\cite{he2017mask}, Faster-RCNN~\cite{fasterrcnn} is modified for \textit{instance segmentation} and Mask-RCNN is introduced, the process of which is illustrated in Fig.~\ref{fig:mask_rcnn}. Unlike semantic segmentation, instance segmentation aims at assigning labels to pixels at an object level as opposed to a class level. Union of instance and semantic level segmentation is called panoptic segmentation~\cite{kirillov2019panoptic}, where each pixel is associated with both instance and class level labels, as shown in Fig.~\ref{fig:segmentations} along with different segmentation types. Readers interested in semantic segmentation can find detailed taxonomy of models, applications, discussion on challenges, and possible future directions in~\cite{hao2020brief, lateef2019survey, garcia2018survey}.

\subsubsection{Image and Video Captioning}
\begin{figure}[t]
    \centering
    \includegraphics[width=0.8\textwidth]{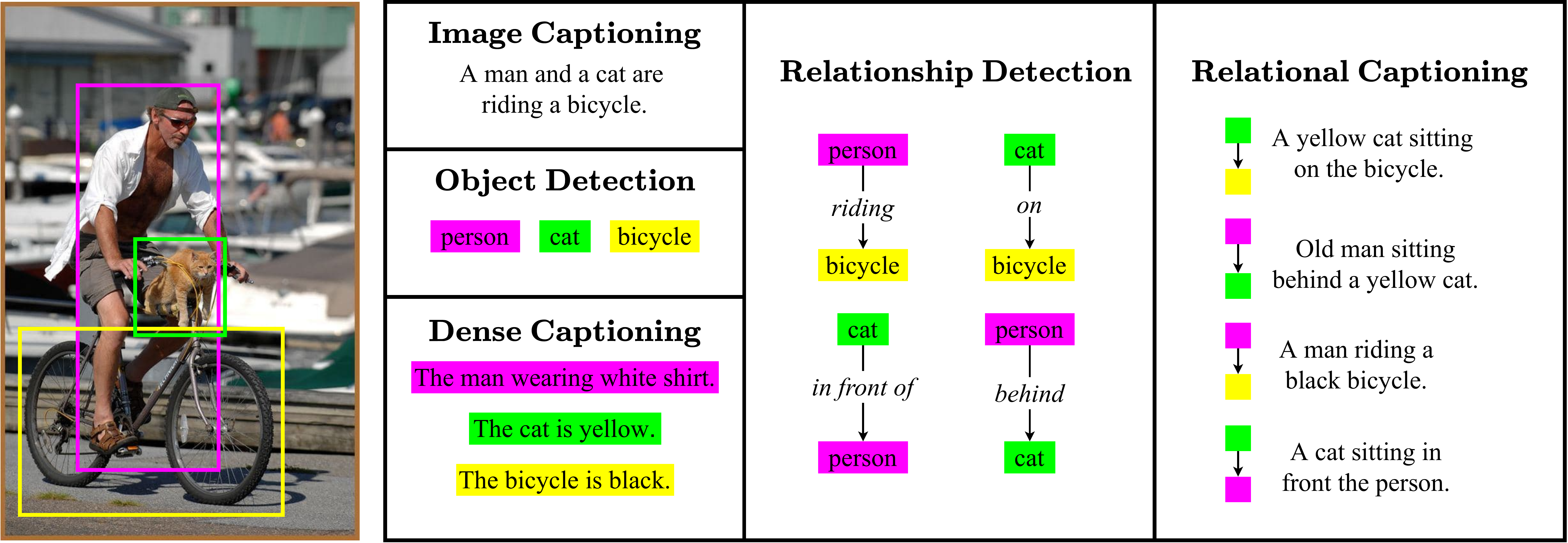}
    \caption{Some semantic transformation examples on the image domain. Image captioning simply generates a textual description of the image. Object detection identifies the objects residing in the image. Dense captioning generates a caption for each salient region or detected object in the image. Relationship detection identifies relations among each object pair. Relational captioning generates a caption for each detected object pair.}
    \label{fig:captioningetc}
\end{figure}

An intuitive way of representing the semantic information embedded in images or videos is to describe them via natural languages (NLs). Image caption or annotation generation is defined as the process of generating textual descriptions for images, which include not only the descriptions of objects within frames but also their interrelations and states. Typical fundamental building blocks of captioning models include CNNs and Recurrent Neural Networks (RNNs). More specifically, a CNN backbone is utilized to extract the visual features and RNNs are used for sequence modeling~\cite{donahue2015long, karpathy2015deep, mao2014explain, donahue2015long, chen2015mind, xu2015show, densecap}. To improve the caption quality, visual attention on CNNs~\cite{xu2015show} or captioning on multiple image regions are proposed~\cite{karpathy2015deep}. In~\cite{densecap}, dense captioning is introduced where object detection and caption generation tasks are tackled jointly in such a way that the detected visual concepts are described with short NL phrases. To overcome localization issues of visual concepts due to overlapping target regions, global image features are fused with region features for dense captioning in~\cite{yang2017dense}. In~\cite{relationalcap}, the dense captioning task is extended to relational captioning where multiple captions are generated for each object pair based on spatial, attentive, and contact relation information~\cite{visualgenome}. Examples of image captioning techniques are shown in Fig.~\ref{fig:captioningetc}.

\begin{figure}[!b]
    \centering
    \includegraphics[width=1\textwidth]{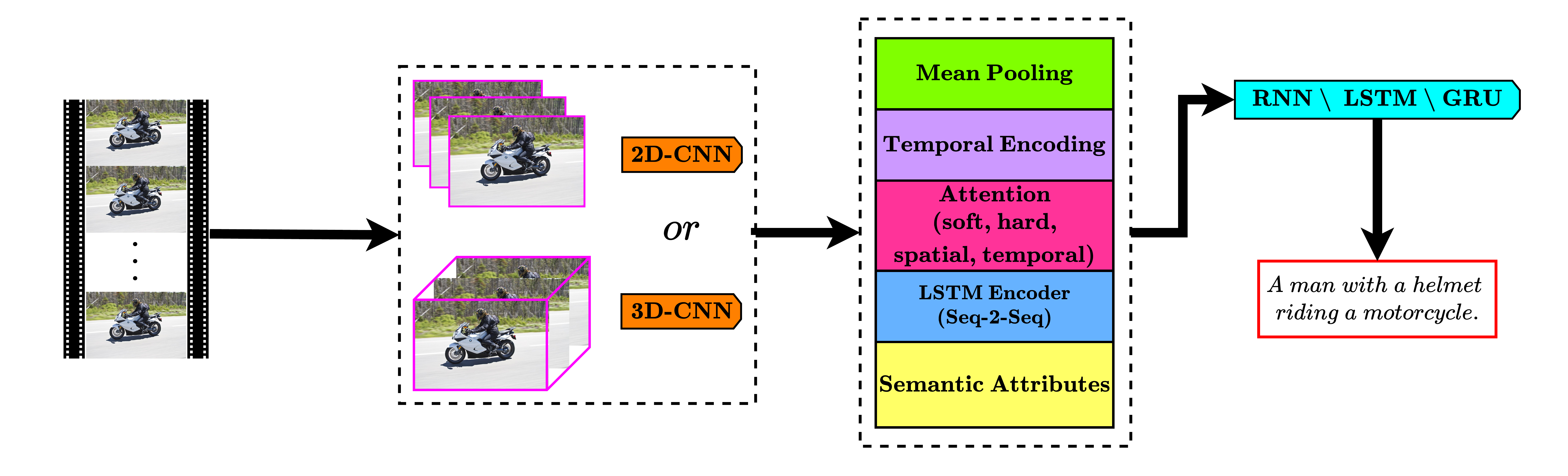}
    \caption{Video captioning pipeline. Often, frames are processed by 2D or 3D convolutions to extract features. Then, extracted features are processed further in the temporal domain by methods like attention, temporal encoding, LSTM, etc. Finally, a caption is generated by feeding recurrent architectures with temporally assessed features.}
    \label{fig:video_cap_pipe}
\end{figure}

Video captioning can be considered as a temporal extension of image captioning. Just like image captioning, the video captioning architectures also use CNNs (2D or 3D) as attention mechanisms or to extract visual semantic content. Then, RNNs or LSTMs are typically used to generate NL text sequences~\cite{aafaq2019video} as illustrated in Fig.~\ref{fig:video_cap_pipe}. Specifically, one of the early works~\cite{venugopalan2014translating}, which is only applicable for videos of short duration, employs mean-pooling to frame representations extracted by a shared CNN and utilizes an LSTM architecture for caption generation. To extend the validity of extracted features to longer durations, recurrent visual encoder architectures are used~\cite{donahue2015long, xu2015multi, venugopalan2015sequence}. In~\cite{yu2016video}, instead of a single caption, multiple captions are generated without their temporal localizations using a hierarchical-RNN architecture. Furthermore, a hierarchical-RNN architecture is used to generate paragraphs for videos in~\cite{krause2017hierarchical}. Dense captioning works in the image domain are also extended to video signals in~\cite{krishna2017dense} where multiple captions are generated for each detected event, and they are temporally localized. Dense video captioning is also referred to as joint event detection and description generation. In~\cite{krishna2017dense} CD3 features are extracted beforehand, and they are fed into a proposal module that uses an attention mechanism. JEDDi-Net proposed in~\cite{xu2019joint} is employed for dense video captioning in an end-to-end fashion without pre-feature extraction. It uses a 3D-CNN to extract the video features and a segment-proposal-network (SPN) to generate candidate segments for events. A comprehensive survey of image and video captioning models can be found in~\cite{hossain2019comprehensive} and~\cite{aafaq2019video}, respectively.

\subsubsection{Scene Graph Generation}
\begin{figure}[ht]
    \centering
    \includegraphics[width=0.7\textwidth]{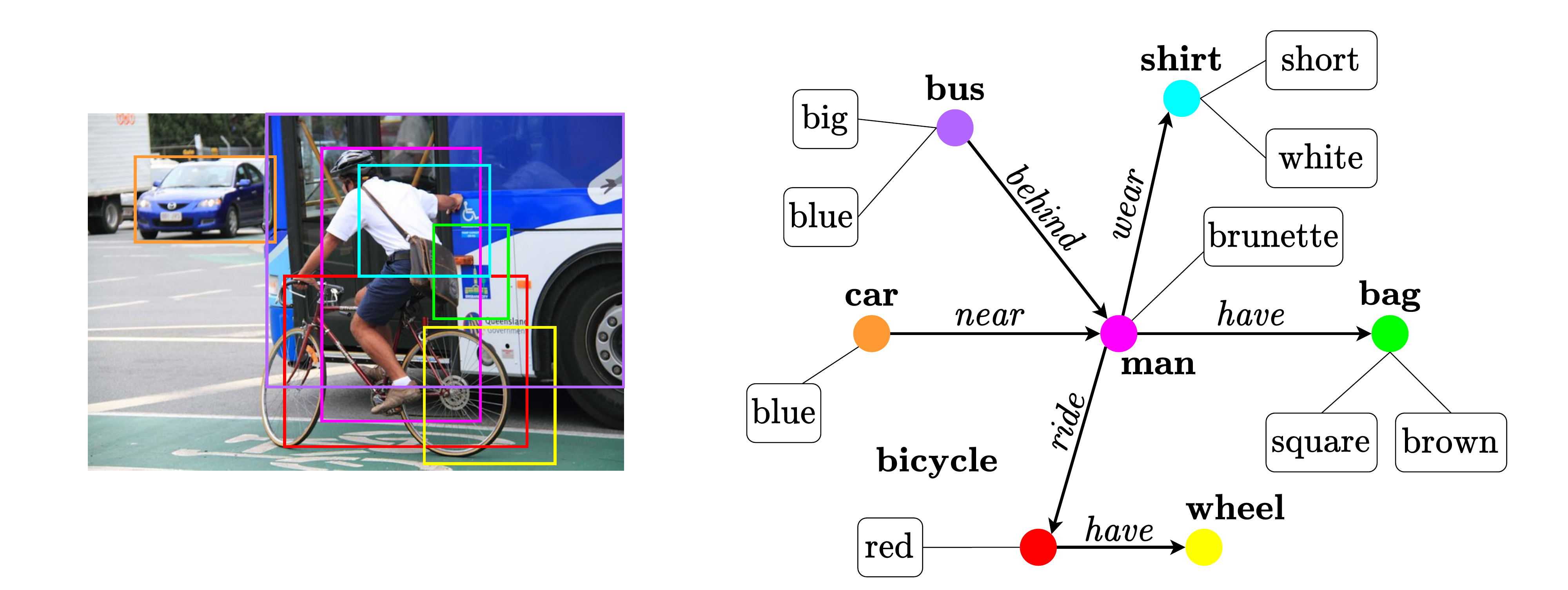}
    \caption{A scene graph example. Detected objects are represented as objects with circular colored nodes. Relationships among object nodes are shown with directed edges. Object attributes are denoted with rectangular uncolored nodes and they are connected to object nodes with undirected edges.}
    \label{fig:scene-graph-ex}
\end{figure}

\begin{figure}[ht]
    \centering
    \includegraphics[width=0.6\textwidth]{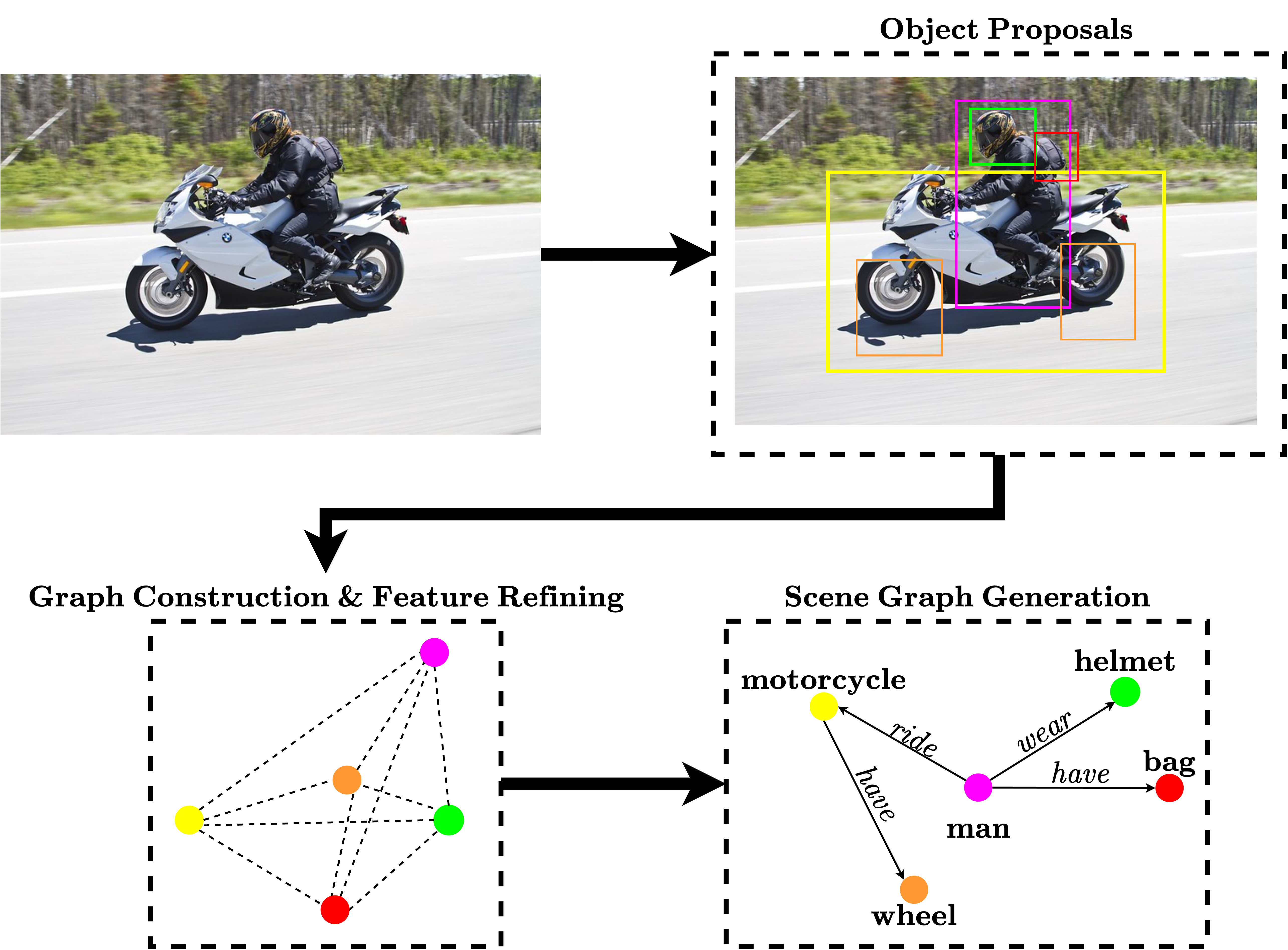}
    \caption{Scene graph generation pipeline. As a first step, object proposal regions are generated via RPN~\cite{fasterrcnn}. The proposals are used to construct a fully connected coarse graph. The initial graph and its features are refined iteratively. Refined features are used to infer node types and relationship types in the final graph.}
    \label{fig:sgg-pipeline}
\end{figure}

A powerful form of semantic transformation is to convert images into graphs that represent a scene and encode the visual relationships presented in the image. Scene graphs are proposed in~\cite{johnson2015image} describe image features and object relationships in an explicit and structured way for image retrieval. Scene graph generation models can capture a higher-level of understanding of the scenes compared to object detection models by additionally identifying object attributes and relationships among them. In essence, scene graph generator models can be considered as image captioning models where they produce parsed versions of succinct captions that are represented in the graph format, instead of NL sentences. Specifically, a scene graph is a graphical data structure that describes the contents of a scene where the nodes represent the detected objects, and edges linking them represent the inter-node relationships. An example scene graph is shown in Fig.~\ref{fig:scene-graph-ex}, while a typical scene graph generation process is illustrated in Fig.~\ref{fig:sgg-pipeline}.

\begin{figure}[ht]
    \centering
    \includegraphics[width=0.7\textwidth]{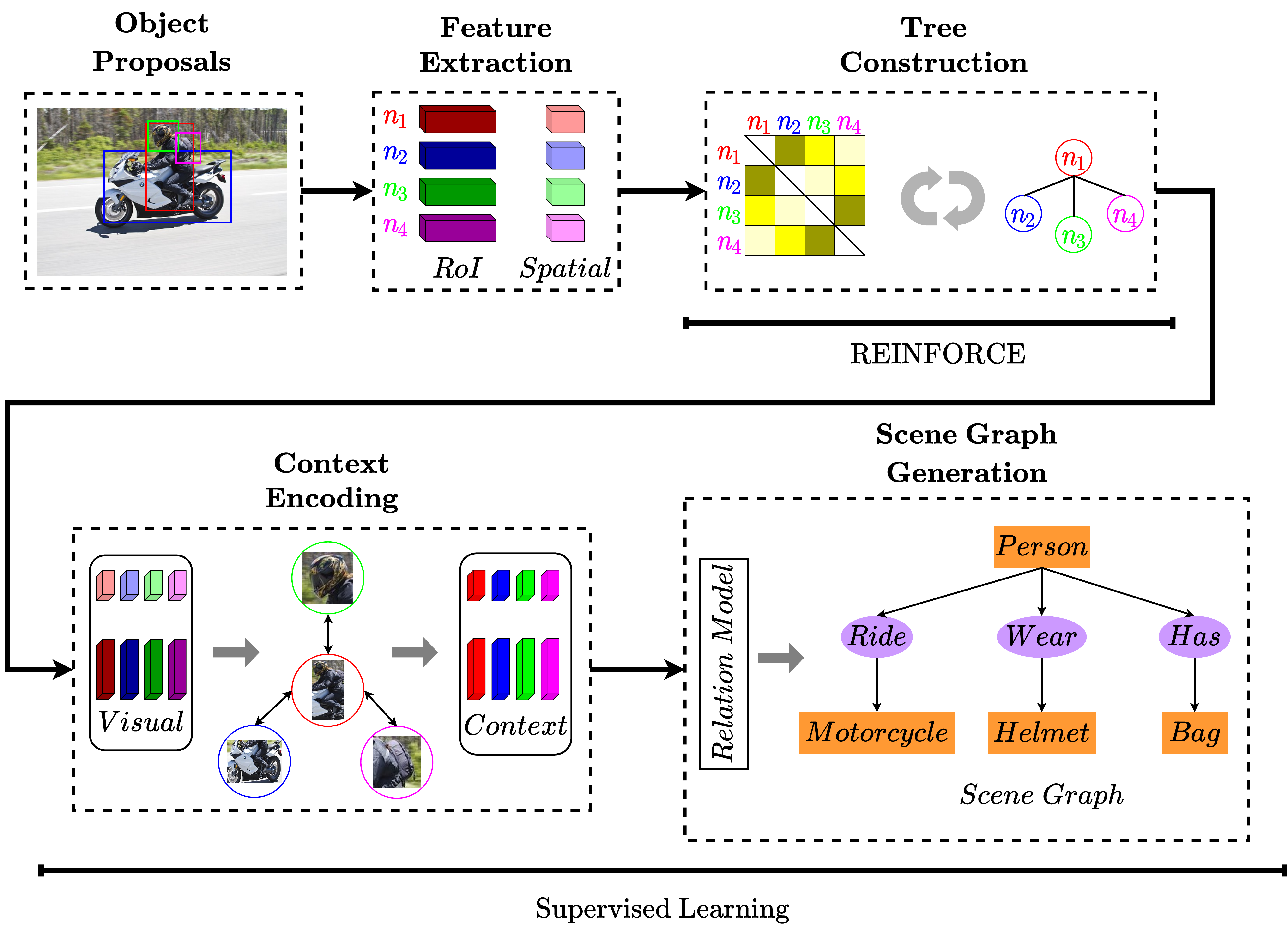}
    \caption{VCTree model proposed in~\cite{tang2019learning}. Visual features are extracted from object proposals. Extracted features are used to compute a scoring matrix. Based on the score matrix, a dynamic tree structure is constructed using REINFORCE algorithm~\cite{williams1992simple}. Finally, visual features are encoded into context features and a scene graph is generated via supervised learning.}
    \label{fig:vctree}
\end{figure}

The scene graph generation process consists of several steps~\cite{chang2021scene}. First, given an image, an object detection module extracts object region proposals and visual features corresponding to these regions. Generally, a Faster-RCNN~\cite{fasterrcnn} is used for the backbone of object detection. The extracted features are used to identify object categories and their attributes. Identified objects along with their extracted features are used as nodes in the initial graph, e.g., a fully connected dense graph. Then, the extracted features along the nodes and edges are iteratively refined, and a final graph is inferred according to these refined features. Some recent papers~\cite{Xu_2017_CVPR, Li_2017_ICCV, li2017vip} also consider joint optimization of object detection and relationship recognition parts. Specifically, Factorizable-Net is proposed in~\cite{Li_2018_ECCV} where an RPN is used to extract object proposals and proposed objects are paired to obtain a fully-connected initial coarse graph. In~\cite{Yang_2018_ECCV}, graph-RCNN is introduced. Graph-RCNN uses relation-proposal-network (RePN) to prune the connections in the initial graph and an Attentional Graph Convolutional Network~\cite{velivckovic2017graph} refines the features on the graph. On the other hand, VCTree model~\cite{tang2019learning} constructs a dynamic tree from a scoring matrix where visual context is encoded into the tree structure. An illustration of the VCTree model is given in Fig.~\ref{fig:vctree}. Furthermore, some recent works~\cite{Xu_2017_CVPR, Li_2017_ICCV} use recurrent architectures for graph inference. Particularly, in~\cite{Xu_2017_CVPR} a feature refining module consisting of edge and node Gated Recurrent Units (GRU), and in~\cite{zellers2018neural}, stacked bi-LSTMs are used. 

There are approaches to incorporate external prior knowledge for scene graph generation tasks as well. In~\cite{lu2016visual}, natural language priors are incorporated and visual relationships and textual relationships are jointly learned and aligned. A linguistic knowledge distillation framework is proposed in~\cite{yu2017visual} where statistics obtained from external texts are used to regularize visual models. In~\cite{zareian2020bridging}, knowledge-graphs are employed as prior information where the generated scene and knowledge graphs are abridged and iteratively refined. 

Scene graphs can also be extracted from video signals. In~\cite{wang2020storytelling}, each frame in the video is converted into a scene graph as an intermediate semantic representation. Then, using frame and cross-frame level relationships of intermediate scene graphs, a \textit{story} of the video is generated. Joint parsing of cross-view videos is introduced in~\cite{qi2018scene} where scene-centric and view-centric graphs are hierarchically generated. For more details on scene graphs, we refer the readers to the pertinent survey papers~\cite{chang2021scene, agarwal2020visual}.

\subsubsection{Automatic Speech Recognition}
Speech is one the most natural sources of semantic information. The most popular application of semantic transformation on speech signals is the conversion of audio signals into NL texts via automatic speech recognition (ASR) systems~\cite{malik2021automatic}. Similar to the image/video captioning applications, the range of target semantic modality is formed by a predetermined portion of the desired NL. The extent of the covered portion of the target natural language is continuously expanding throughout the years from digits~\cite{davis1952automatic} to tens of thousands of words~\cite{chiu2018state}. 

\begin{figure}[ht]
    \centering
    \includegraphics[width=1\textwidth]{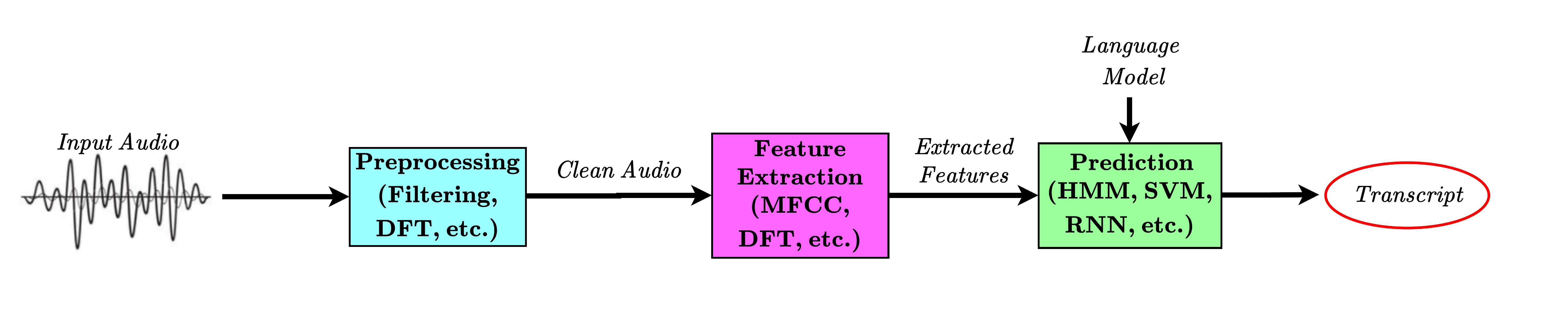}
    \caption{Automatic speech recognition pipeline. The semantic features of a preprocessed speech audio signal are extracted and fed into a predictor for text generation. The prediction module is designed for a target language model.}
    \label{fig:asr_pipe}
\end{figure}

Even though the requirements of an ASR system may vary for different applications (e.g., speaker dependency, vocabulary size, and utterance-awareness), most of the approaches conform with the process depicted in~Fig.~\ref{fig:asr_pipe}, as described in~\cite{malik2021automatic}. In Fig.~\ref{fig:asr_pipe}, the input audio signal is first preprocessed via framing, filtering, Discrete Fourier Transform (DFT), denoising, or normalization, etc. Then, using the preprocessed version of the audio signal, some spectral or temporal features can be extracted via Mel-frequency Cepstral Coefficients (MFCC)~\cite{chiu2018state, korba2008robust, collobert2016wav2letter} or  Discrete Wavelet Transform (DWT)~\cite{polikar1996wavelet, anusuya2011comparison, ranjan2010discrete}. The extracted features are passed through a prediction module that employs Hidden Markov Models (HMMs)~\cite{juang1991hidden, birkenes2009penalized}, SVMs~\cite{tang2010initial, solera2007svms, kruger2005speech}, RNNs~\cite{wang2020attention, islam2019speech, shewalkar2019performance}, or CNNs~\cite{makino2019recurrent, palaz2015analysis, park2017multiresolution, ganapathy20183}, among others, to obtain the text equivalence in the desired language restricted by a predefined vocabulary and grammar rules. More details about ASR can be found in the pertinent survey papers~\cite{malik2021automatic, besacier2014automatic}.

\subsection{State-of-the-Art Semantic and Goal-Oriented Communications}
With the tremendous advances in machine learning, particularly, on DNNs, there has been a renewed interest in semantic communications in recent years. As a recent example, joint learning and communication problem has been studied for multi-agent reinforcement learning (MARL) framework in~\cite{tung2021joint}, where the authors develop a systematic structure for collaborative MARL over noisy channels while taking into account the learning and the communication issues. In particular, a multi-agent partially observable Markov decision process (MA-POMDP) is studied in which agents can communicate with one another over a noisy channel to increase their shared long-term average reward.
The agents not only learn to collaborate through a noisy link but also figure out how to communicate \textit{effectively}, which is achieved by introducing the channel as a part of the environment dynamics. Thus, the authors of~\cite{tung2021joint} illustrate the increased performance of joint learning and communication framework compared to separate treatment of communications and the MARL problem. The \textit{effectiveness problem} addressed partly by~\cite{tung2021joint} was previously discussed in Section~\ref{sec:SemanticAndGoalOrientedInformation}. 

Another interesting study on semantic communications is performed in~\cite{xie2020deep}, where the authors focus on joint semantic channel coding by designing a deep-learning enabled semantic communication (DeepSC) system for text transmission. Unlike the traditional approaches, the objective of this study is to minimize the semantic errors in the transmitted text, as opposed to bit errors, while maximizing the system capacity. The performance metrics introduced by~\cite{xie2020deep} rigorously formulate the semantic information at a sentence-level, and the corresponding semantic errors. The proposed \textit{semantic level} communication architecture is shown to be more robust compared to the traditional \textit{technical level} approaches, especially in the low signal-to-noise ratio (SNR) regime. In~\cite{xie2020lite}, a lite version of this distributed semantic communication system for text
transmission is studied, where the authors consider an affordable IoT network by pruning the model redundancies and reducing the model size. 

Transmission of speech data for semantic communications is considered in~\cite{weng2021semantic}. The authors propose a deep learning-enabled semantic communication system for speech signals (DeepSC-S). DeepSC-S extracts essential semantic information from speech audio signals by using squeeze-and-excitation (SE) networks~\cite{hu2018squeeze}. Then, selective weighting is applied to emphasize the more essential information during DNN training.
A joint design of speech encoder/decoder and the channel encoder/decoder is performed to learn and extract the speech features that mitigate the channel distortion.

In~\cite{Strinati2020}, the authors propose using semantic and goal-oriented communications in the next generation (6G) networks, which may significantly improve the effectiveness and sustainability of the system. The authors argue that instead of focusing on the bit-level recovery of the transmitted data, one may identify the relevant meaning, which can increase the effectiveness of the network without increasing the usage of communication resources such as bandwidth and energy.
In~\cite{Kountouris2007}, authors point out possible advantages of a goal-oriented joint consideration of information generation, transmission, and usage resulting in semantic-empowered communications, which is defined by transmitting the most informative data samples, enabling the end-user to attain the most current and essential information for its goal. Especially for future massive and intelligent systems, semantics-empowered communications are expected to improve the network resource utilization. Other studies on semantic communications are presented in~\cite{guler2018semantic, shi2021new,uysal2021semantic, gunduz2020communicate}.

Inspired by the previous work on semantic information, signal processing, and communications, we introduce a formal definition of the proposed goal-oriented semantic language and the semantic signal processing framework in the next section.

\section{Proposed Goal-Oriented Semantic Language and Signal Processing Framework}
\label{sec:ProposedLanguage}

In this section, we propose and rigorously define a graph-based semantic language to represent any type of signal in terms of a well-organized and easy-to-parse structure. We then describe a typical semantic signal processing framework and its basic building blocks in detail.

\begin{figure}[ht]
	\centering
	\includegraphics[width=0.6\textwidth]{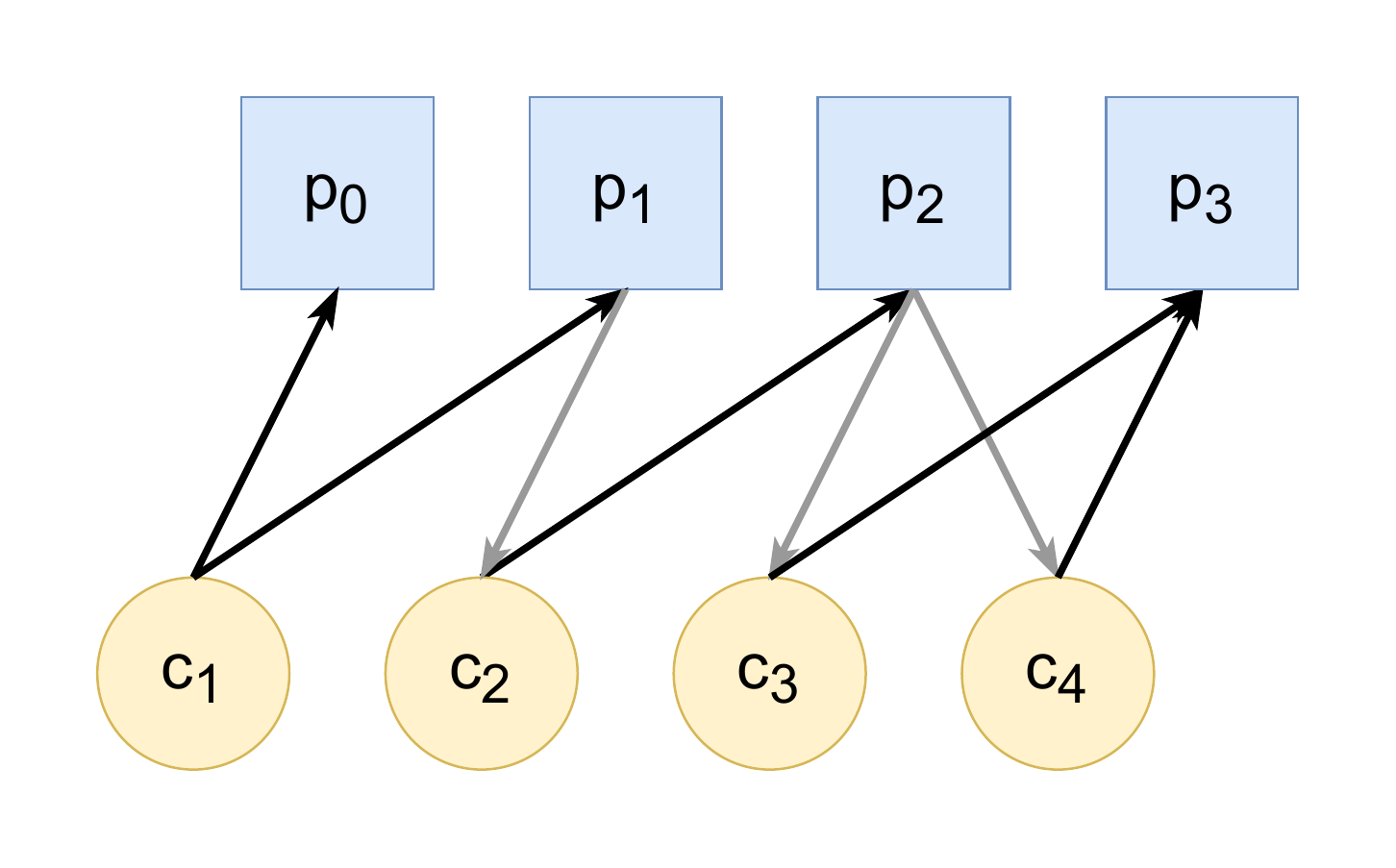}
	\caption{Proposed bipartite graph structure with signal components and predicates.}
	\label{fig:graph_example}
\end{figure}

We propose a directed bipartite graph, shown in Fig.~\ref{fig:graph_example}, as a semantic representation of a signal that consists of components with semantic connotations. In this representation, the signal components are first classified into a predefined set of classes (represented as circles in Fig.~\ref{fig:graph_example}). Then, their states (e.g., $c_1 \rightarrow p_0$) and relationship with other components (e.g., $c_1 \rightarrow p_1 \rightarrow c_2$) are labelled with appropriate predicates, again chosen from a predefined set (shown as squares in Fig.~\ref{fig:graph_example}). As illustrated in the bipartite graph structure in Fig.~\ref{fig:graph_example}, the signal components are depicted as nodes with labels $c_i$, and their semantic states/relationships are shown as nodes with labels $p_i$. Note that such a representation of raw sensor data might require a preprocessing stage that can involve transformation to an appropriate domain prior to the classification of signal components and identification of their semantic relationships. We further note that, unlike the general graph structures where extraction, pattern matching, etc., can be challenging, especially as the graphs get more complex, the proposed hierarchical bipartite structure allows for a complete but much simpler semantic description of a signal, and therefore, it is much simpler to generate and operate on.

The proposed semantic description of a signal (e.g., a sensor output) includes a multi-graph description and corresponding attribute sets for each node in each graph. This structure of information in a signal enables easy parsing for goal-oriented storage, processing, and communications. In the following subsections, the multi-graph description, the companion attribute sets, and the goal-filtering process are presented in detail.

\subsection{Multi-Graph Semantic Description}
The node sets for components and predicates in the proposed bipartite graph structure are defined as follows:
\begin{align}
C &= \{c_1, c_2, \ldots, c_{N_c}\}, \\
P &= \{p_0,p_1, p_2, \ldots, p_{N_p-1}\},
\label{eq:CP}
\end{align}
where $N_c$ and $N_p$ are the numbers of component and predicate classes in the graph language, respectively. Note that a separate $p_0$ class is defined as the first element of the predicate classes. If a detected component has no other semantic connection to another component, then that component is connected to the predicate $p_0$ to form a bipartite graph. This is to ensure that a detected component never becomes an isolated node in a graph and always has a non-zero edge weight in the corresponding biadjacency matrix representation. Moreover, the predicate $p_0$ and the similarly defined predicates in the set $P$ can be considered as describing states of each component without describing any relation to the other components. Note that state describing predicates like $p_0$ have a single directed connection to a component, while relation predicates can have different polarities of connections depending on their definition (e.g., $c_i \rightarrow p_j \rightarrow c_k$ or $c_i \rightarrow p_j \leftarrow c_k$), as shown in Fig.~\ref{fig:graph_example}.

Using the class definitions from~\eqref{eq:CP}, a \textit{multi-graph class description} of a signal at time $t$ can be defined as a union of connected bipartite graphs as
\begin{equation}
\mathcal{S}_t=\{S_{t,1}, S_{t,2}, \ldots, S_{t,N}\},
\label{eq:multigraphclass}
\end{equation}
where $N$ is the number of \textit{class atomic graphs} representing the disjoint bipartite graphs in the data, and 
\begin{equation}
S_{t,i} = ( C,P,E_{t,i} ),
\label{eq:Sti}
\end{equation}
is a connected directed bipartite graph with node sets $(C,P)$ and edges (connections) at time $t$, $E_{t,i}$. Due to the bipartite nature of the proposed graph formalism, only connections from the component classes to the predicate classes and vice versa are allowed. Therefore, each connected graph $S_{t,i}$ in~\eqref{eq:multigraphclass} can also be represented by the following biadjacency matrix:
\begin{equation}
\boldsymbol{S}_{t,i} = \begin{bmatrix}
\boldsymbol{0} & \boldsymbol{S}^{CP}_{t,i} \\
\boldsymbol{S}^{PC}_{t,i} & \boldsymbol{0}
\end{bmatrix},
\label{eq:Adjacency}
\end{equation}
where $\boldsymbol{S}^{CP}_{t,i}$ and $\boldsymbol{S}^{PC}_{t,i}$ represent the connections from the components to the predicates and vice versa, respectively. In other words,
\begin{equation}
\boldsymbol{S}^{CP}_{t,i}[m,n] =
\begin{cases}
   1,        & \text{if } \exists \text{ connection from } c_m \text{ to } p_n, \\
   0,        & \text{otherwise,}
\end{cases}
\label{eq:SCP}
\end{equation}
\begin{equation}
\boldsymbol{S}^{PC}_{t,i}[m,n] =
\begin{cases}
   1,        & \text{if } \exists \text{ connection from } p_m \text{ to } c_n,  \\
   0,        & \text{otherwise.}
\end{cases}
\label{eq:SPC}
\end{equation}

Note that in the multi-graph definition given in~\eqref{eq:multigraphclass},\eqref{eq:Sti} the nodes $C$ and $P$ are class definitions that may have multiple detections of their particular members in a signal. This class-only representation is a higher level of abstraction of the signal, and it is useful when the goals are defined similarly over classes and not over individual instances of components or predicates. For the case when the goals are defined over individual members, we introduce a lower level of abstraction again in the form of bipartite graphs to represent only the detected instances from classes $C$ and $P$. For the $i$-th atomic class graph $S_{t,i}$ at time $t$, we define the \textit{detected component} and \textit{detected predicate} sets as follows:
\begin{equation}
C^D_{t,i} = \{(c_j,k) \mid k\in [1,N_{c_j}^{t,i}] \text{ and } \exists m \text{ } \boldsymbol{S}^{CP}_{t,i}[j,m] = 1 \text{ or } \boldsymbol{S}^{PC}_{t,i}[m,j] = 1\},
\label{eq:Dti_C}
\end{equation}
\begin{equation}
P^D_{t,i} = \{(p_j,k) \mid k\in [1,N_{p_j}^{t,i}] \text{ and } \exists m \text{ } \boldsymbol{S}^{CP}_{t,i}[m,j] = 1 \text{ or } \boldsymbol{S}^{PC}_{t,i}[j,m] = 1\},
\label{eq:Dti_P}
\end{equation}
where $N_{c_j}^{t,i}$ and $N_{p_j}^{t,i}$ are the number of detected instances of component $c_j$ and predicate $p_j$ classes in the atomic graph $S_{t,i}$, and $k$ is the unique detection index for each class instance. Using the \textit{detected} component and predicate sets, we can now define a \textit{multi-graph instance representation} as 
\begin{equation}
\mathcal{D}_t=\{D_{t,1}, D_{t,2}, \ldots, D_{t,N}\},
\label{eq:multigraphdetections}
\end{equation}
\begin{equation}
D_{t,i} = ( C^D_{t,i},P^D_{t,i},E^D_{t,i} ),
\label{eq:Dti}
\end{equation}
where $D_{t,i}$ are the \textit{instance atomic graphs}. The corresponding connection lists $E^D_{t,i}$ are defined as triplets indicating a connection from $(c_a,k_a)$ to $(p_b,k_b)$ to $(c_c,k_c)$ as
\begin{equation}
E^D_{t,i} = \left\{ \left( (c_a,k_a), (p_b,k_b), (c_c,k_c) \right) \right\}.
\label{eq:ED_ti}
\end{equation}
Note that in~\eqref{eq:ED_ti}, the last entry of each triplet can be an empty element, as there can be a single connection from a component to a predicate, but not vice versa. This is in line with the definition of the $p_0$ predicate in~\eqref{eq:CP}. The biadjacency matrix representations $\boldsymbol{D}_{t,i}$, $\boldsymbol{D}^{CP}_{t,i}$, $\boldsymbol{D}^{PC}_{t,i}$ for $D_{t,i}$ can be defined similarly to~\eqref{eq:Adjacency}--\eqref{eq:SPC}. 

As noted before, the multi-graph sets $\mathcal{S}_t$ and $\mathcal{D}_t$ provide a hierarchical bipartite representation that is much simpler to generate and operate on than a general single graph structure. In typical machine-type applications, the number of components and predicates scale according to the complexity and the processing capabilities of the sensors or devices. For example, a simple IoT sensor may not have more than a few components and predicates, which---combined with the multi-graph representations---makes pattern matching and goal-filtering operations extremely simple implementations of the available methods in the literature~\cite{fan2010, fan2013, serratosa2014, sun2005}.

\subsection{Attribute Sets in Multi-Graph Semantic Descriptions}
The class definitions and detected class instances do not necessarily constitute a complete semantic description of a signal. As such, to fully capture additional properties and features, we use associated attribute sets for each component or predicate node that is present in each atomic graph $D_{t,i}$. For each $D_{t,i}$ in the multi-graph description of the signal, we define an attribute superset $A_{t,i}$ as
\begin{equation}
    A_{t,i} = \{\Theta_{t,i}(n_j,k) \mid (n_j,k) \in {C^D_{t,i} \cup P^D_{t,i}}\},
\label{eq:Ati}
\end{equation}
where $(n_j, k)$ corresponds to a node (component or predicate) and instance index pair. The operator $\Theta_{t,i}(.)$ is defined as the multi-level attribute set of its argument nodes as 
\begin{equation}
\Theta_{t,i}(n_j,k) = \{\boldsymbol{\theta}_{t,i}^{(1)}(n_j,k), \boldsymbol{\theta}^{(2)}_{t,i}(n_j,k), \ldots, \boldsymbol{\theta}^{(L_{n_j})}_{t,i}(n_j,k) \},
\label{eq:theta_ti}
\end{equation}
where $L_{n_j}$ is defined as the number of levels in attributes for node-$j$, which may be defined separately for each type of component or predicate, or kept the same for all nodes for simplicity. The introduction of levels of attributes enables a hierarchical organization of attributes from a simpler to a more complex representation, which in turn makes goal-oriented filtering easier and more intuitive. The exact definition of~\eqref{eq:theta_ti} depends on the application, the nature of internal/external goals, and the type of signal processing to be performed on the obtained semantic information. Concrete examples of how attribute sets can be defined are given via several use-cases in Section~\ref{sec:ApplicationsOfSPExtraction}.

The companion attribute set for the object multi-graph description $\mathcal{D}_t$ is defined as 
\begin{equation}
\mathcal{A}_t = \{A_{t,1}, A_{t,2}, \ldots, A_{t,N}\}.
\label{eq:A_t}
\end{equation}
Finally, the full semantic description of the sensor data at time $t$ is defined as
\begin{equation}
Y_t = (\mathcal{S}_t, \mathcal{D}_t, \mathcal{A}_t),
\label{eq:Y}
\end{equation}
where $\mathcal{S}_t$, $\mathcal{D}_t$, and $\mathcal{A}_t$  correspond to the multi-graph class description, multi-graph instance description, and the corresponding attribute sets, respectively.

\subsection{Semantic Goals and Goal-Oriented Filtering}

We now define a goal operator using the proposed semantic framework to parse and filter the extracted semantic information from a signal. Note that beyond this point, the time index subscript $t$ is dropped for simplicity. A goal-filtering operator $\mathcal{G}$ acting on semantic description of the sensor data $Y$ is defined as 
\begin{gather}
\mathcal{G}(G^S, G^D, \boldsymbol{l}_g, \boldsymbol{l}_a)\{Y\} = \overset{\sim}{Y},
\label{eq:Goal}\\
\overset{\sim}{Y} = (\mathcal{\overset{\sim} S}, \mathcal{\overset{\sim} D}, \mathcal{\overset{\sim} A}),
\label{eq:goalFilteredY}
\end{gather}
where the terms inside the parentheses in~\eqref{eq:Goal} are the arguments of the goal operator $\mathcal{G}$. $\overset{\sim} Y$ is the filtered semantic description, and $(\mathcal{\overset{\sim} S}, \mathcal{\overset{\sim} D}, \mathcal{\overset{\sim} A})$ correspond to the goal-filtered versions of $(\mathcal{S}, \mathcal{D}, \mathcal{A})$, respectively. $G^S = (C,P,E^S)$ and $G^D = (C^D,P^D,E^D)$ are the bipartite graphs that correspond to the queried patterns in the class-level graph $\mathcal{S}$ and the instance-level graph $\mathcal{D}$, respectively. The last two arguments are the graph complexity vector $\boldsymbol{l}_g$ and attribute complexity vector $\boldsymbol{l}_a$ that define the desired level of complexity for the semantic output in terms of the multi-graph representation and the attribute sets, respectively. 

The distinction between class-level and instance-level queries for $\mathcal{G}$ enables the goals to be defined at different levels of abstraction; namely, \textit{global goals} with a typical interest on class-level information, and \textit{local goals} with a typical interest on detected instances and their relations. Hypothetically, the global goals can be defined first and change slowly over time, while local goals can be defined following the detection of a component of interest and change more rapidly in time.

The graph complexity vector $\boldsymbol{l}_g = \{l_g \in \mathbb{N}\}$ is a vector of natural numbers $l_g$, defining an $l_g$-hop neighborhood around components of each queried graph pattern in $\mathcal{G}$. This allows goals to search for a specific pattern with $G^S$ or $G^D$, while being interested in other relationships among the components in those specific patterns without explicitly defining them.

\begin{figure}[ht]
\centering
\begin{subfigure}[b]{.24\textwidth}
  \centering
  \includegraphics[width=1\linewidth]{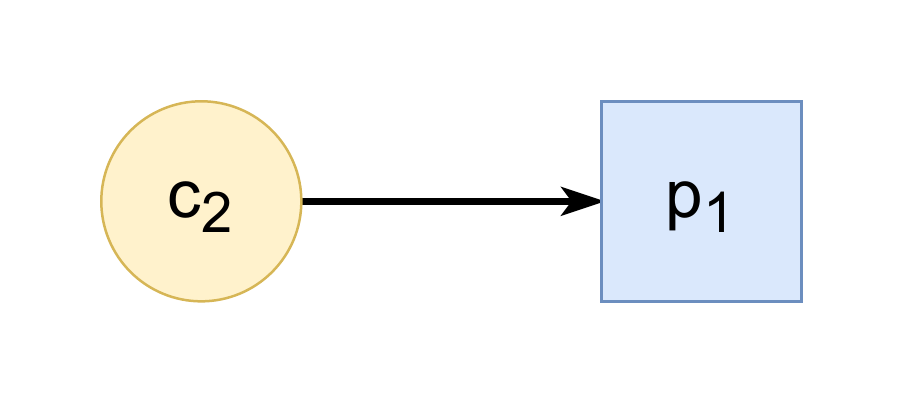}
  \caption{Goal Pattern}
  \label{fig:lHop_a}
\end{subfigure}%
\begin{subfigure}[b]{.33\textwidth}
  \centering
  \includegraphics[width=1\linewidth]{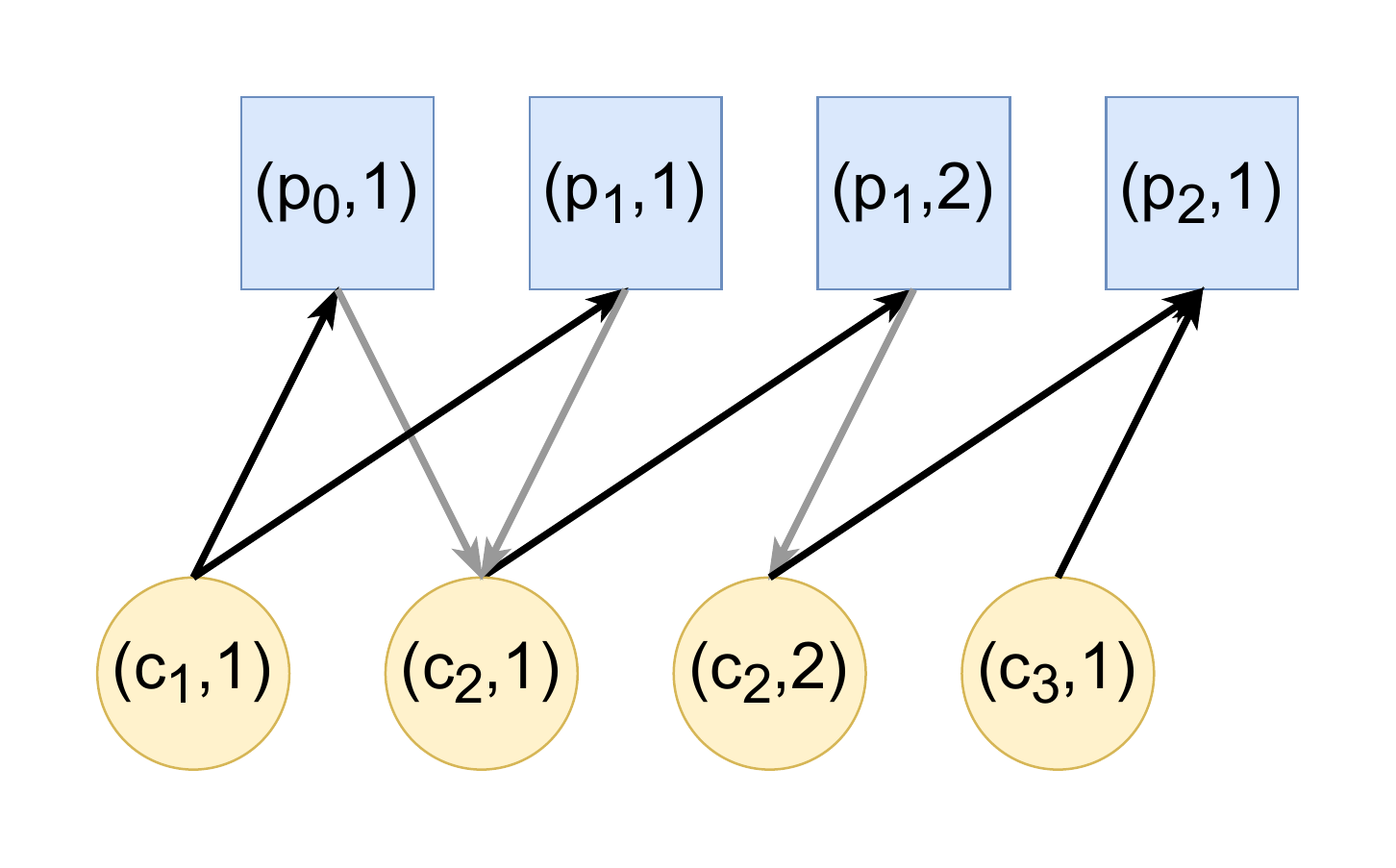}
  \caption{Semantic output $\mathcal{D}$}
  \label{fig:lHop_b}
\end{subfigure}
\begin{subfigure}[b]{.4\textwidth}
  \centering
  \includegraphics[width=1\linewidth]{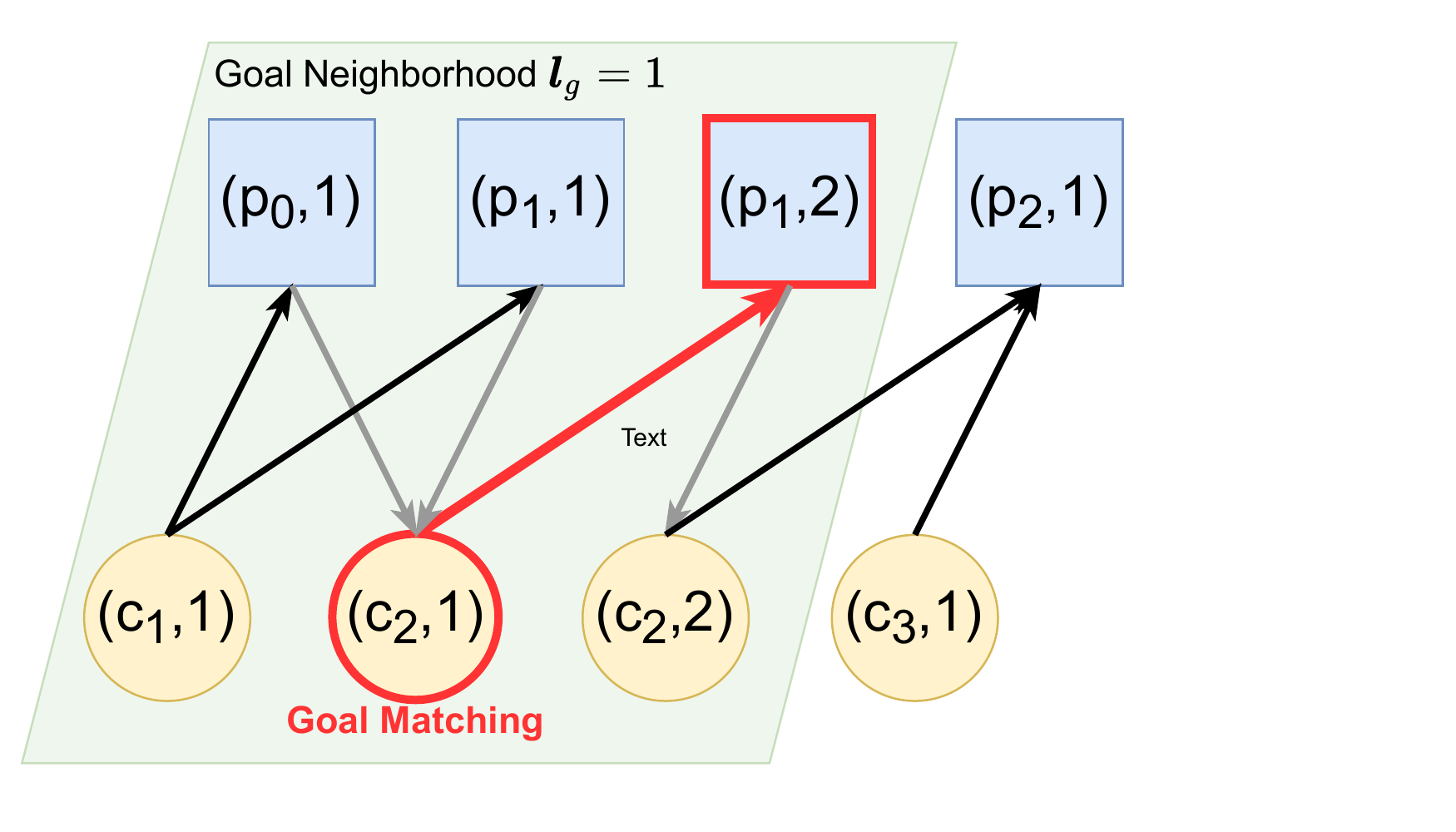}
  \caption{Goal-filtered output}
  \label{fig:lHop_c}
\end{subfigure}
\caption{Goal filtering example. Note that $l_g = 1$ covers neighboring components that are connected to the matched pattern by a single predicate path.}
\label{fig:lHop}
\end{figure}

Using $G^S$, $G^D$, and $\boldsymbol{l}_g$, the output semantic graphs $\mathcal{\overset{\sim} S}$, $\mathcal{\overset{\sim} D}$ can be defined as
\begin{align}
\mathcal{\overset{\sim} S} &= \{\overset{\sim} S_i \mid \overset{\sim} S_i \subset S_i, G^S \subset \overset{\sim} S_i \text{ and } \overset{\sim} S_i \text{ includes } l_g \text{-hop neighborhood of } G^S \text{ within } S_i\},
\label{eq:goalFilteredS} \\
\mathcal{\overset{\sim} D} &= \{\overset{\sim} D_i \mid \overset{\sim} D_i \subset D_i, G^D \subset \overset{\sim} D_i \text{ and } \overset{\sim} D_i \text{ includes } l_g \text{-hop neighborhood of } G^D \text{ within } D_i\}.
\label{eq:goalFilteredD}
\end{align}
Note that the specific pattern matching and $l$-hop neighborhood search algorithms employed to generate~\eqref{eq:goalFilteredS} and \eqref{eq:goalFilteredD} can be chosen from many available alternatives in the literature~\cite{fan2010, fan2013, serratosa2014, sun2005}. However, we emphasize again that the proposed multi-graph structure allows for a simple and complete description of a signal, compared to a general single-graph structure. Therefore, the computational complexity of the pattern matching and search algorithms is relatively low and they scale according to the complexity of the sensor or device. An illustration of the goal filtering with a given graph complexity of $l_g=1$ is given in Fig.~\ref{fig:lHop}.
A simple class-level goal ($c_2 \rightarrow p_1$) is applied to the instance representation in Fig.~\ref{fig:lHop_b}, with $l_g = 1$. In Fig.~\ref{fig:lHop_c}, the exact matching of the goal is shown with red edges, and the requested neighborhood around the matched pattern is shown inside the shaded parallelogram. Note that the graph complexity parameter is defined over neighboring components. Therefore, the output pattern includes components separated by at most $l_g$ predicates with the exact-matched pattern.

After defining the filtered graph outputs $\mathcal{\overset{\sim} S}$ and $\mathcal{\overset{\sim} D}$, their corresponding attribute sets are filtered with the attribute complexity vector $\boldsymbol{l}_a = \{l_a \in \mathbb{N}\}$ that requests only the first $l_a$ levels within the hierarchy of attribute sets of each instance in the graphs as
\begin{equation}
\mathcal{\overset{\sim} A} = \{\overset{\sim} \Theta_i \in \overset{\sim} D_i \mid \Theta_i = \{\boldsymbol{\theta}^{(1)}, \boldsymbol{\theta}^{(2)}, \ldots, \boldsymbol{\theta}^{(L_{min})}\} \},
\label{eq:goalFilteredA}
\end{equation}
where $L_{min} = min(l_a, L)$ is the minimum of the desired and available levels of complexity.

\newpage
\subsection{Goal-Oriented Semantic Signal Processing Framework}

\begin{figure}[t]
\centering
  \includegraphics[trim={0 2cm 0 2cm},clip, width=1\linewidth]{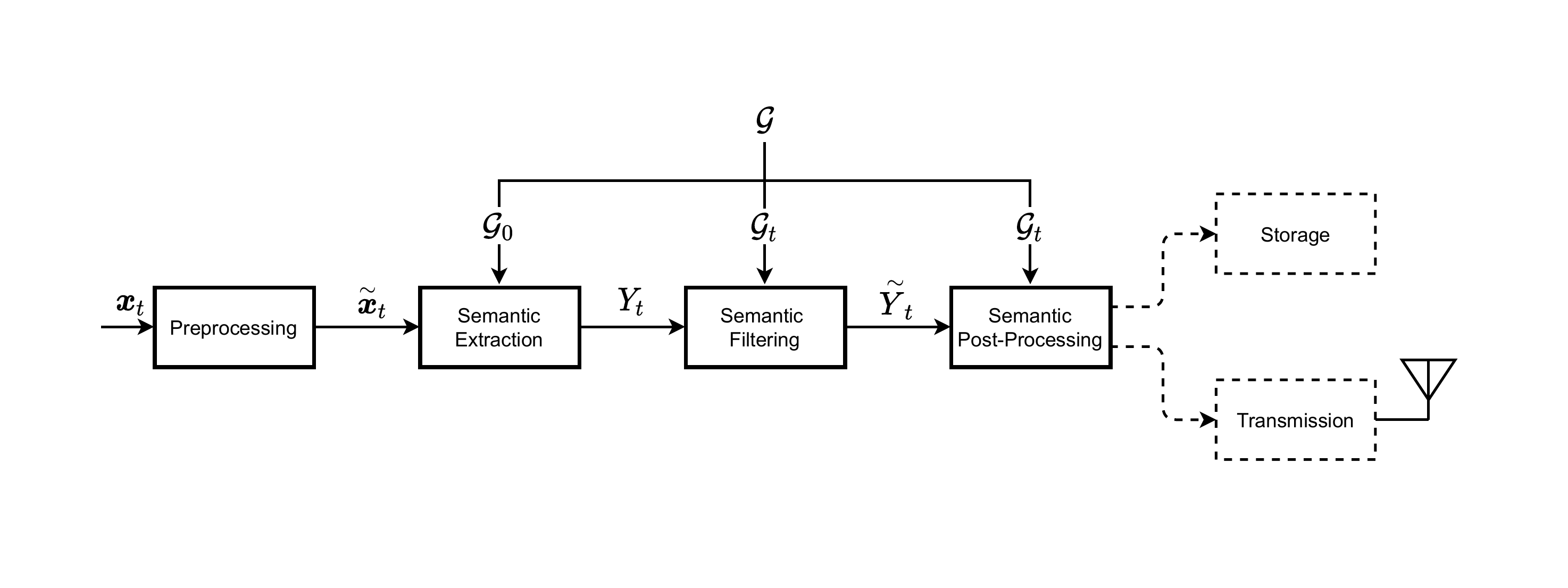}
  \caption{The proposed goal-oriented semantic signal processing framework.}
  \label{fig:SSPframework}
\end{figure}

Based on the goal-oriented semantic language framework defined above, the proposed semantic signal processing framework in its generic form is illustrated in Fig.~\ref{fig:SSPframework}. In the following, we give brief descriptions for each conceptual block in Fig.~\ref{fig:SSPframework}, before discussing its potential implementations at the goal and the hardware levels. 

The proposed goal-oriented semantic signal processing framework consists of the conceptual blocks of preprocessing, semantic extraction, semantic filtering, semantic post-processing, and storage or transmission, depending on the application. In the \textit{preprocessing} block, following a prefiltering for noise and interference reduction, the input signal $\boldsymbol{x}_t$ is transformed to an appropriate domain where detection and classification of components and predicates can be performed efficiently. For one-dimensional (1-D) input signals, decomposition of the data for graphical representation may not be as self-evident as in the case for higher-dimensional input data. In such cases, commonly used domain transformation approaches such as the time-frequency and the time-scale domain processing~\cite{cohen2012time, chan1994wavelet, torrence1998practical}, can be used to segment and detect distinct signal components via pattern recognition and computer vision techniques ~\cite{tufekci2005}. For image and video applications, transformation to another domain may not be necessary, as component detection and classification in these domains can be efficiently implemented on the input signals directly. 

The \textit{semantic extraction} block is where the multi-graph description and corresponding attribute sets are generated from the input data. In the case of semantic representations of 2-D signals such as images or 3-D data such as video streams of consecutive slices of MRI data, components or objects can be first identified by image segmentation. For this purpose, scene graph extractors can be used~\cite{Li_2017_ICCV, Xu_2017_CVPR, Li_2018_ECCV, Yang_2018_ECCV, Gu_2019_CVPR, Wang_2019_CVPR}. The predicates can be identified based on the spatial relationships among the objects in an image, or their spatio-temporal relationships in a video stream. Note that a global goal $\mathcal{G}_0$ can be incorporated at this point. If a constant or slowly changing goal is defined either internally, or it is received from an external agent, the semantic extraction algorithm can adapt accordingly. For example, if the global goal is only related to a small subset of potential components that can be extracted, the algorithm efficiency can be increased by limiting the output classes of the classification algorithm implemented as a DNN. 

In the next block, \textit{semantic filtering} operations are performed. Once the scene descriptions $Y_t$ are generated, local and time-varying goals $\mathcal{G}_t$ can be applied to filter \textit{interesting} semantic information within $Y_t$. In a typical application, local goals $\mathcal{G}_t$ are either generated internally via decision-making algorithms or are received from external agents. This block typically employs graph signal processing algorithms to match and extract goal queries from the graph-based representation of the signal. Note that the semantic output $Y_t$, generated by the proposed language yields relatively simple graphs due to their bipartite and goal-oriented definitions; hence, the computational complexity of any graph operation is considered to be more tractable compared to general graph structures. 

The \textit{semantic post-processing} block is a conceptual block where further processing of the extracted and goal-filtered semantic data can be performed. Note that at this point, the semantic data are strictly organized and only include goal-filtered components and their corresponding attributes. These attributes allow the implementation of a wide spectrum of processing tasks. For instance, since the location information of the components is a probable part of the attribute sets, it can allow for tracking of components or groups of components by using standard tracking techniques. Furthermore, the detailed attributes of components such as wavelet representation of its time-scale characterization facilitate the application of more specific signal processing algorithms on the components of interest such as further denoising or statistical modeling. The semantic post-processing block can also incorporate local goals $\mathcal{G}_t$, to schedule transmission and storage operations according to the level of desired semantic information.

Finally, the \textit{storage} and the \textit{transmission} are tentative blocks that can be added to the framework depending on the application. In either case, an encoding scheme is utilized to compress the goal-filtered graphs and attributes. Then, the compressed semantic information can either be stored internally or passed forward to a transmission block where further channel coding operations can be performed prior to transmission.

The semantic signal processing framework described above can be adopted in the next generation of devices and communication networks. Although the types and implementation strategies of goals are previously discussed, the generation and dissemination of semantic goals warrant further discussion. Typically, global goals $\mathcal{G}_0$ for a semantic device can be defined either by human operators at the language level or by manufacturers at a hardware level. Alternatively, in a hierarchical network such as a Smart City application~\cite{hall2000vision, su2011smart}, global goals can be defined at the highest level, and then they can be disseminated by mapping the global goals to the low-level device language by intermediate devices such as base stations. On the other hand, the generation of rapidly varying local goals $\mathcal{G}_0$ requires decision-making algorithms to be implemented at intermediate levels to optimize throughput and goal implementation. Possible approaches to the goal generation and dissemination problems address the so-called \textit{effectiveness problem} described in Section~\ref{sec:SemanticAndGoalOrientedInformation}, and constitute a future research direction for goal-oriented semantic signal processing. 

The adoption of the proposed semantic signal processing framework in Fig.~\ref{fig:SSPframework} will also have repercussions at a hardware level. The preprocessing and the semantic extraction stages can be implemented as part of a sensor subsystem designed to acquire signals of the desired modality such as audio or video signals. Such a sensor subsystem must provide its output in the proposed semantic structure so that the rest of the semantic processing chain can be implemented on its output. Therefore, the proposed semantic signal processing framework may form a basis for sensor standardization efforts for future applications of semantic signal processing and communication systems.

In the following sections, detailed examples of generating and processing semantic information within the proposed framework as well as possible encoding/transmission strategies are discussed in detail. 
 
\section{Applications of Semantic Signal Extraction with the Proposed Framework}
\label{sec:ApplicationsOfSPExtraction}

In this section, we provide concrete examples of the proposed goal-oriented semantic signal processing framework described above. First, we present a real-time computer vision application for a video stream input. Next, in contrast to the higher dimensional nature of video signals, the proposed language is implemented on a generic 1-D sensor output on which we also implement a novel and efficient non-uniform sampling strategy. We then provide a brief discussion on the potential applications of the proposed framework on heterogeneous multi-sensor networks. It is important to note that these case studies are provided only as a proof-concept of the generality of the proposed framework, which has a great potential to be customized for many different applications outside of what is presented here.   

\subsection{Real-Time Semantic Processing of Video-Stream Data}
\label{subsec:caseStudy_video}

\begin{figure}[t]
    \centering
    \includegraphics[width = 0.9\textwidth]{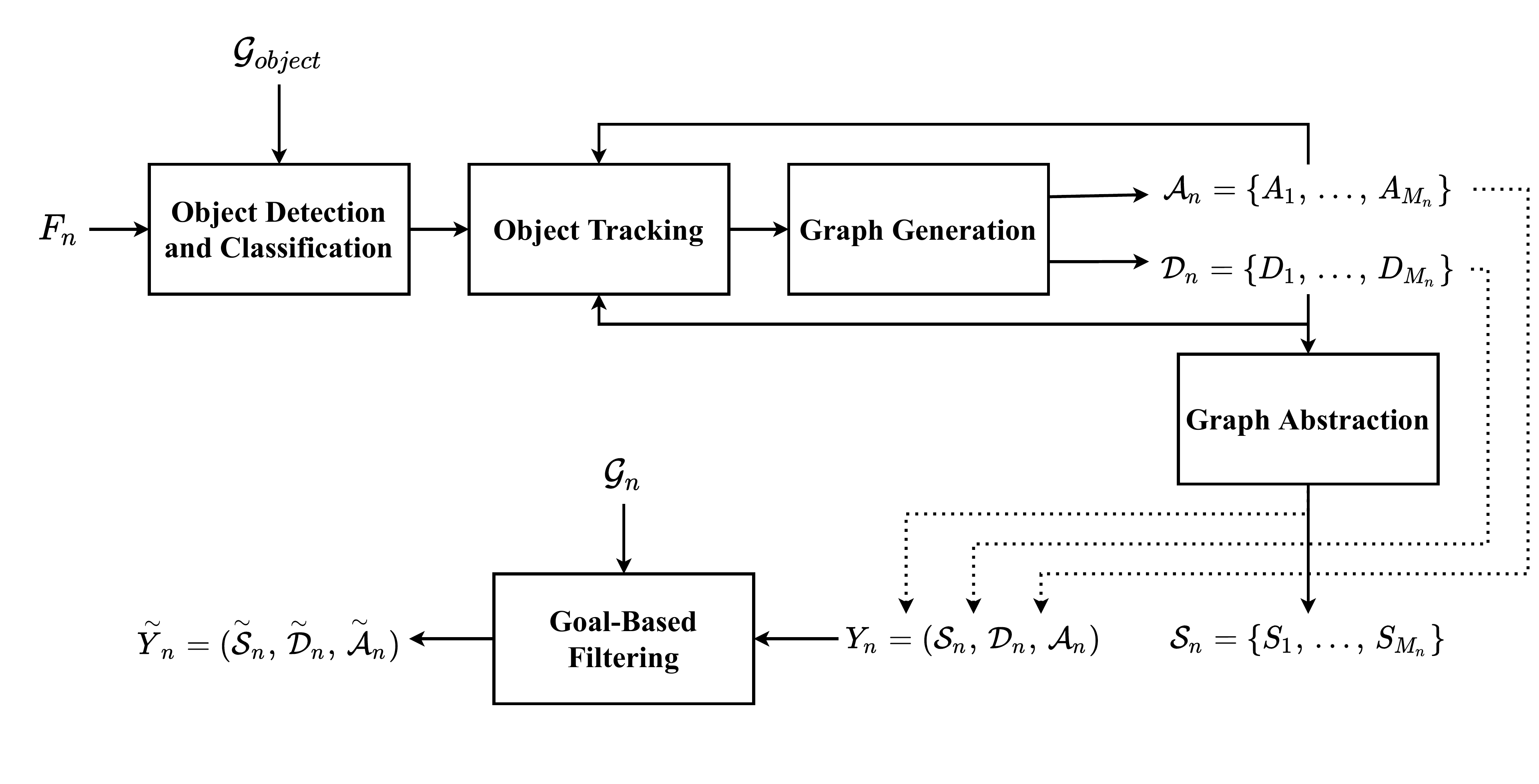}
    \caption{Semantic processing architecture for real-time video signals. Objects are detected and classified at each frame $F_n$ by YOLOv4-CSP~\cite{scaledyolov4}. For temporal extension, detected objects are tracked across the frames using DeepSORT~\cite{deepsort}. $\mathcal{S}_n$ is obtained by performing a summarization of $\mathcal{D}_n$, i.e., simply drawing a crude version of $\mathcal{S}_n$ by only keeping the class information of the nodes. Finally, the complete semantic description $Y_n$ consists of the generated graph $\mathcal{D}_n$, its abstract version $\mathcal{S}_n$ and its attribute set $\mathcal{A}_n$. The resulting semantic description $Y_n$ is then filtered via local goals $\mathcal{G}_n$.}
    \label{fig:structure}
\end{figure}

As a first application scenario, we demonstrate the use of the proposed semantic signal processing framework on real-time video signals via the architecture shown in Fig.~\ref{fig:structure}. An input sequence of frames denoted as $F_n$, with $n$ as the discrete time index, is fed into a detection and classification block, and the detected entities in each frame are filtered out via a global goal $\mathcal{G}_{object}$. The remaining \textit{globally-interesting} objects and extracted predicates among them are tracked across frames using their position and velocity information provided by the object tracking block. Based on the tracks, a \textit{multi-graph instance representation} $\mathcal{D}_n = \left \{ D_1, \ldots, D_{M_n} \right \}$ composed of $M_n$ disconnected subgraphs and a corresponding attribute set $\mathcal{A}_n = \left \{ A_1, \, \ldots, \, A_{M_n} \right \}$ are generated. The detailed semantic representation for each frame is updated in accordance with the tracking unit's updates. In this specific application, the \textit{multi-graph class representation} $\mathcal{S}_n$ is constructed via a post-processing step as a higher-level abstraction and summary of $\mathcal{D}_n$ in the Graph Abstraction block. Finally, the complete semantic description $Y_n = (\mathcal{S}_n, \, \mathcal{D}_n, \, \mathcal{A}_n)$ is generated. The local goals $\mathcal{G}_n$ prune the semantic description $Y_n$ to generate the goal-oriented semantic description $\overset{\sim}{Y}_n$. Note that, in this example, we are considering fast algorithms that are available for real-time implementation. Alternatively, for delay-tolerant (offline) applications, scene graph generator models~\cite{agarwal2020visual} can be used directly in place of the first three blocks. A detailed discussion of each block and the algorithms employed are discussed in the following.

\begin{figure}[ht]
    \centering
    \includegraphics[width=0.45\textwidth]{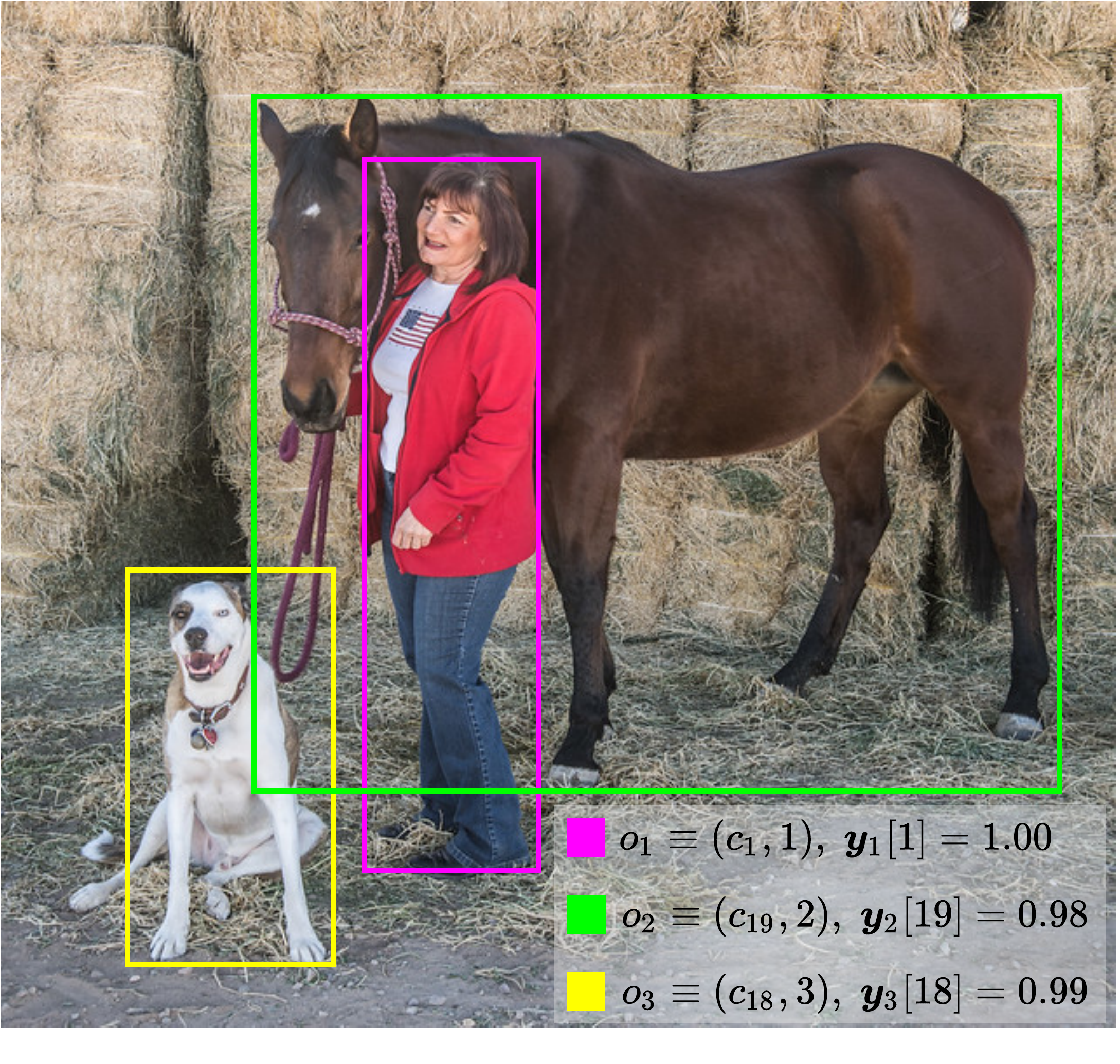}
    \caption{An example of detected objects by YOLOv4-CSP model. Predicted bounding boxes $\boldsymbol{b}_i$ are shown and maximum element of class confidence scores $\text{max}\{ \boldsymbol{y}_i[c] \}$ are listed.}
    \label{fig:objectdetect}
\end{figure}

Taking into account the real-time processing requirements, we use a pretrained YOLOv4-CSP~\cite{scaledyolov4} model on COCO dataset~\cite{mscoco} for our object detection module since it achieves greater accuracy compared to the other models~\cite{efficientdet, spinenet} under the latency requirements of a typical video signal with 24 frames-per-second. Consequently, our semantic graph language contains $N_C = 80$ semantic component classes, i.e., $C = \left \{ c_1, \ldots , c_{80} \right \}$, corresponding to the object categories in the COCO dataset ($c_1 :$ person, $c_2 :$ bike, $c_{18} :$ dog, etc.). Omitting the time index $n$ for simplicity, YOLOv4-CSP provides each detected object in the form $o_i = ( \boldsymbol{b}_i, \boldsymbol{y}_i )$, where $\boldsymbol{b}_i \in \mathbb{R}^4$ contains the position of the object according to its bounding box and $\boldsymbol{y}_i \in \mathbb{R}^{N_C}$ contains the category confidence scores. Here, we determine the classes of detected semantic instances via $\boldsymbol{y}_i$ and represent each object in the form $o_i \equiv (c_a, i)$. This representation simply indicates that the object $o_i$ is an instance of component class $c_a$ with a unique identifier index $i$ where
\begin{equation}
    a=\text{arg}\underset{c=1,\ldots,80}{\text{max}} \, \boldsymbol{y}_i[c].
\end{equation}
To put it another way, $(c_a, i)$ is the pointer of object $o_i$ so that $C^{D}$ as defined in~\eqref{eq:Dti_C} stores the detected objects as
\begin{equation}
    C^{D} = \left \{ (c_a, i) \equiv o_i \, \bigg\vert \, c_a \in C \right \}.
\end{equation}
An example output of object detection/classification module is shown in Fig.~\ref{fig:objectdetect}. The three detected objects belong to the semantic component classes $c_1 :$ person, $c_{18} :$ dog, and $c_{19} :$ horse, i.e., the set of detected objects is simply $C^{D} = \left \{ o_1 \equiv (c_1, 1), \, o_2 \equiv (c_{19}, 2), \, o_3 \equiv (c_{18}, 3) \right \}$.

In this example, we define our predicate set as $P = \left \{p_0, p_1, p_2 \right \}$ where $p_0:$ null, $p_1:$ moving-together, and $p_2:$ conjunct. The \textit{null} predicate $p_0$ is connected to isolated objects by default (see the discussion on avoiding non-zero adjacency matrices in Section~\ref{sec:ProposedLanguage}). The predicate $p_1:$ \textit{moving-together} is placed between adjacent objects moving towards to the same direction with similar velocities, whereas $p_2:$ \textit{conjunct} predicate is placed between objects which have significantly overlapping bounding boxes and possibly possessive relationships, e.g., a person with their bike or bag. Note that all predicates are defined in such a way that all the edges start from the detected components and end at the predicates. This is due to the symmetric nature of the predicates in this specific application. Moreover, it should be noted that these predicates are only defined due to the simplicity of their inference, and the predicate set can be enriched further by extracting other relationships among the detected instances. For instance, if a \textit{person} and a \textit{bicycle} are moving towards the same direction with a significant overlap between their bounding boxes, then the predicate between these objects can be set to \textit{riding}. With a similar reasoning for objects \textit{person} and \textit{car}, a \textit{driving} predicate can be generated. Analogous to the objects, we denote the predicates as $e_k \equiv (p_w, k)$ with $k$ being a unique identifier index. Predicates $e_k$ are placed between some objects $o_i \equiv (c_a, i)$ and $o_j \equiv (c_b, j)$, where the set $P^{D}$ as defined in~\eqref{eq:Dti_P} stores the detected predicates as
\begin{equation}
    P^{D} = \left \{ (p_w, k) \equiv e_k \, \bigg\vert \, p_w \in P \right \}.  
\end{equation}

\begin{figure}[t]
    \centering
    \includegraphics[width=0.9\textwidth]{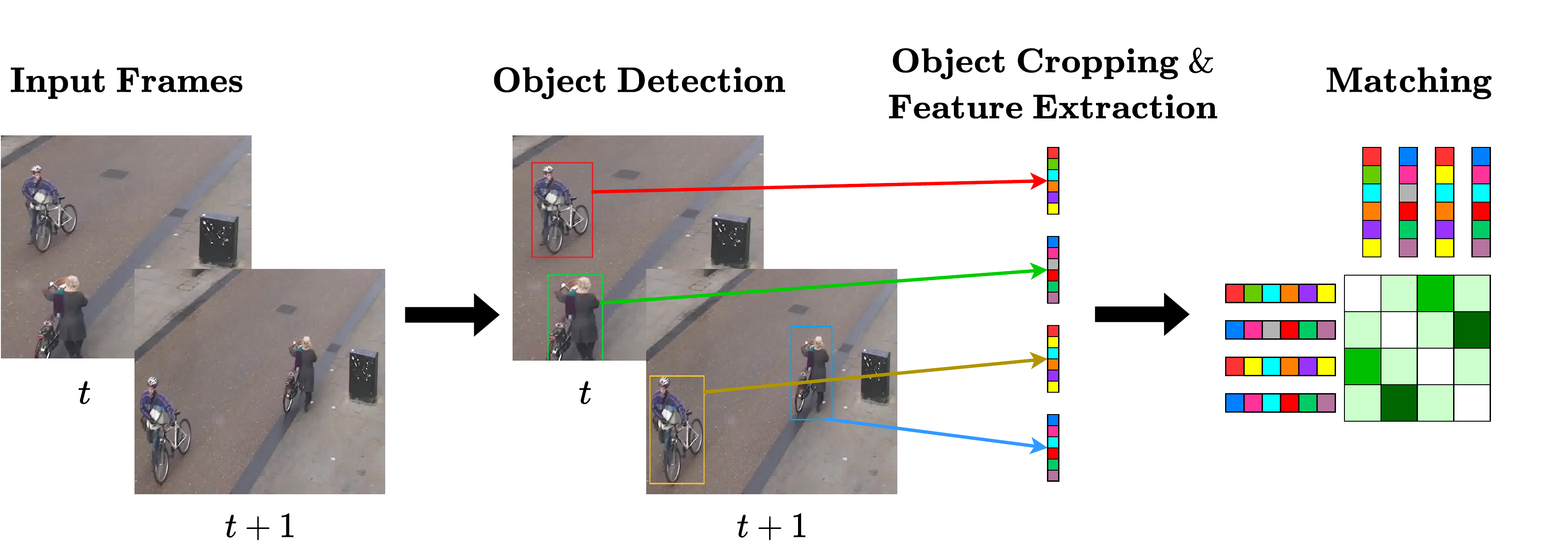}
    \caption{DeepSORT~\cite{deepsort} compares feature vectors of detected objects and tracks them for track assignment. Detected objects are cropped and fed into a CNN for feature extraction. Features are utilized as additional information for track/detection matching.}
    \label{fig:deepsort_simple}
\end{figure}

\begin{figure}[t]
    \centering
    \includegraphics[width=0.9\textwidth]{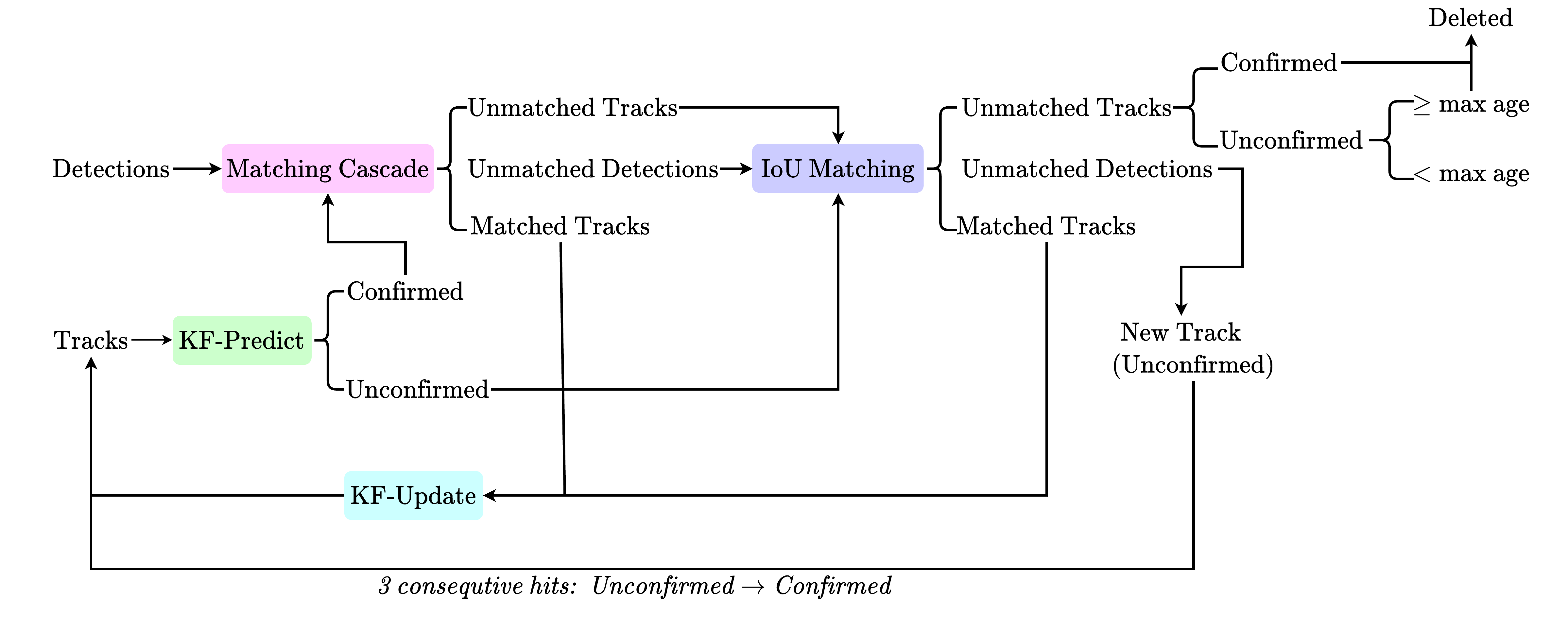}
    \caption{The DeepSORT tracking algorithm. At each frame, states of tracked objects are updated via a Kalman filter. Matching Cascade module matches detections with existing tracks using both extracted features and predictions of the Kalman filter. Specifically, detections and tracks are matched using Hungarian assignment~\cite{jonker1986improving}. The cost matrix of the Hungarian algorithm consists of the smallest cosine distances between extracted feature vectors and Mahalanobis distances between bounding boxes (see~\cite{deepsort} for details).}
    \label{fig:deepsort_algo}
\end{figure}

To identify the temporal and spatial changes within frames we perform tracking on the detected objects. Just like the object detection module, we require a tracker with a fast inference capability. Therefore, we use the DeepSORT~\cite{deepsort} algorithm for tracking individual objects in the video stream. Specifically, DeepSORT utilizes a Kalman filter~\cite{kalmanfilter} to recursively predict future positions of the objects. The Kalman filter models the state transitions using a linear constant velocity model. We denote the state vector of each object $o_i \equiv (c_a, i)$ with $\boldsymbol{m}_i$, which contains the parameters of its bounding box $\boldsymbol{b}_i$ and its respective velocities as
\begin{equation}
    \boldsymbol{m}_i = \left [ x^c_i, y^c_i, w_i, h_i, \Dot{x}^c_i, \Dot{y}^c_i, \Dot{w}_i, \Dot{h}_i \right ],
\end{equation}
where $x^c_i, y^c_i, w_i, h_i$ are the center coordinates, the width, and the height of the bounding box $\boldsymbol{b}_i$, respectively. To improve the tracking performance under challenging scenarios such as occlusion, non-stationary camera and multiple viewpoints, DeepSORT further utilizes deep cosine association metric~\cite{deepsort, deepcosmetric}. Particularly, each detected object $o_i \equiv (c_a, i)$ is cropped from the original frame based on the bounding boxes $\boldsymbol{b}_i$ and passed through a CNN to be represented as unit-norm vectors $\boldsymbol{r}_i \in \mathbb{R}^{128}, \, \norm{\boldsymbol{r}_i}_2 = 1$, as illustrated in Figs.~\ref{fig:deepsort_simple}~and~\ref{fig:deepsort_algo}. We refer to $\boldsymbol{r}_i$'s as the feature vectors. In fact, for each tracked object $o_i \equiv (c_a, i)$, we store multiple feature vectors across frames $\left \{ \boldsymbol{r}_i^{(l)} \right \}_{l=1}^{T_i}$ where $\boldsymbol{r}_i^{(l)}$ stands for the $l$-th feature vector of the tracked object $o_i \equiv (c_a, i)$, and $T_i$ is number of times the object $o_i$ is observed. 

\begin{figure}[ht]
    \centering
    \begin{subfigure}[c]{0.32\textwidth}
        \centering
        \includegraphics[width=\textwidth]{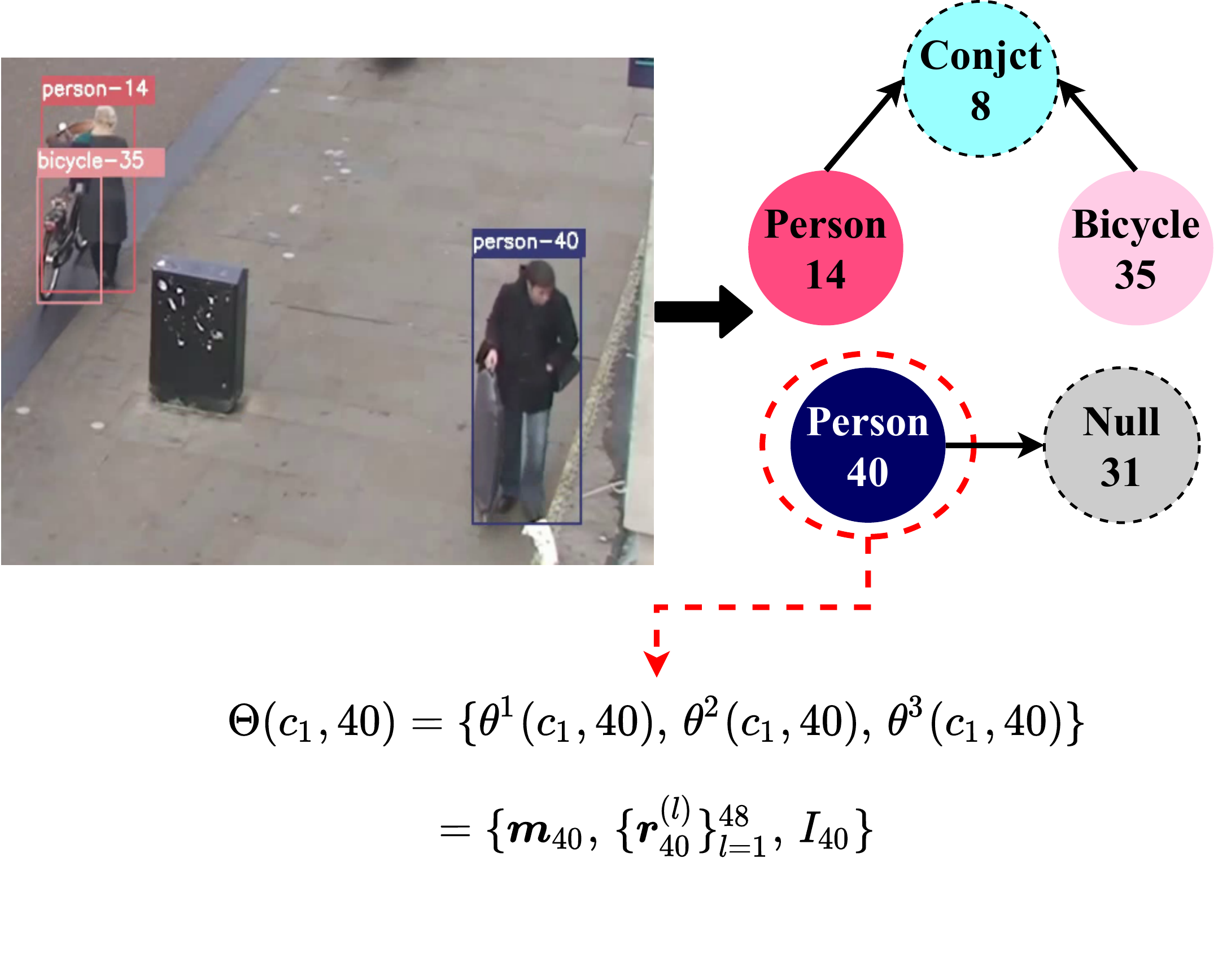}
        \caption{Frame under process, generated instance graph, and corresponding attribute set for Person-$40$.}
        \label{fig:object_attr}
    \end{subfigure}
    \hfill
    \begin{subfigure}[c]{0.32\textwidth}
        \centering
        \vspace{0.8cm}
        \includegraphics[width=\textwidth]{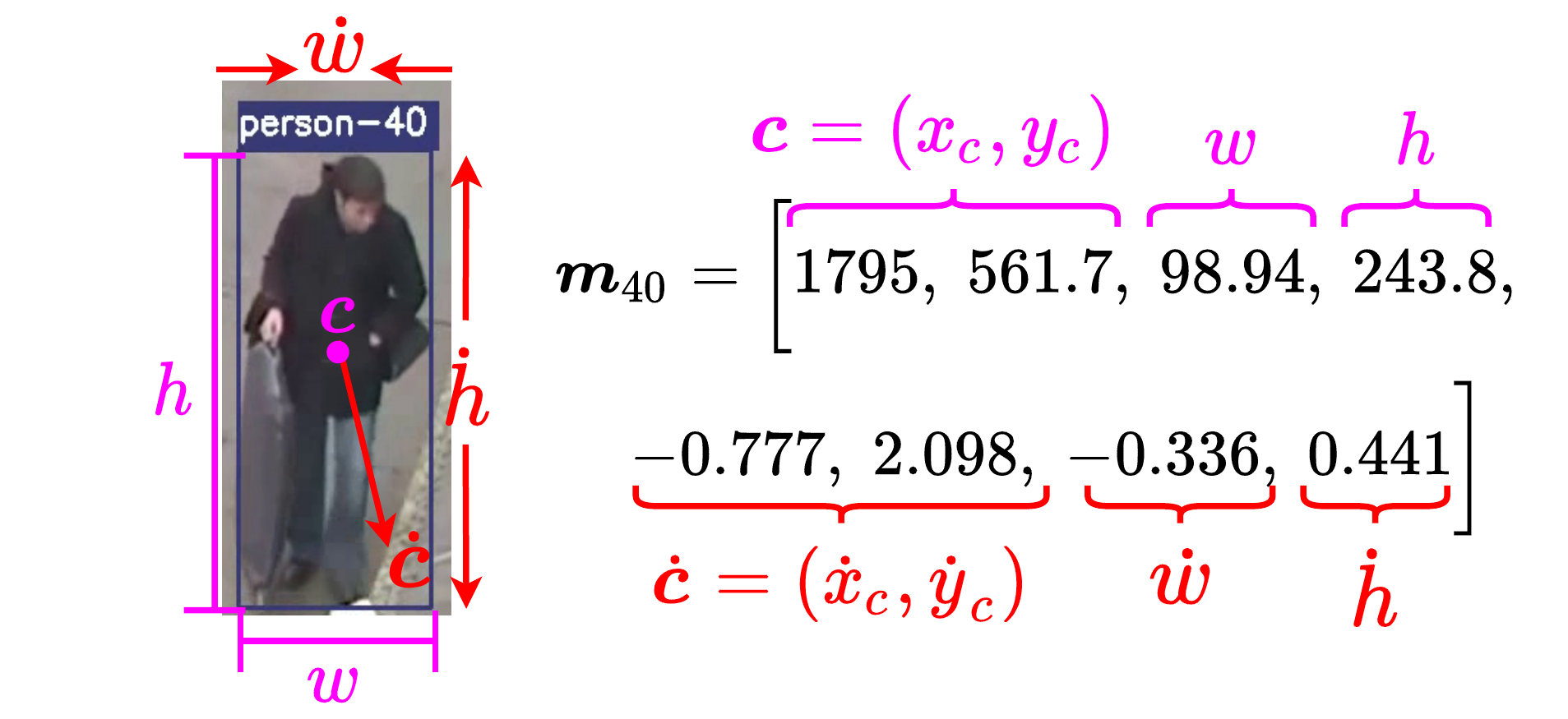}
        \vspace{0.8cm}
        \caption{Lowest level attribute set corresponding to the state vector $\boldsymbol{m}$ of Person-$40$. Units are in pixels for position and pixels-per-frame for velocities.}
        \label{fig:attr_level1}
    \end{subfigure}
    \hfill
    \begin{subfigure}[c]{0.32\textwidth}
        \centering
        \includegraphics[width=\textwidth]{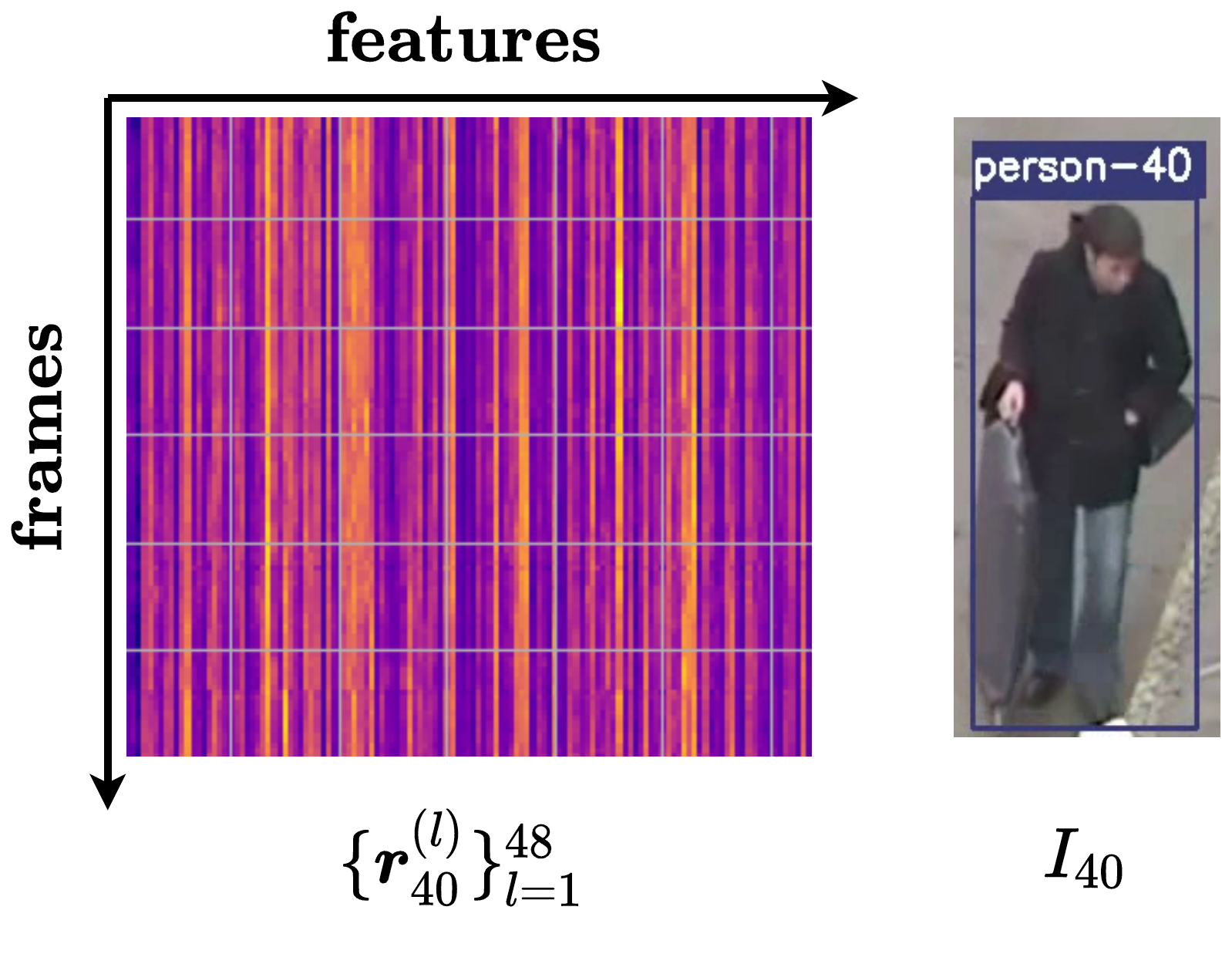}
        \caption{Second and third level attributes of Person-$40$. The second level includes a set of collected feature vectors across $48$ frames with $T_{40}=48$. The third level includes the cropped RGB image $I_{40}$, although alternative encoded versions of $I_{40}$ can also be used.}
        \label{fig:attr_level23}
    \end{subfigure}
    \caption{Illustration of attributes for a \textit{person} component instance.} 
    \label{fig:attribute_vis}
\end{figure}

We employ the state $\boldsymbol{m}_i$, feature vectors $\left \{ \boldsymbol{r}_i^{(l)} \right \}_{l=1}^{T_i}$ and the cropped RGB image $I_i\in \mathbb{R}^{w_i\times h_i \times 3}$ as the multi-level attributes of object $o_i \equiv (c_a, i)$ with $L=3$ number of levels as defined in~\eqref{eq:theta_ti}, i.e.,
\begin{equation}
    \Theta(c_a, i) = \left \{ \boldsymbol{\theta}^{(1)}(c_a, i), \, \boldsymbol{\theta}^{(2)}(c_a, i), \, \boldsymbol{\theta}^{(3)}(c_a, i) \right \} =  \left \{\boldsymbol{m}_i, \, \{ \boldsymbol{r}_i^{(l)} \}_{l=1}^{T_i}, \, I_i \right \}.
\end{equation}
Note that, it is possible that different categories can have a different number of attribute levels $L_a$; however, for our experiments, we set $L_a=3, \; \forall a$, for simplicity. It is important to note that, even though in our experiments,  $\boldsymbol{\theta}^{(3)}(c_a, i)$ is set to the raw data $I_i$ itself, it can also be chosen as any multi-scale representation of $I_i$ such as its wavelet decomposition or discrete cosine transform. As an example, attributes for a detected \textit{person} object is illustrated in Fig.~\ref{fig:attribute_vis}.

The exploitation of object positions and velocities enables us to determine types of predicates between different object pairs. We identify the type of a predicate between objects $o_i \equiv (c_a, i)$ and $o_j \equiv (c_b, j)$ by using their states $\boldsymbol{m}_i$ and $\boldsymbol{m}_j$. The following three metrics are defined for predicate detection
\begin{align}
    & d^{(1)}(i, j) = \frac{ \norm{(x_i^c, y_i^c) - (x_j^c, y_j^c)}_2^2}{\sqrt{w_i \cdot h_i \cdot w_j \cdot h_j}}, \\
    & d^{(2)}(i, j) = \frac{(\dot{x}_i^c, \dot{y}_i^c)^\top (\dot{x}_j^c, \dot{y}_j^c)}{\norm{(\dot{x}_i^c, \dot{y}_i^c)}_2 \,  \norm{(\dot{x}_j^c, \dot{y}_j^c)}_2} = cos \left ( (\dot{x}_i^c, \dot{y}_i^c),\,(\dot{x}_j^c, \dot{y}_j^c) \right ), \\
    & d^{(3)}(i, j) = IoU\left (\boldsymbol{b}_i, \boldsymbol{b}_j \right ).
\end{align}
Here, $d^{(1)}(i,j)$ corresponds to scaled Euclidean (2-norm) distance between center coordinates of the positions, $d^{(2)}(i,j)$ 
is the cosine similarity between the velocities, while $d^{(3)}(i,j)$ is the intersection-over-union of bounding boxes of objects $o_i$ and $o_j$. Specifically, for a predicate $e_k \equiv (p_1, k)$, i.e., \textit{moving-together} between objects $o_i \equiv (c_a, i)$ and $o_j \equiv (c_b, j)$, across multiple frames we require
\begin{equation}
    \mathds{1} \left( d^{(1)}(i, j) \leq z^{(1)}\right) \cdot \mathds{1} \left ( d^{(2)}(i, j) \geq z^{(2)} \right ),
\end{equation}
and similarly, for $e_k \equiv (p_2, k)$, i.e., $conjunct$,  we require
\begin{equation}
    \mathds{1} \left( d^{(1)}(i, j) \leq z^{(1)}\right) \cdot \mathds{1} \left ( d^{(2)}(i, j) \geq z^{(2)} \right ) \cdot \mathds{1} \left ( d^{(3)}(i, j) \geq z^{(3)} \right ),
\end{equation}
for predetermined thresholds $z^{(1)}$, $z^{(2)}$ and $z^{(3)}$, where $\mathds{1}(\cdot)$ denotes the indicator function. Furthermore, even though it is not crucial for our application, we define attributes for a predicate $e_k \equiv (p_w, k), \, p_w \in \mathcal{P} \setminus \{ p_0 \}$ as
\begin{equation}
    \Theta(p_w, k) = \left \{ \boldsymbol{\theta}^{(1)}(p_w, k) \right \} = \left \{ \bigg[d^{(1)}(i,j), \, d^{(2)}(i,j), \, d^{(3)}(i,j) \bigg] \right \},
\end{equation}
with $L_w = 1, \; \forall w$. Here, $\Theta(p_w, k)$ consists of a single-level attribute, i.e., a vector that contains metric evaluations between objects $o_i$ and $o_j$. Note that unlike the other predicates belonging to classes in $\mathcal{P} \setminus \{ p_0 \}$ which are connected to two objects $o_i \equiv (c_a, i)$ and $o_j \equiv (c_b, j)$, a null predicate $e_l \equiv (p_0, l)$ is only connected to $o_u \equiv (c_f, u)$, for an isolated object $o_u$. For null predicates, the attribute set is simply defined as the the empty set, i.e., $\Theta(p_0, l) = \varnothing$. 

Given the set of detected objects $C^{D}$ and predicates $P^{D}$, we denote our detected connection set $E^{D}$ with triplets and pairs as
\begin{equation}
    E^{D} = \left \{ (o_i, \, e_k, \, o_j)  \; \bigg\vert \; o_i, o_j \in C^{D}, \, e_k \in P^{D}, \, p_w \in P \setminus \{p_0\} \right \} \;
    \bigcup \; \left \{ (o_u, \, e_l) \; \bigg\vert \; o_u \in C^{D}, \, e_l \equiv (p_0, l) \right \}.
\end{equation}
In other words, the detected connection set $E^{D}$ simply consists of \textit{(object, predicate, object)} triplets and (\textit{unaccompanied object, null predicate}) pairs. Consequently, the graph $D$ is generated by the detected object, predicate, and edge sets as
\begin{equation}
    D = \left ( C^{D}, \, P^{D}, \, E^{D} \right ).
\end{equation}
We then define the companion attribute set $A$ of graph $D$ as
\begin{equation}
    A = \left \{ \Theta(c_a, i) \; \bigg\vert \; o_i \equiv (c_a, i) \in C^{D} \right \} \, \bigcup \, \left \{ \Theta(p_w, k) \; \bigg\vert \; e_k \equiv (p_w, k) \in P^{D} \right \}, 
\end{equation} 
which stores attributes for both object and predicate instances.

\begin{figure}[b]
     \centering
     \begin{subfigure}[c]{0.3\textwidth}
         \centering
         \vspace{0.5cm}
         \includegraphics[width=\textwidth]{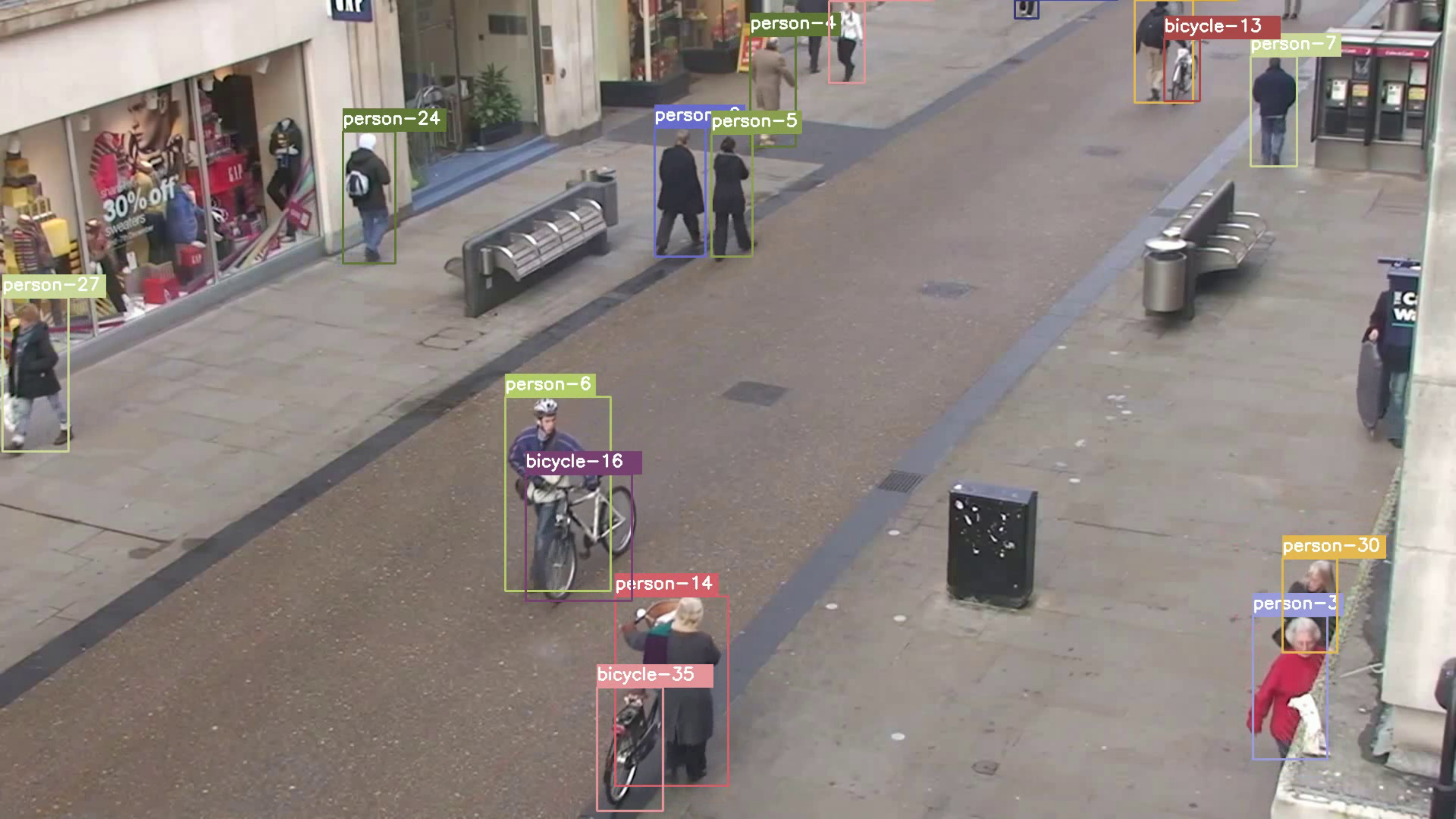}
         \vspace{0.5cm}
         \caption{Frame under process with detections.}
         \label{fig:frame22}
     \end{subfigure}
     \hfill
     \begin{subfigure}[c]{0.35\textwidth}
         \centering
         \includegraphics[width=\textwidth]{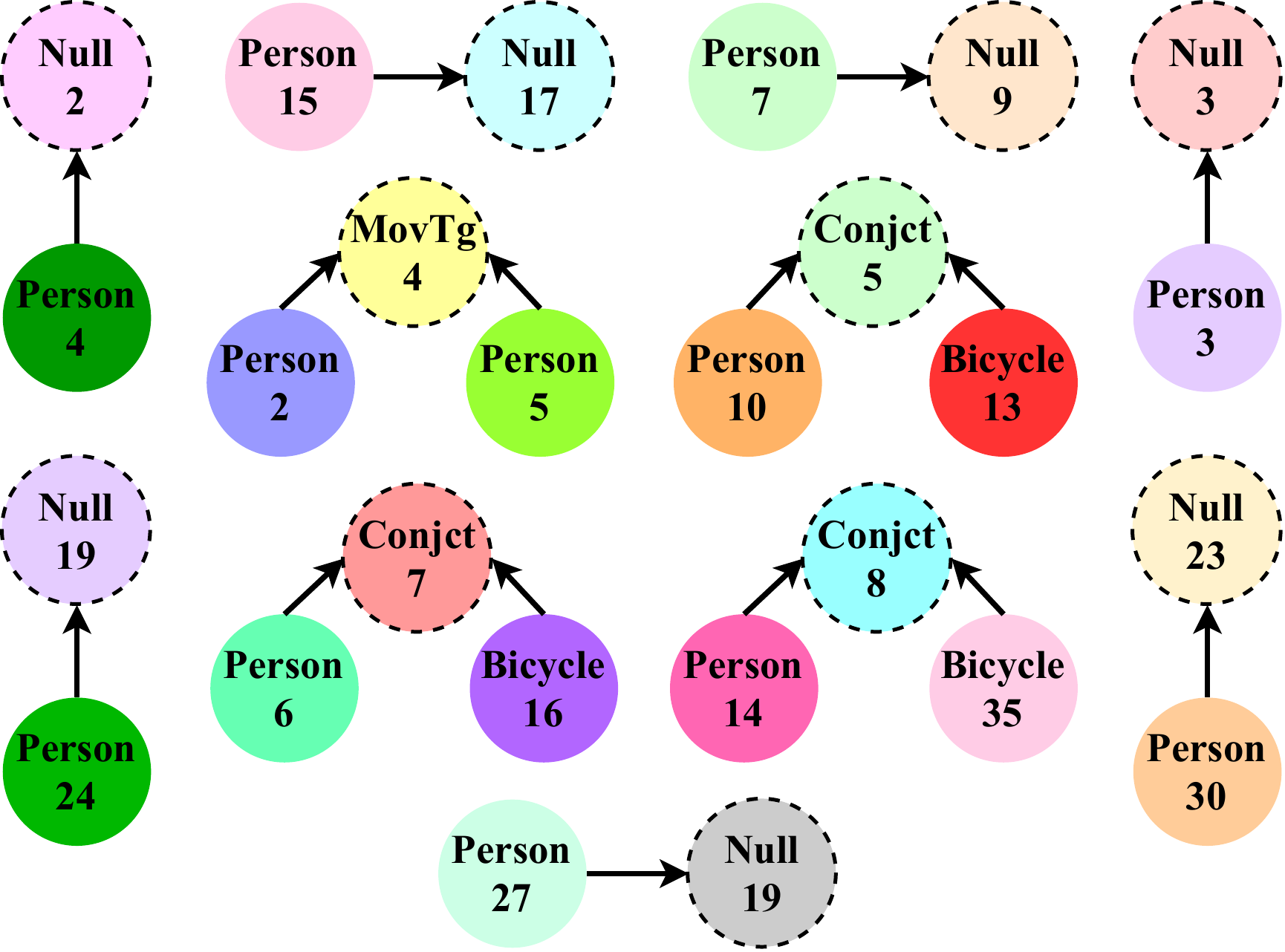}
         \caption{Instance-level graph $D$. Object nodes are denoted with solid circles, predicates are denoted with dashed circles. Isolated objects are connected to the \textit{null} predicate by default.}
         \label{fig:frame22_graph}
     \end{subfigure}
     \hfill
     \begin{subfigure}[c]{0.30\textwidth}
         \centering
         \includegraphics[width=0.9\textwidth]{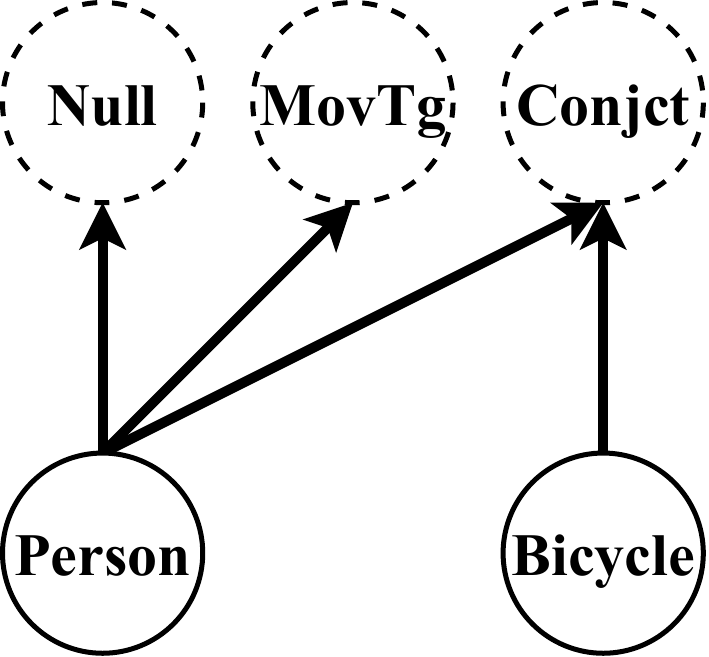}
         \caption{Class-level graph $S$.}
         \label{fig:frame22_full_abs}
     \end{subfigure}
    \caption{Illustration of the instance-level and class-level graph representations.}
    \label{fig:frame22_fulls}
\end{figure}

Furthermore, we generate the class-level representation of instance graph $D$ as defined in~\eqref{eq:Sti}. The class description graph $S$ is defined as
\begin{equation}
    S = (C^{S}, P^{S}, E^{S}),
    \label{eq:Sti_example}
\end{equation}
where 
\begin{align}
    & C^{S} = \left \{ {c_a} \; \bigg\vert \; o_i \equiv (c_a, i) \in C^{D} \right \}, \\
    & P^{S} = \left \{ {p_w} \; \bigg\vert \; e_k \equiv (p_w, k) \in P^{D} \right \}, \\
    & E^{S} = 
    \begin{aligned}
        &&\bigg \{(c_a, \, p_w, \, c_b)  \; \bigg\vert (o_i, \, e_k, \, o_j) \equiv ((c_a, i), \, (p_w, k) \,  (c_b, j)) \in E^{D} \bigg \}  \\
        && \bigcup \, \bigg \{ (c_f, p_0) \; \bigg\vert \; (o_u, \, e_l) \equiv ((c_f, u), \, (p_0, l)) \in E^{D} \bigg \}.
    \end{aligned}
\end{align}
To facilitate a better understanding of the high-level abstraction provided by~\eqref{eq:Sti_example}, we provide an illustrative example in Fig.~\ref{fig:frame22_fulls}. The class level graph generation is a summarization of the components and predicates in the instance level graph, i.e., it is a surjective function mapping $D$ to $S$.

\begin{figure}[t]
    \centering
    \begin{subfigure}[t]{0.48\textwidth}
        \centering
        \includegraphics[width=\textwidth]{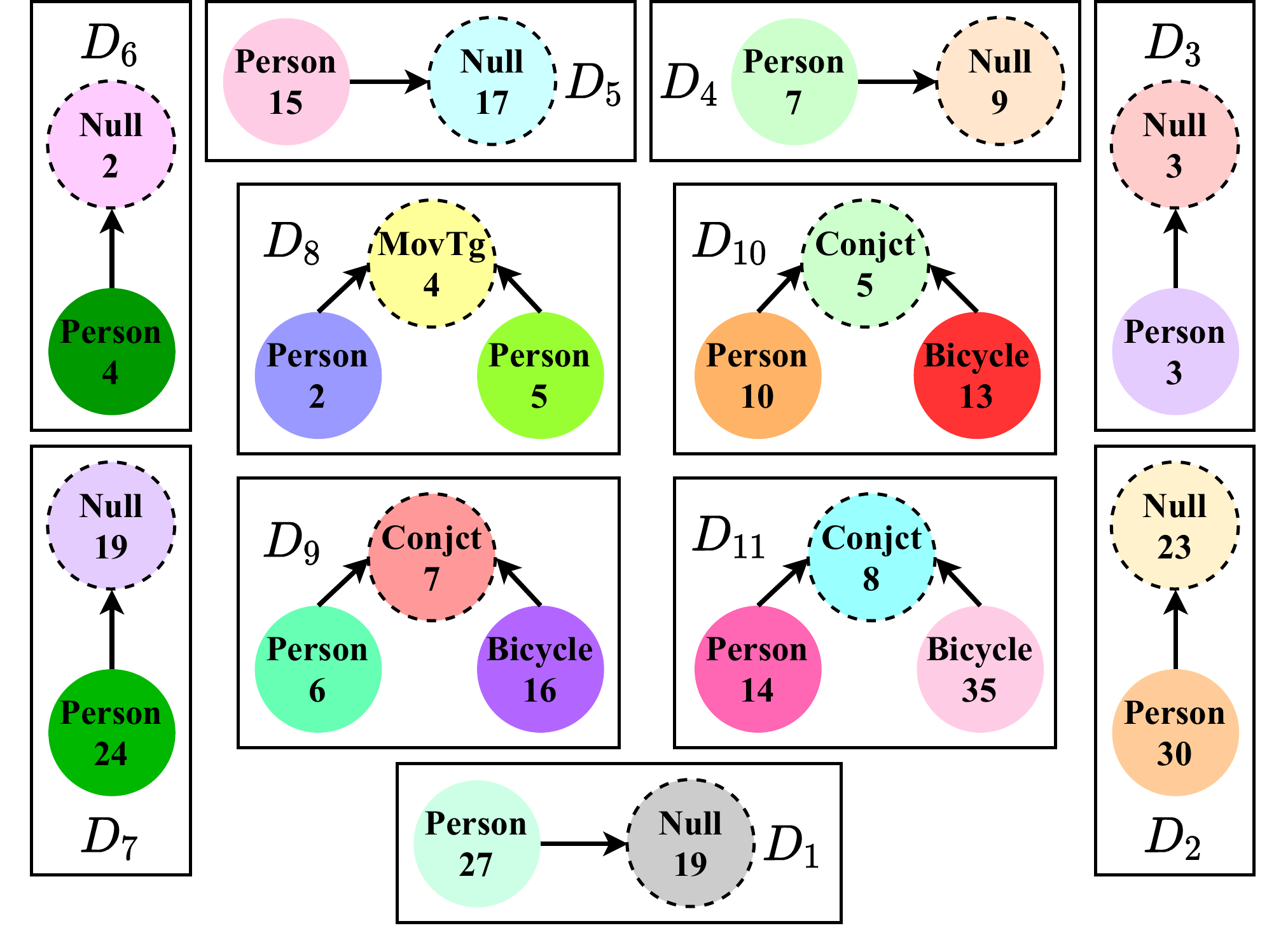}
        \caption{Instance-level graph splitting at post-processing. Graph $\mathcal{D}$ in Fig.~\ref{fig:frame22_graph} is split into its atomic components $D_m$ such that $ \mathcal{D} = \{ D_1, \, \ldots, \, D_{11} \} $. Corresponding attributes are split to atoms $ \mathcal{A} = \{ A_1, \, \ldots, \, A_{11} \}$ as well.}
        \label{fig:frame22_split}
    \end{subfigure}
    \hfill
    \begin{subfigure}[t]{0.48\textwidth}
        \centering
        \includegraphics[width=\textwidth]{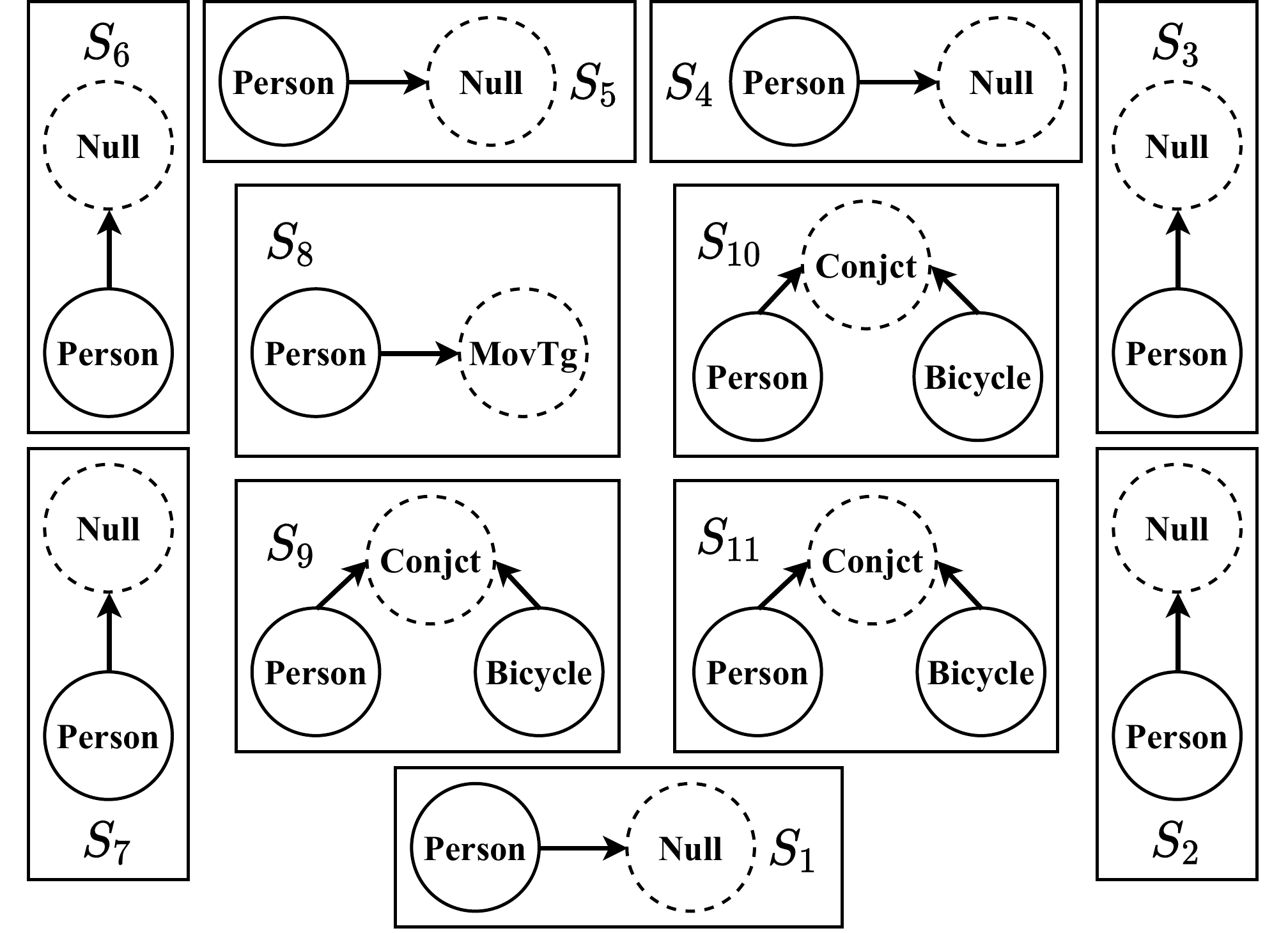}
        \caption{Class-level abstractions are performed on disconnected subgraphs similar to the one in Fig.~\ref{fig:frame22_full_abs} so that we obtain $\mathcal{S} = \{ S_1, \, \ldots, \, S_{11} \} $. Together with original graph $\mathcal{D}$ and attribute set $\mathcal{A}$, abstraction $\mathcal{S}$ forms the complete semantic description $Y$. Global goal-based filtering is performed using these high-level graphs.}
        \label{fig:frame22_split_abs}
    \end{subfigure}
    \caption{Decomposing the instance-level and class-level graphs into their atomic components.}
    \label{fig:frame22_splited}
\end{figure}

Finally, we define the semantic description $Y$ as a triplet consisting of the class-level graph, the instance-level graph, and attribute supersets as
\begin{equation}
    Y = (\mathcal{S}, \mathcal{D}, \mathcal{A}).
\end{equation}
It should be noted that, while in Section~\ref{sec:ProposedLanguage}, an instance level graph $\mathcal{D}$ is defined as union of disjoint subgraphs $\mathcal{D} = \left \{ D_1, \ldots, D_m, \ldots D_M\right \}$, we use a single combined graph $D$ to represent the whole scene. It is desirable to split $\mathcal{D}$ into \textit{atomic graphs} for an easy processing of goals, as discussed in Section~\ref{sec:ProposedLanguage}. That is, splitting of $\mathcal{D}$ into its disjoint subgraphs $D_m$ (also called atomic graphs), can be performed during post-processing for this specific scenario. A similar splitting operation can also be applied to the attribute set $\mathcal{A} = \{A_1, \ldots, A_m, \ldots, A_M\}$ and class-level graph $\mathcal{S} = \{S_1, \ldots, S_m, \ldots, S_M \}$ where $S_m$ denotes the abstraction of $D_m$. The splitting operation on $\mathcal{D}$ and its class level counterpart is illustrated in Fig.~\ref{fig:frame22_splited}.

\begin{figure}[t]
    \centering
    \includegraphics[width=0.95\textwidth]{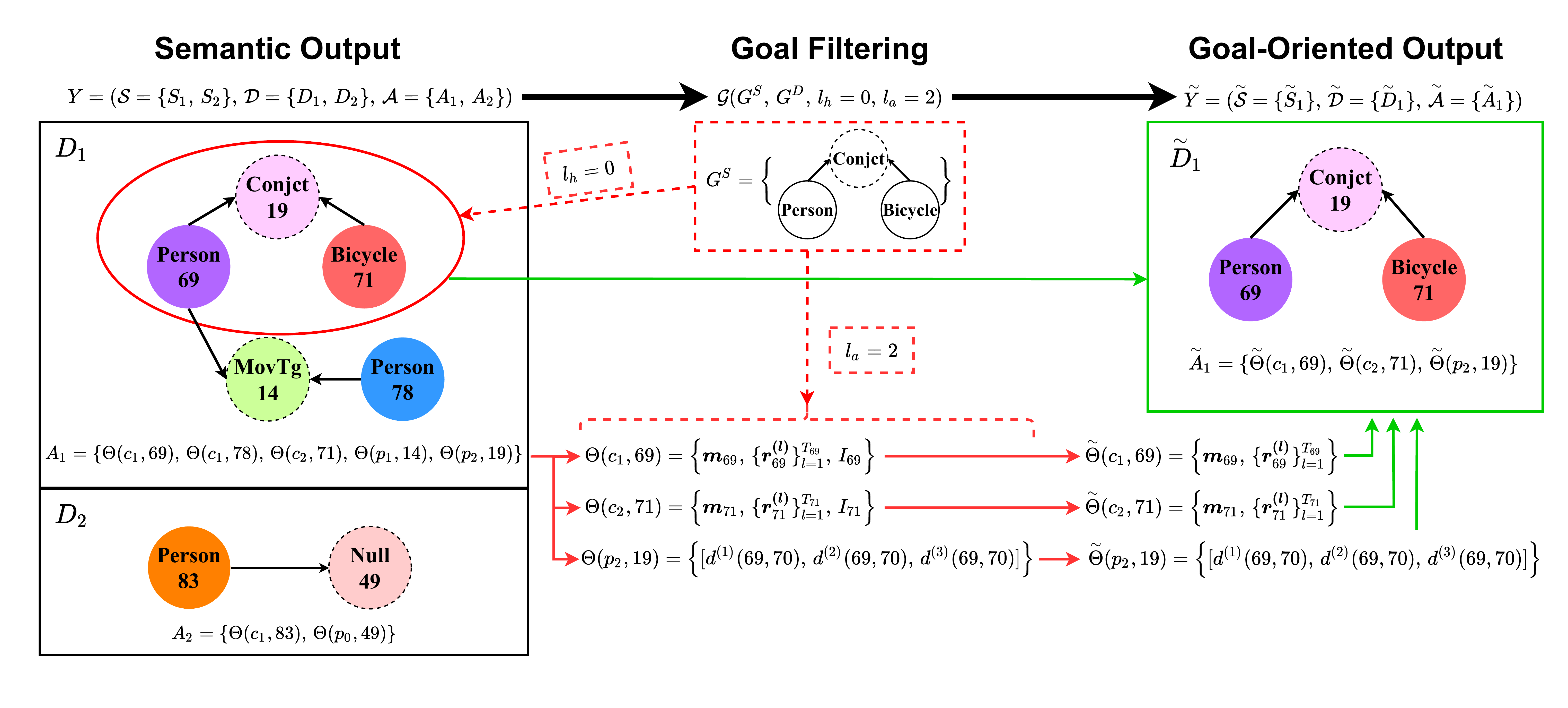}
    \caption{A goal-based filtering example. The filtering operation yields goal-oriented semantic description $\overset{\sim}{Y}$. Dashed red lines denote the \textit{queries/goals} and solid green lines denote the goal-oriented returns. For brevity, we do not visualize class-level graphs $S$. The goal pattern $G^{S}$ is searched in $S_1$ and $D_1$ to provide $\overset{\sim}{S}_1$, and $\overset{\sim}{D}_1$. Graph complexity is chosen as $l_h=0$ so that only the exact matchings are returned. The corresponding attributes $A_1$ are distilled into $\overset{\sim}{A}_1$ according to the given attribute complexity parameter $l_a = 2$.}
    \label{fig:goalfilter}
\end{figure}

We can also illustrate the goal-based filtering defined in Section~\ref{sec:ProposedLanguage} using the same scenario. Goal-based filtering operation allows for further distillation of the semantic information with respect to a specified goal and it enables reasoning over the semantic graph. In a sense, the filtering operation acts as a semantic parsing operator. The goal filtering operation for a small semantic graph with only two subgraphs is illustrated in Fig.~\ref{fig:goalfilter}. It should be observed that the queried graph patterns $G^{S}$ and $G^{D}$ as defined in~\eqref{eq:Goal}, both include simple motifs with a single graph. Consequently, graph complexity and attribute complexity vectors reduce to scalars.

As illustrated in this part, the proposed goal-oriented semantic graph language and the signal processing framework can be incorporated into video signals for real-time applications. The semantic language allows for a structured and complete representation of the \textit{meaningful} and \textit{interesting} information embedded within the signal and allows for easy parsing of the information according to the desired goal. 

\subsection{Energy Efficient Sampling and Semantic Modeling of a Scalar Sensor Output}

Energy efficiency is a critical research area for sensors, especially for the next generation of communication networks where the number of IoT devices is expected to rapidly increase. The proposed goal-oriented semantic language and the signal processing framework can enable dramatic reductions in transmitted information, especially when the sampled phenomenon is changing slowly with respect to the goals defined either internally or externally, leading to significant energy savings.

Another avenue of improvement for the efficiency of sensors can be introduced by employing smart non-uniform sampling strategies. For scalar sensors, if the signal output can be modeled as a random process whose statistical characterization is available, the sampling times can be adjusted according to desired \textit{detection} and \textit{missed detection} probabilities of certain events, e.g., a threshold crossing. 

In this subsection, we present a novel sampling strategy for a generic scalar sensor output, in conjunction with a goal-oriented semantic representation to improve the energy efficiency of sampling and transmission operations. First, we introduce a novel and optimal sampling strategy for scalar sensor outputs that conform to an auto-regressive AR($p$) Gaussian model, and present statistical results on the improved performance. Next, we implement the proposed goal-oriented semantic language definition in Section~\ref{sec:ProposedLanguage}, to illustrate its validity even for relatively simple signal modalities.

\subsubsection{Optimal Sampling Strategies for Gaussian Auto-Regressive Models}
\label{subsec:optimalsampling}

We start with a discrete time input signal $x_n$ over a horizon $H$, i.e., $n = 0:H-1$, where $x_n$ is a Gaussian AR($p$) process with standard normal innovations and initial condition. For $p=1$, we have an AR($1$) process described by
\begin{align}
\begin{split}
\label{eq:AR1}
    x_n &= \alpha x_{n-1}+\sqrt{1-\alpha^2}w_{n-1} \text{ for } n\geq 1,\\
\end{split}
\end{align}
where $x_0\sim \mathcal{N}(0,1)$ and $w_i \sim \mathcal{N}(0,1)$. For a two-threshold event detection problem, we define $\tau \triangleq \inf \Big\{n>0:\,x_n\not\in (l,u)\Big\}$ where $u$, $l$ are the upper and lower thresholds, respectively. Assume that we take the first sample at time $n_1 = 1+k$ for some non-negative $k$, and denote its probability of missing a threshold crossing as $P_{err}(k,l,u)$. We can relate this error probability with the CDF of $\tau$ as
\begin{align}
        P_{err}(k,l,u) &= \mathbb{P}(\tau\leq k) \nonumber \\ 
        &= 1-\mathbb{P}\left(\left\{\sup_{s\leq k}x_s< u\right\}\bigcap \left\{\inf_{s\leq k}x_s> l\right\}\right) \nonumber \\
        &= 1-\mathbb{P}\left(x_s\in (l,u)\quad \forall s\in \{1,2,\ldots,k\}\right) \nonumber\\
        &= 1-\psi_k,
\end{align}   
where $\psi_k \triangleq \mathbb{P}\left(x_s\in (l,u)\quad \forall s\in \{1,2,\ldots,k\}\right)$ with $\psi_0 = 1$ as a convention.
    
Given a tolerable missed detection probability $p_{m}$, we choose the largest $k\in\mathbb{N}$ such that $\psi_k\ge 1-p_{m}$ and take the first sample at $n_1 = 1+k$. Since $x_n$ follows an AR($p$) process, calculation of $\psi_k$ can be done by using the Markov Property~\cite{markov1954theory}, i.e., factorization of the joint probability using conditional independence. 
    
Next, we illustrate the derivation of the sampling algorithm for an AR($1$) model. Generalization to a $p$-th order AR process can be performed similarly. We consider the model in~\eqref{eq:AR1} assuming that $x_0\in(l,u)$ is observed. Therefore, we have $\mathbb{E}\left[x_n \mid x_0 \right] = \alpha^n\,x_0$, $\mathbb{V}\left[x_n \mid x_0 \right] = 1-\alpha^{2\,n}$, and $\mathbb{C}\left[x_n,\,x_{n-1} \mid x_0 \right] = \alpha\,(1-\alpha^{2(n-1)})$, i.e., the mean is strictly decreasing towards $0$ while the variance and covariance are strictly increasing towards $1$ and $\alpha$, respectively. 
    
To be able to calculate the next optimal sampling instance, we define the probability of the $i$-th sample being inside the two thresholds given that the previous sample was also inside the thresholds as
\begin{equation}
    h_i \triangleq \mathbb{P}\left(x_i \in (l,u)\,\vert\,x_{i-1}\in (l,u) \right),
    \label{eq:hi}
\end{equation}
with an initial condition of $h_1 = \mathbb{P}\left(x_1\in(l,u)\right)$. Calculation of~\eqref{eq:hi} requires only the first and second order statistics of $x_{i-1}$ and $x_i$, which can be computed numerically. Finally, the next optimal sampling time can be calculated as 
\begin{equation}
        n_1 = \inf \left\{n\in \mathbb{Z}_+\,\bigg\vert\, \psi_n < 1-p_m \right\},
\label{eq:sampling}
\end{equation}
where $\psi_k$ is defined using the Markov Property as
\begin{equation}
    \psi_k = \prod_{i=1}^k h_i.
    \label{eq:psi_k}
\end{equation}
Once we observe the process at time $n_1$, we proceed to calculate the next sampling time $n_2$, and for any $k\in\mathbbm{Z}^{+}$, the algorithm uses $x_{n_k}$ to determine the next optimal sampling time $n_{k+1}$. 
    
\begin{figure}[t]
    \centering
    \includegraphics[width = 0.6\textwidth]{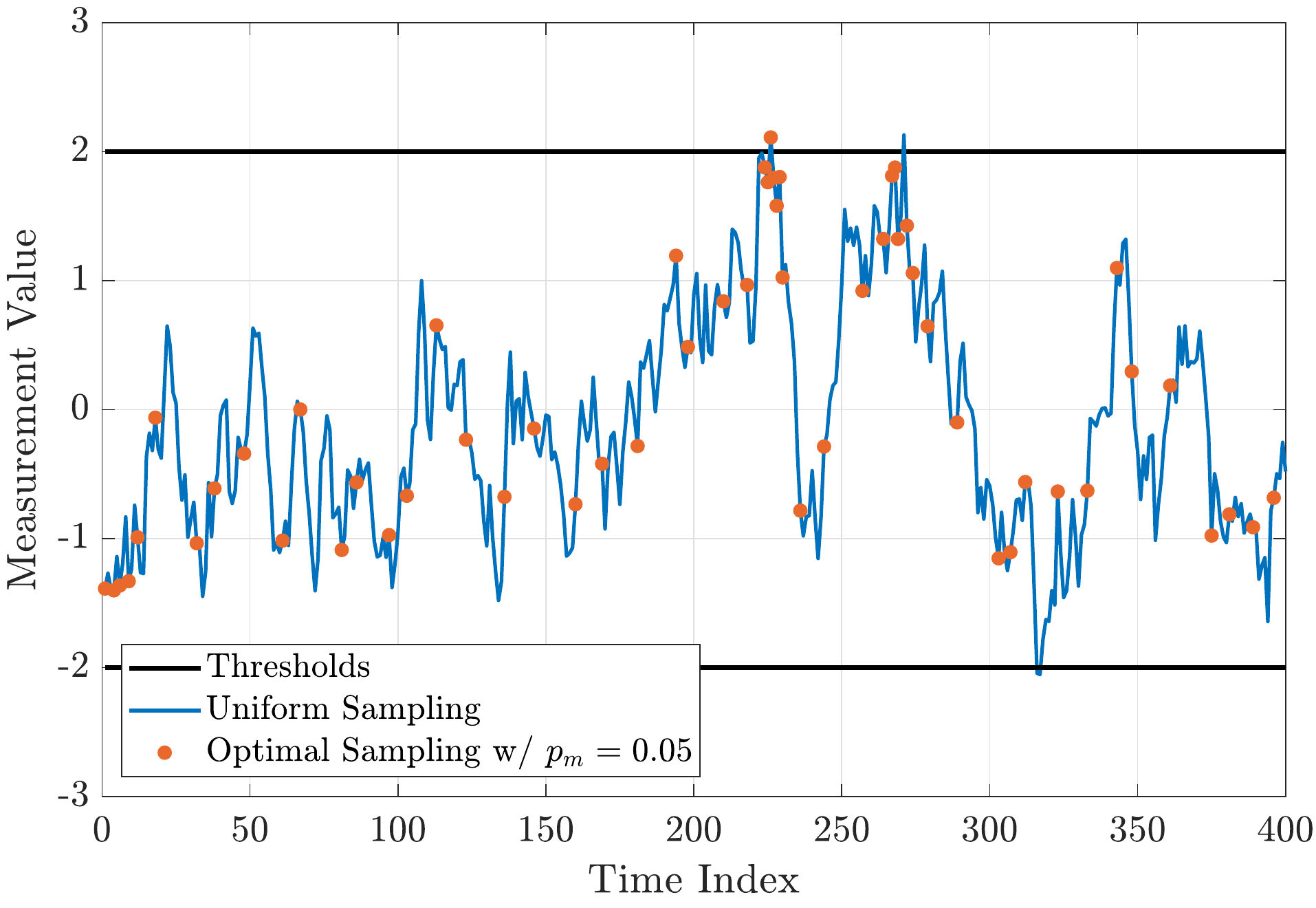}
    \caption{Optimal sampling strategy applied to an AR($1$) process with two-thresholds.}
    \label{fig:1D_horizon}
\end{figure}

In Fig.~\ref{fig:1D_horizon}, we illustrate the proposed sampling algorithm for a first-order Gaussian AR process. Our aim is to detect crossings of thresholds $u = 2$ and $l=-2$. The desired maximum probability of missing the threshold crossings is set at $0.05$. As expected, when the minimum distance to thresholds ($\min \{|x_n-l|,|u-x_n|\}$) decreases, the sampling rate increases. 

We now investigate various statistical properties of~\eqref{eq:sampling} under the assumption
\begin{align}
    x_0\sim\mathcal{N}(0,1) \text{ is within the boundaries, i.e., } x_0\in(l,u).
    \label{eq:x0assumption}
\end{align}
The distribution of the optimal sampling times requires calculation of~\eqref{eq:psi_k}, i.e., the products of dependent random variables $h_i$, under the assumption given in~\eqref{eq:x0assumption}. The expectation of $n_1$ has significance since it represents the \textit{average sampling period} of the algorithm. The expressions for the CDF and the expectation of $n_1$ are as follows:
\begin{align}
    \label{eq:cdfn1}
    \mathbb{P}\left(n_1\leq k\,\vert\,x_0\in(l,u)\right) =  \mathbb{P}\left(\prod_{i=1}^k\,h_i< 1-p_m\right),\\
    \label{eq:expectation}
      \mathbb{E}\left[n_1\,\vert\,x_0\in(l,u)\right] = 1+\sum_{k=1}^\infty \mathbb{P}\left(\prod_{i=1}^k\,h_i\ge 1-p_m\right), 
\end{align}
which can be computed numerically for given threshold levels $l,u$ and missed detection probability $p_m$. 
   
Next, we present two simulation results. In the first simulation, we investigate the expected sampling period for various missed detection probabilities for AR(1) models. We use a first-order Gaussian AR process as in~\eqref{eq:AR1} with $\alpha = 0.95$, i.e.,
\begin{gather}
    x_n = 0.95\, x_{n-1}+\sqrt{1-0.95^2}\,w_{n-1} \text{ for } n\geq 1, \label{eq:AR1_095}\\
    x_0\sim \mathcal{N}(0,1) \text{ and } w_i \sim \mathcal{N}(0,1). \nonumber
\end{gather}
    
\begin{figure}[t]
    \centering
    \includegraphics[width = 0.6\linewidth]{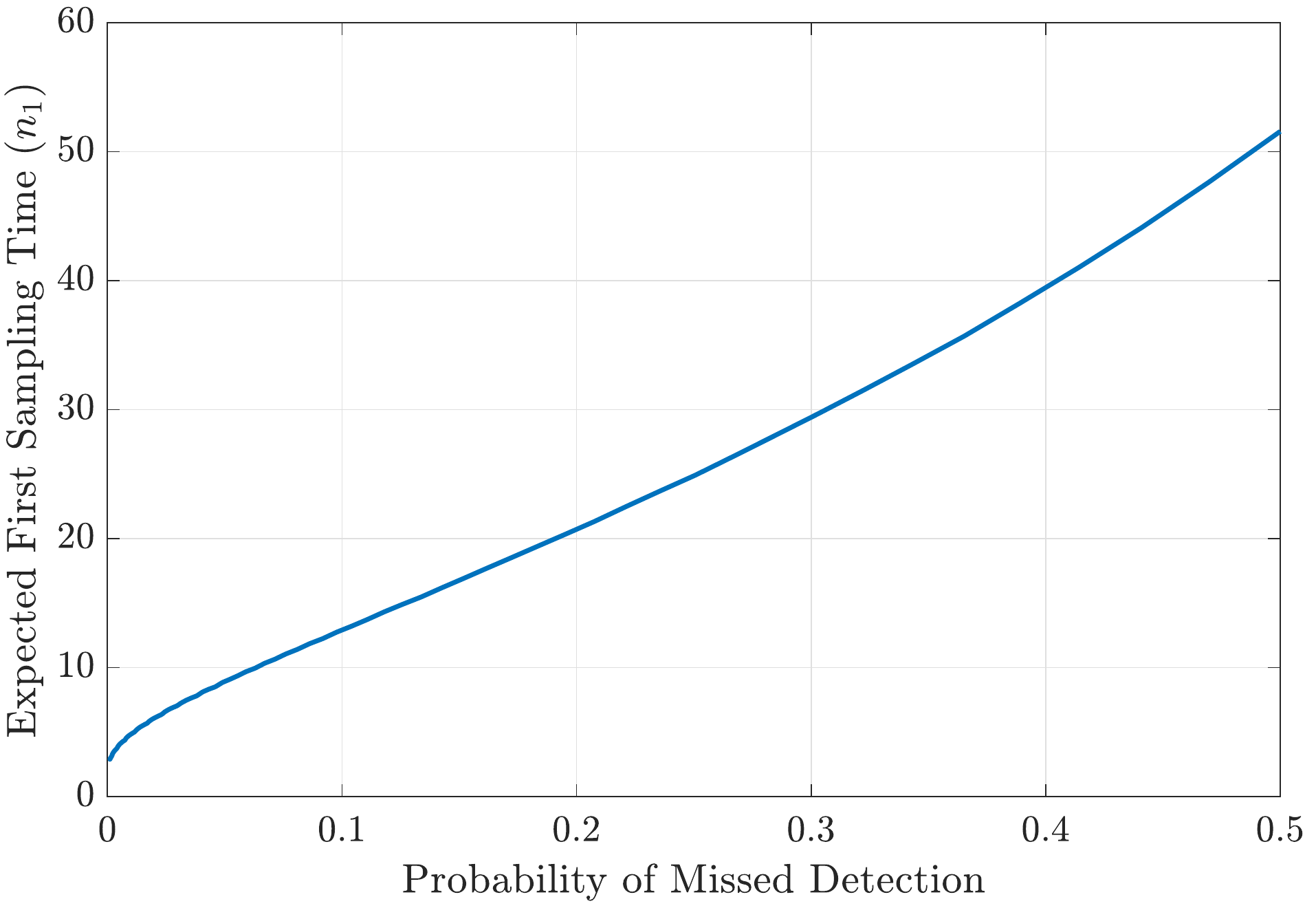}
    \caption{Expected first sampling time vs. probability of missed detection.}
    \label{fig:1D_expectedsampling}
\end{figure}

Setting the thresholds $l=-2,\,u = 2\,$, we plot the expected first sampling time $\mathbb{E}\left[n_1\,\vert\,x_0\in(l,u)\right]$ versus the tolerated maximum probability of missed detections under the assumption that the current sample is within the boundaries in Fig.~\ref{fig:1D_expectedsampling}. As expected, the sampling period is positively correlated with the missed detection probability. The relationship is observed to be almost linear for this example.

\begin{figure}[t]
\centering
\includegraphics[width = 0.6\linewidth]{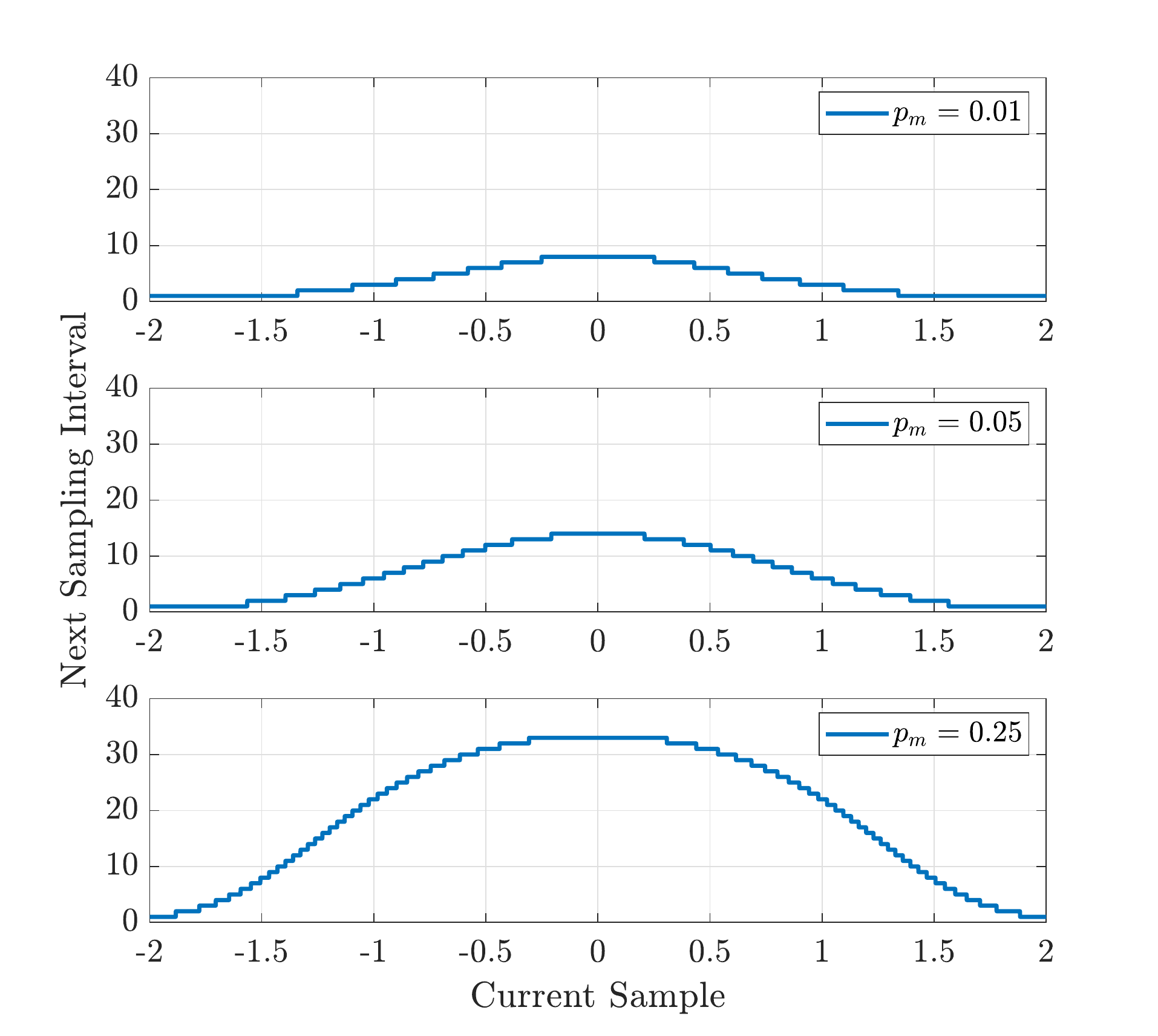}
\caption{Next sampling time interval vs. current sample value for $p_m \in \{0.01,0.05,0.25\}$.}
\label{fig:1D_nextsampling}
\end{figure}

In Fig.~\ref{fig:1D_nextsampling}, we present the effect of the current sample value on the next sampling time interval for different missed detection probabilities $p_m$. The underlying assumptions and the model are the same as those in the previous simulation setup.
We observe that as the minimum distance to the thresholds ($\min \{|x_n-l|,|u-x_n|\}$) increases, the constraint on the next sampling time that satisfies the missed detection probability requirement relaxes.

\subsubsection{Semantic Modeling Example for One-Dimensional Signals}
We can model the threshold detection problem studied in the previous subsection using the proposed semantic language introduced in Section~\ref{sec:ProposedLanguage}. In a typical threshold detection problem, the semantic information extraction may include the detection of the region in which the signal resides at a given moment, as well as the descriptions of threshold crossing events. Using the proposed formulation in Section~\ref{sec:ProposedLanguage}, the components can be defined as the regions separated by the upper and lower thresholds that we denote as $u$ and $l$.
\begin{align}
    c_1 &= \{x\in \mathbbm{R}\,\vert\,x>u \} \label{eq:c1}\\
    c_2 &= \{x\in \mathbbm{R}\,\vert\,l \leq x\leq u \} \label{eq:c2}\\
    c_3 &= \{x\in \mathbbm{R}\,\vert\,x<l \} \label{eq:c3}
\end{align}
We can easily extend the regions to a $k-$threshold model for $k>2$ as well. Let $t_1<t_2<\ldots<t_k$ denote the thresholds in an ascending order. Then, the components can be defined as 
\begin{align}
    c_1 &= \{x\in \mathbbm{R}\,\vert\, x>t_k\},\\
    c_i &= \{x\in \mathbbm{R}\,\vert\, t_{k-i+1}\leq x\leq t_{k-i+2}\},\quad \forall i\in \{2,\ldots,k\},\\
    c_{k+1} &= \{x\in \mathbbm{R}\,\vert\, x<t_1\}.
\end{align}

The predicate set for this problem can be defined as descriptors of threshold crossing events as
\begin{equation}
  P = (U,D,I),
  \label{eq:1D_p}
\end{equation}
where $U,D,I$ denote \textit{upward crossings}, \textit{downward crossings}, and \textit{idling} (staying in the same region), respectively. Note that for this particular case, the \textit{class multi-graph description} $\mathcal{S}_t$ and the \textit{instance multi-graph description} $\mathcal{D}_t$ are defined as equivalent, since a higher level abstraction of a scalar signal is not required. 

For the multi-level attribute set, a suitable choice for the first level is the current time index and amplitude of the process, and for each level $l$ we can store the previous $(l-1)$-th time and amplitude information as
\begin{align}
    \mathcal{A}_t &= \{\Theta_{t,i}\}\text{ where } \Theta_{t,i} = \{\theta^{(1)},\theta^{(2)},\ldots,\theta^{(r)}\},  \label{eq:1D_At} \\
    \theta^l &= (n-l+1,\,x_{n-l+1}), 
    \label{eq:1D_Theta}
\end{align}
where $r$ is the total number of levels in the attribute sets. Depending on the specific application, alternative attribute levels including desired transform domain representations of the signal can be defined. 

For a two-threshold case, and using the same Gaussian AR(1) process input signal in~\eqref{eq:AR1_095} with $l = -2$ and $u = 2$, we illustrate a region of interest around sampling indices $\tilde{n} \in [38, 45]$ in Fig.~\ref{fig:1D_RoI}. Note that sampling points $\tilde{n}$ are not the same as the time index $n$, as we employ the optimal sampling strategy described in Section~\ref{subsec:optimalsampling} to improve the energy and processing efficiency of the sensor. In Fig.~\ref{fig:1D_RoI}, the samples marked with circles, upward-pointing triangles, and downward-pointing triangles represent \textit{idling}~(\textit{I}), \textit{upward crossing}~(\textit{U}), and \textit{downward crossing}~(\textit{D}) events, respectively. In Fig.~\ref{fig:1D_semantic}, class-level graph representations of the samples in Fig.~\ref{fig:1D_RoI}, at $\tilde{n}=41$ and $\tilde{n}=42$, corresponding to $n=225$ and $n=226$ in Fig.~\ref{fig:1D_horizon}, are illustrated with the previously defined semantic graph representation in~\eqref{eq:c1}--\eqref{eq:1D_p}. Additionally, for the semantic graphs shown in Fig.~\ref{fig:1D_semantic}, the attribute sets as defined in~\eqref{eq:1D_At},\eqref{eq:1D_Theta} for $r=2$ are listed in Table~\ref{tb:1D_attributes}.
\begin{figure}[t]
    \centering
    \includegraphics[width = 0.6\textwidth]{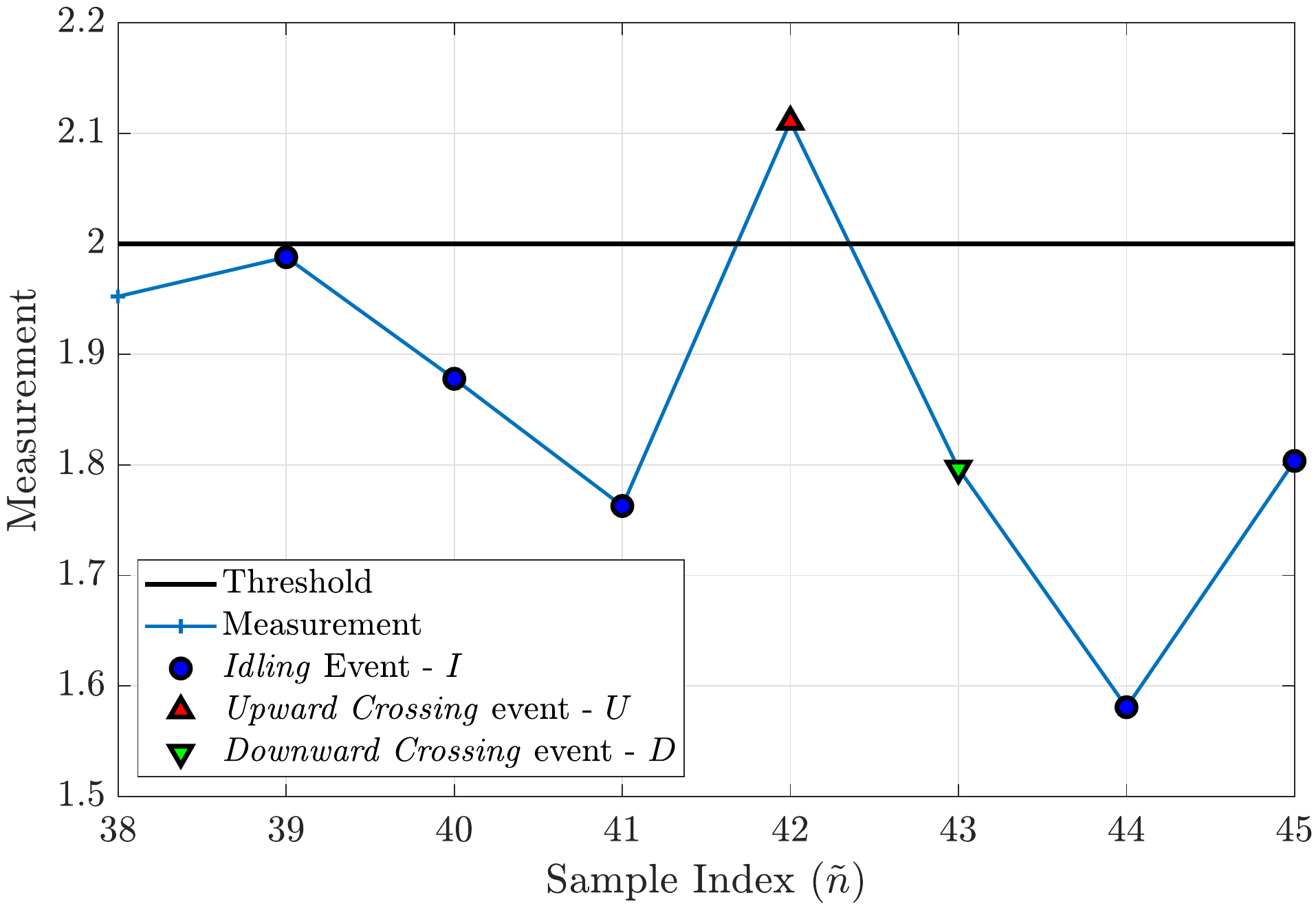}
    \caption{Illustration of \textit{idling}, \textit{upward crossing}, and \textit{downward crossing} events for the AR(1) signal given in Fig.~\ref{fig:1D_horizon}.}
    \label{fig:1D_RoI}
\end{figure}
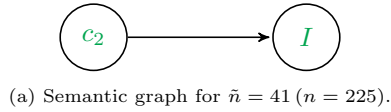
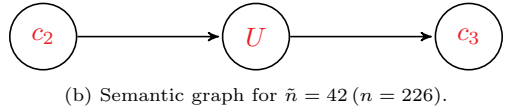
\begin{figure}[H]
    \centering
    \begin{subfigure}[t]{0.48\textwidth}
        \centering
        \begin{tikzpicture}[->,>=stealth',shorten >=1pt,auto,node distance=2.8cm, semithick]
        \tikzstyle{every state}=[text=Green]
        \node[state] (A) {$c_2$};
        \node[state] (C) [right of=A] {$I$};
        \path (A) edge (C);
        \end{tikzpicture}
        \caption{Semantic graph for $\tilde{n}=41 \, (n = 225)$.}
        \label{fig:1D_semantic_a}
    \end{subfigure}
    \vfill
    \vspace*{.5cm}
    \begin{subfigure}[t]{0.48\textwidth}
        \centering
        \begin{tikzpicture}[->,>=stealth',shorten >=1pt,auto,node distance=2.8cm, semithick]
        \tikzstyle{every state}=[text=Red]
        \node[state] (A) {$c_2$};
        \node[state] (C) [right of=A] {$U$};
        \node[state] (B) [right of=C] {$c_3$};
        \path (A) edge (C);
        \path (C) edge (B);
        \end{tikzpicture}
        \caption{Semantic graph for $\tilde{n}=42 \, (n = 226)$.}
        \label{fig:1D_semantic_b}
    \end{subfigure}
\caption{Semantic graph descriptions for the events depicted in Fig.~\ref{fig:1D_RoI}. (a) The \textit{idling} event. (b) The \textit{upward crossing} event.}
\label{fig:1D_semantic}
\end{figure}
\begin{table}[H]
\caption{The attribute sets for samples taken for time instants $225$ and $226$, generated as companions to graphs given in Fig.~\ref{fig:1D_semantic}.}
\centering
\begin{tabular}{c|c|c}
           & $\tilde{n}=41 \, (n = 225)$ & $\tilde{n}=42 \, (n = 226)$ \\ \hline
$\theta^1$ & $(225,\,1.76)$ & $(226,\,2.11)$ \\ \hline
$\theta^2$ & $(224,\,1.88)$ & $(225,\,1.76)$ \\ 
\end{tabular}
\label{tb:1D_attributes}
\end{table}
\noindent As seen in Fig.~\ref{fig:1D_RoI}, the signal is \textit{idling} in Region-2 (between thresholds) at $\tilde{n}=41$; hence, the corresponding component is $c_2$ and its predicate is $I$. At $\tilde{n}=42$, the signal exceeds the upper threshold; therefore, the semantic description includes a flow from $c_2$ to $c_3$ with the predicate $U$ (\textit{upward crossing}).

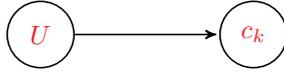
\begin{figure}[t]
    \centering
    \begin{tikzpicture}[->,>=stealth',shorten >=1pt,auto,node distance=2.8cm, semithick]
    \tikzstyle{every state}=[text=Red]
    \node[state] (C)  {$U$};
    \node[state] (B) [right of=C] {$c_k$};
    \path (C) edge (B);
    \end{tikzpicture}    
\caption{Goal pattern defined for an interest in detecting largest threshold crossings.}
\label{fig:1D_goal}
\end{figure}

With the adoption of the semantic language in Section~\ref{sec:ProposedLanguage}, we can easily define specific goals to reduce the storage or transmission rates for this particular application. In a $k-$threshold crossing detection problem, a typical goal might be the detection of exceeding the largest threshold, i.e., $x>t_k$. This means that we are interested in a \textit{upward crossing} predicate connected to $c_k$, hence the goal pattern can be defined as shown in Fig.~\ref{fig:1D_goal}.
With the goals as defined in Fig.~\ref{fig:1D_goal}, the transmission events (or, storage events for offline applications) can be reduced considerably. Coupled with the introduction of optimal sampling strategies in Section~\ref{subsec:optimalsampling}, the overall energy and processing efficiency of scalar sensors can be improved dramatically.

\subsection{Discussion on the Applications of the Proposed Framework on Heterogeneous Multi-Sensor Networks}

In this section, we presented detailed application examples using the proposed semantic signal processing framework on two extremes of signal complexity: namely, the real-time computer vision on video streams and 1-D scalar sensor outputs. The potentials of the proposed framework can be extended to many applications, including heterogeneous sensor networks that work on a dedicated task. A good example of the many fields that can benefit from semantic signal processing is agriculture. 

For intelligent crop and plantation monitoring applications, a heterogeneous network of sensors (cameras, temperature/humidity sensors, etc.) provides information on critical events such as the crop yield and flowering status, temperature and humidity of the soil, existence of pests~\cite{baggio2005wireless,srbinovska2015environmental}. In this application, a global goal regarding the ultimate objectives of the farm can be defined for a common customized language, and then it can be mapped to the respective capabilities of each sensor in the network. Based on goal-oriented returns from the sensors, appropriate actions can be taken. For example, an appropriate pesticide can be recommended and applied automatically to eliminate insects, or in the case of plant growth monitoring, ripe plants can be harvested or infected plants can be exterminated. 

There are countless many other applications such as elderly fall detection, robot navigation, event detection in sports, animal monitoring, farm automation, traffic condition analysis, etc., that can employ the proposed framework or semantic approaches in general. We strongly believe that future research on signal processing should include a focus on different adaptations of a semantic framework for these applications.

\section{Transmission of Goal-Oriented Semantic Information}
\label{sec:TXOfSPExtraction}    
The proposed semantic signal processing framework represents signals in a very organized and easy-to-parse structure, which enables goal-oriented filtering of the data to achieve very high compression rates. This is especially desirable in the next generation of machine-type communications where a huge amount of raw information will be generated by a plethora of IoT devices that needs to be transmitted throughout massive networks. 

In this section, we first showcase the potential compression rates that can be achieved by semantic representation and goal-filtering through a simple example. Then, we discuss efficient coding and compression strategies for the proposed multi-graph and attribute set descriptions for storage and transmission, as well as strategies for transmission over noisy channels.  

\subsection{Data Compression Using Semantics and Goal-Filtering}
\begin{figure}[t]
        \centering
        \includegraphics[width=0.7\textwidth]{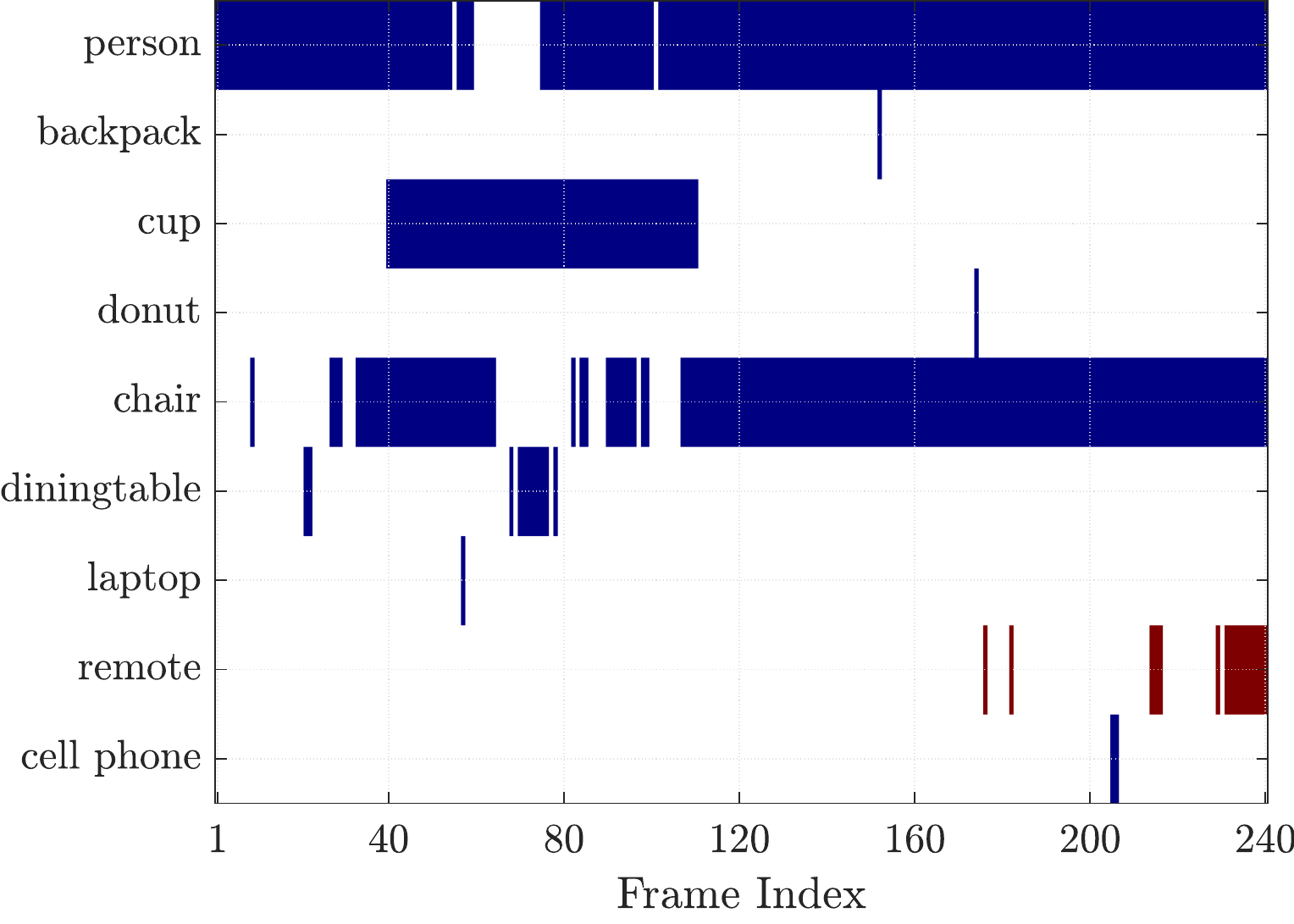}
        \caption{Detected component classes using YOLOv4 on a 240-frame long video. The \textit{interesting} detections from class \textit{remote} are shown in red.}
        \label{fig:detected_class_instances}
\end{figure}

Depending on the application and the nature of the signals of interest, the amount of data to be stored or transmitted can be greatly reduced using the proposed semantic signal processing framework. To illustrate this point via a simple example, we have simulated an object detection problem using YOLOv4~\cite{scaledyolov4} on a 240-frame-long prerecorded video. The object set of YOLOv4 is defined as our component set $C$, while the predicate set is defined only with a $p_0 :$~\textit{exists} predicate, similar to the \textit{null} predicate definition in Section~\ref{subsec:caseStudy_video}. Therefore, in this example, each detection from a frame can be represented as a single component-predicate connection (i.e., $c_i \rightarrow p_0$). The detected class instances throughout the 240-frame video are illustrated in Fig.~\ref{fig:detected_class_instances}.

In accordance with the semantic language definition in Section~\ref{sec:ProposedLanguage}, the detector outputs in Fig.~\ref{fig:detected_class_instances} correspond to a \textit{multi-graph instance representation} $\mathcal{D}$. The corresponding attribute set $\mathcal{A}$ for this experiment is chosen as the cropped bounding box images of each detection, with a single level of complexity (i.e., $L=1$). The class information is then encoded using Huffman coding~\cite{huffman1952method} with a predefined historical occurrence rate. The bounding box images and the full frames are encoded using JPEG compression~\cite{jpeg} with a constant compression rate of 10:1. Note that the specific coding schemes discussed here are selected only to give a general idea of the possible data throughputs, and alternative encoding schemes can be used for different applications.

A typical application for this object detection problem could be the transmission of \textit{interesting} objects to an external agent. For the demonstration of the goal-filtering capabilities of the proposed framework, we assume that an external agent is interested in the detections of \textit{remote} class, and may or may not require the bounding box images of pertinent detections. 

\begin{figure}[t]
        \centering
        \includegraphics[width=0.7\textwidth]{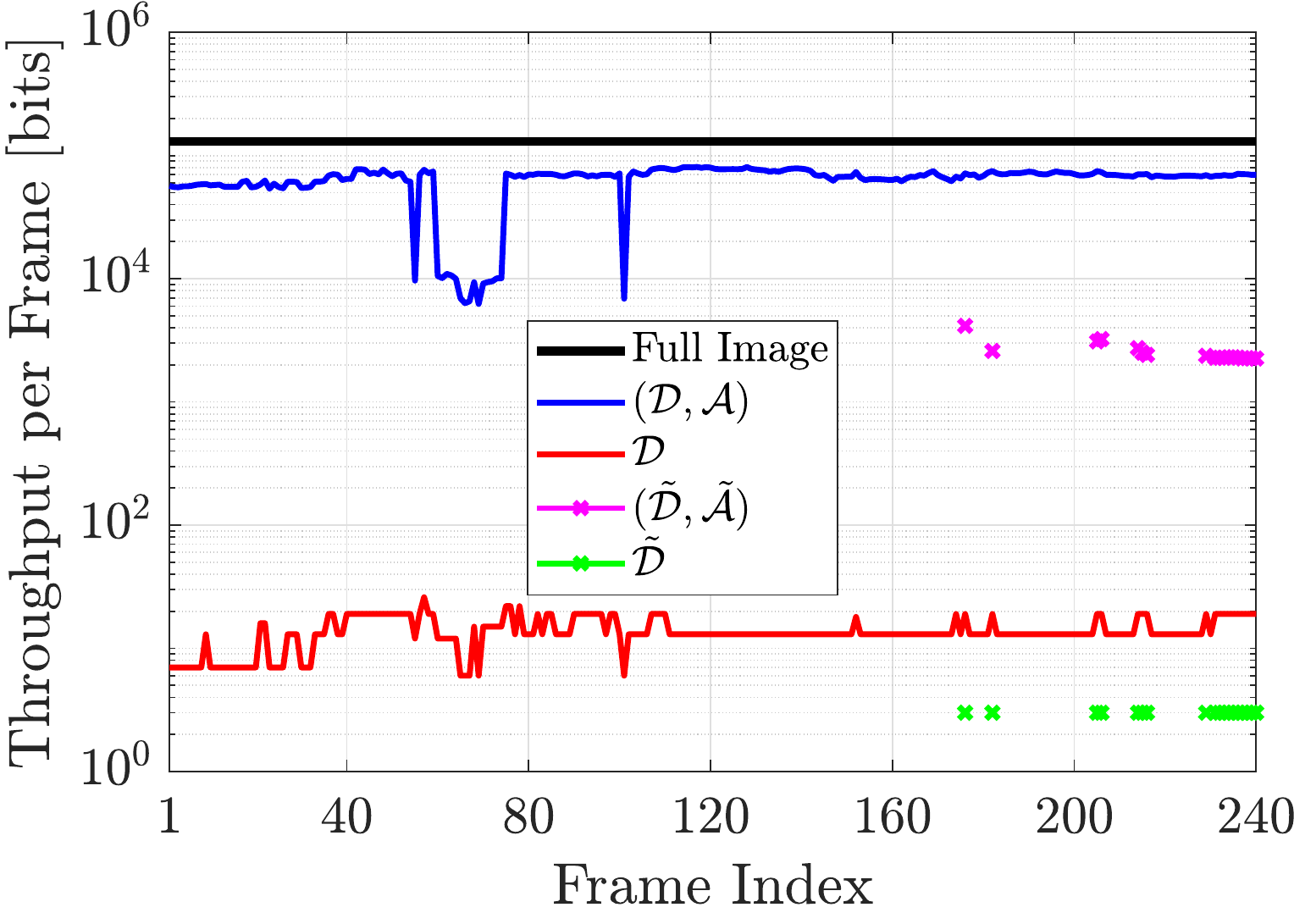}
        \caption{Data throughput using different transmission strategies.}
        \label{fig:GOSC_datacompression}
\end{figure}

With the above configuration of the experiment and the input video stream given in Fig.~\ref{fig:detected_class_instances}, the following transmission strategies are investigated:
\begin{itemize}
    \item Full-Image Transmission: each full frame is sent,
    \item  $(\mathcal{D},\mathcal{A})$: the semantic graph outputs and the bounding box images are sent,
    \item  $\mathcal{D}$: only the semantic graph outputs are sent,
    \item  $(\overset{\sim}{\mathcal{D}},\overset{\sim}{\mathcal{A}})$: the goal-filtered semantic graph outputs and the corresponding bounding box images are sent,
    \item $\overset{\sim}{\mathcal{D}}$: only the goal-filtered semantic graph outputs are sent.
\end{itemize}
The data throughput per video frame using the above transmission strategies is given in Fig.~\ref{fig:GOSC_datacompression}.
As seen in Fig.~\ref{fig:GOSC_datacompression}, the amount of data generated and transmitted can be reduced dramatically by organizing the data in a semantic framework (see $\mathcal{D}$, $\mathcal{A}$). Even further reductions are possible by introducing goals, and filtering out unwanted signal components (see $\overset{\sim}{\mathcal{D}}$, $\overset{\sim}{\mathcal{A}}$). With a goal-oriented approach, transmissions are only sent when events of interest happen. Therefore, the amount of data reduction is dependent on the rate of innovation of \textit{interesting} patterns in the signal. A summary of the results in Fig.~\ref{fig:GOSC_datacompression} is given in Table~\ref{tb:DataCompression}, where the total throughput of the 240-frame video and the corresponding data compression rates are listed. As reinforced by Table~\ref{tb:DataCompression}, extremely high reductions in data rates are possible with the proposed goal-oriented approach.
\begin{table}[ht]
\renewcommand{\arraystretch}{2}
\footnotesize
\centering
\caption{Total throughput and compression rates using the semantic representation of a video stream. Compression rates are given with respect to JPEG frame throughput.} \label{tb:DataCompression}
\begin{tabular}{c|c|c|c|c|c|}
\cline{2-6}
 &
   JPEG Frames &
   $(\mathcal{D},\mathcal{A})$ &
   $\mathcal{D}$ &
   $(\overset{\sim}{\mathcal{D}},\overset{\sim}{\mathcal{A}})$ &
   $\overset{\sim}{\mathcal{D}}$ \\ \hline
\multicolumn{1}{|c|}{\textbf{Total Throughput {[}bits{]}}} &
  $3.1 \times 10^7$ &
  $1.6 \times 10^7$ &
  $3316$ &
  $4.6 \times 10^4$ &
  $54$
   \\ \hline
\multicolumn{1}{|c|}{\textbf{Compression Rate}} &
  \cellcolor[HTML]{C0C0C0} &
   $2 : 1$ &
   $9468 : 1$ &
   $684 : 1$ &
   $580000 : 1$
   \\ \hline
\end{tabular}
\end{table}

As explained above, the encoding schemes used for this experiment are selected for demonstration purposes and ease of implementation. Throughout the rest of this section, we discuss efficient coding schemes to represent multi-graph representations with attribute sets, and possible transmission strategies for semantic information.

\subsection{Efficient Transmission of Proposed Multi-Graph Representations and Attribute Sets}

As described previously, graphs can be used to describe the goal-oriented semantic information acquired by different sensors. This description can be in the form of multi-graph class representations by using class atomic graphs $S_{t,i}$'s as given in \eqref{eq:Sti}, or more detailed information can be obtained through object multi-graph representations via object atomic graphs $D_{t,i}$'s as given in \eqref{eq:Dti}. As such, the storage and/or transmission of this graphical data warrant further discussion. 

In order to store or transmit the graphical data, one may use the biadjacency matrix representation in~\eqref{eq:Adjacency}. As extensively studied in the literature, an adjacency matrix can be described by  $\mathcal{O}(n^2)$ bits where $n$ is the number of vertices. The number of required bits can be reduced to $\mathcal{O}(n \log m + m \log n)$ by using adjacency list representation, where $m$ is the number of edges~\cite{besta2018survey}.  Since $n$ and $m$ can be reasonably small for practical scenarios, the resulting complexity becomes relatively low when the adjacency list representation is used. 
In addition, we can further compress the adjacency matrices by lossless compression techniques to increase the storage and transmission efficiencies. 
One way to compress graphs is to employ Huffman coding for the adjacency lists which is studied in~\cite{suel2001compressing} with a focus on web-graph structures. The compression rate of Huffman coding for this graph structure can be further increased by considering the similarities in shared links~\cite{adler2001towards}.
These studies can be easily extended to the proposed model in Section~\ref{sec:ProposedLanguage} to decrease the storage requirement and transmission time of the semantic description. However, Huffman coding uses the input probability distribution during the encoding process. Hence, we may need certain assumptions on the input probability distribution or implement a statistics gathering phase to efficiently compress the graphs in our language model. 

Another line of research to compress the adjacency matrices includes universal codes for positive integers when the input probability distribution is not known as discussed in~\cite{besta2018survey}. 
For instance, one may use Elias-$\gamma$ codes to encode the elements of the adjacency matrix which require $2 \lceil \log x \rceil + 1$ bits to represent a positive integer $x$. This approach is especially preferred when the upper bound of the integers is not known beforehand~\cite{elias1975universal}. This representation can be improved with Elias-$\delta$ coding with $ \lceil \log x \rceil + 2\lceil \log \lceil \log x \rceil +1 \rceil +1$ bits, which is an asymptotically optimal universal code for positive integers~\cite{elias1975universal}.

The storage and transmission of adjacency matrices of graphical data require careful handling to utilize the communication resources effectively. Nonetheless, the communication and storage requirements in the proposed framework are dominated by the multi-level attribute sets of the nodes as defined in \eqref{eq:theta_ti}. A multi-level attribute set contains multiple levels of attributes. As previously discussed, an attribute set may carry information in varying levels of complexity, e.g., video stream applications may include raw image frames, extracted subfeatures of the components, along very simple scalar data. In the following, we focus on the transmission of high-rate data for semantic and goal-oriented communications.

\begin{figure}[hbt]
	\centering
	\includegraphics[trim={0cm 0cm 0cm 0cm},clip,width=0.6\textwidth]{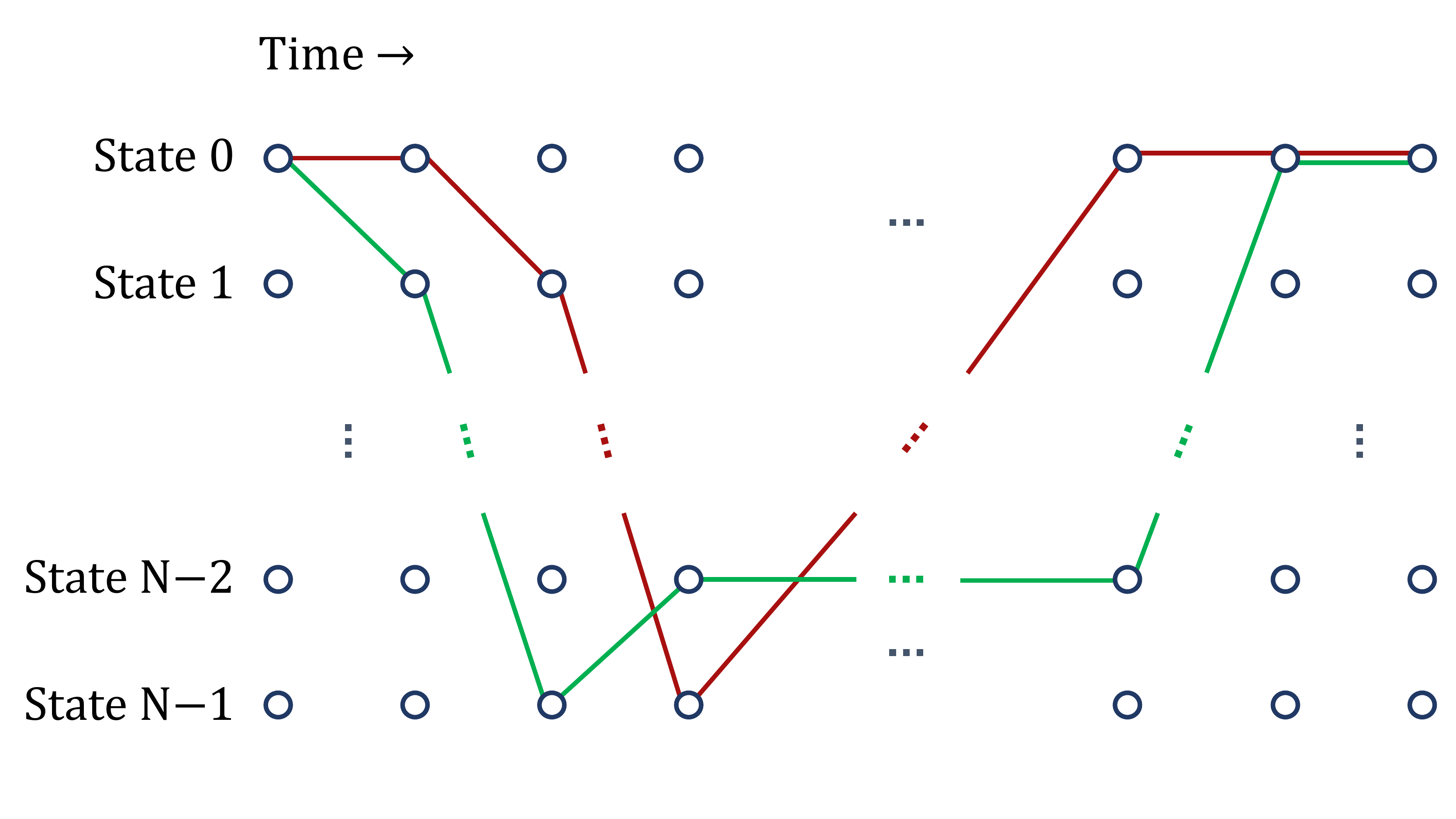}
	\caption{An illustrative trellis diagram. The path with the lowest cost is selected to determine the quantizer output.}
	\label{fig:trellis}
\end{figure}

Let us consider the $l$-th attribute of node-$j$ which can be a subfeature vector of the corresponding component in the scene graph as previously discussed in Section~\ref{subsec:caseStudy_video}, and assume that $ \boldsymbol{\theta}^{(l)}_{t,i}(n_j,k) \in \mathbb{R}^d$ for $l \in \{1, \cdots, L_{n_j}\}$ and $\boldsymbol{\theta}^{(l)}_{t,i}(n_j,k) \in [0, \, 1]$, where the dimension of the vector depends on the design of the semantic extractor. 
To store or transmit these subfeatures efficiently, we can apply lossy compression techniques to map each instance to a finite number of levels, i.e., we can perform quantization. One approach is to perform scalar quantization by addressing a single element of a vector at a time. Another approach is to employ vector quantization through a joint treatment of the entire vector at once, which is proven to be more effective than the scalar quantization~\cite{elementsofInfo}.
A highly efficient and well-performing vector quantization algorithm is the Trellis Coded Quantization (TCQ) proposed in~\cite{marcellin1990trellis}. 
TCQ uses an expanded signal set and coded modulation with set partitioning. To transmit a signal from the codebook of size $2^m$, the original $2^m$-point constellations are doubled into $2^{m+1}$ points. A set partitioning applied to $2^{m+1}$ points to obtain $2^{\tilde{m}+1}$ subsets 
where $\tilde{m}$ is a positive integer which is less than $m$. Then, $\tilde{m}$ input bits are encoded by an $\tilde{m}/(\tilde{m}+1)$ rate convolutional code while the remaining bits are used to determine the codeword from the selected subset. Viterbi decoding~\cite{seshadri1994list} is employed to determine the sequence of codewords for a given input vector by minimizing the mean squared error (MSE) between the input and the output codeword. 
In TCQ, as depicted in Fig.~\ref{fig:trellis}, the paths through the trellis form the quantization codebook, and the path with the lowest cost is selected as the quantizer output for a given input sequence. 

We aim at designing a lossy compression algorithm to store or transmit the semantic information with the smallest possible change on our inference.
As a possible solution, cross-entropy can be considered as a cost metric since it quantifies the difference between two probability distributions. However, it is not easy to find the optimal quantizer output using the cross-entropy cost function for the given attribute, since the cost is not additive, which makes it unsuitable for use as a metric in Viterbi decoding. Instead, one may use other cost metrics such as \textit{MSE, total variation distance (TVD), log-cosh cost}, and \textit{quantile loss} defined as
\begin{itemize}
    \item MSE cost: 
    \begin{equation}
        L_{MSE} = \frac{1}{d} \sum_{b=1}^d \left(\theta_b-\hat\theta_b\right)^2,
    \end{equation}
    \item TVD cost:
     \begin{equation}
        L_{TVD} = \frac{1}{2} \sum_{b=1}^d \left|\theta_b-\hat\theta_b\right|,
    \end{equation}
        \item Log-cosh cost:
     \begin{equation}
        L_{LC} = \frac{1}{d} \sum_{b=1}^d \log\left(\cosh \left(\hat\theta_b - \theta_b \right)\right),
    \end{equation}
            \item Quantile loss with parameter $\gamma \in [0, 1]$:
     \begin{equation}
        L_{Q} = \frac{1}{d} \sum_{b=1}^d \max \left( \gamma \left(\hat\theta_b - \theta_b \right), (\gamma -1) \left(\hat\theta_b - \theta_b \right)\right).
    \end{equation}
\end{itemize}
Note that for ease of notation, we drop the subscripts and superscripts denoting the parameters in the scene graph and denote the $b$-th element of the attribute vector and its quantized version by $\theta_b$ and $\hat\theta_b$, respectively, throughout this section. We also note that the TVD is a distance metric for probability distributions since the vectors to be compared are probability mass functions (PMFs) in the standard construction. In our study, we employ TVD for the subfeatures without any normalization, which proves to be a good metric for our purposes. Furthermore, as an extension to previously mentioned cost metrics, we also define the $\mathit{l}_p$-norm cost as
\begin{equation}
     L_{p} = \frac{1}{d} \left(\sum_{b=1}^d \left(\theta_b-\hat\theta_b\right)^p\right)^{\frac{1}{p}},
\end{equation}
which can be considered as a generalization of the MSE for $p=1$, and the mean absolute error (MAE) for $p =1$.

To summarize, while the construction of the proposed language to describe the semantic information is a significant foundation for goal-oriented signal processing, it is also crucial to address the efficient storage or transmission techniques of the generated semantic information. The adjacency matrices corresponding to the graph representations can be compressed by applying well-known methods in the literature, e.g., Huffman coding or universal codes. Furthermore, to decrease the bandwidth consumption of the dominant components in the attribute sets, one may employ TCQ based quantization, where it is also possible to use different cost metrics to optimize the system performance.  

\subsubsection{A Simple TCQ Based Compression of Semantic Information}
\label{subsubsec:TCQ}
To demonstrate the performance of the semantic information compression, we use the same case study described in Section~\ref{subsec:caseStudy_video}, where the sensors operate on video streams, and the language consists of 80 semantic components. To assess the performance of TCQ based quantization on a learning setup, we use the COCO dataset~\cite{mscoco}. Employing the described CNN model in Section~\ref{subsec:caseStudy_video}, the resulting feature vector of each detected object is stored as an element of the attribute set, i.e., $\boldsymbol{\theta} = \mathbf{r}_i \in \mathbb{R}^{128}$. These attributes can be stored locally or transmitted to a base station, where it is desired to protect semantic information (i.e., meaning) with a lossy compression scheme as much as possible. 
That is, we would like to compress the subfeature vector without changing the corresponding output of the classifier. To compress the subfeature vectors, we use TCQ based quantization with different cost metrics and a few bits, e.g., 2, 3, and 4 bits per element. Here, we do not follow Ungerboeck’s construction for TCQ, which is described in~\cite{marcellin1990trellis}; instead, we employ an $m/(m+1)$ code rate convolutional code for $m$-bit quantization without set partitioning and map each vector entry to $m$ bits. 
For $2$-bit quantization, a rate $2/3$ convolutional code with two connected shift register banks is used, where the constraint lengths of each shift register bank are three and two, respectively. In this architecture, the first output is connected to the first shift register array, the second one is connected to both shift register arrays, while the third output is connected to the last array only (see \cite{ryan2009channel} for the shift register construction of convolutional codes). The corresponding octal code generator matrix specifying the output connections for each input bit is given by
\begin{align}
   G_{2/3} = \begin{bmatrix}
2 & 1 & 4\\
0 & 1 & 2
\end{bmatrix}.
\end{align}
Similarly, for the $3$-bit vector quantization example, we use a rate $3/4$ convolutional code where the constraint lengths of register banks are $[5, 4, 4]$ with the generator matrix
\begin{align}
   G_{3/4} =  \begin{bmatrix}
23 & 35 & 0 & 0\\
0 & 5 & 13 & 0\\
0 & 0 & 6 & 13
\end{bmatrix}.
\end{align}
Finally, for the $4$-bit quantization example, rate $4/5$ convolutional code is employed where the constraint lengths of register banks are $[5, 4, 4, 4]$ with the generator matrix
\begin{align}
   G_{4/5} =  \begin{bmatrix}
23 & 35 & 0 & 0 & 0\\
0 & 5 & 13 & 0 & 0\\
0 & 0 & 6 & 13 & 0\\
0 & 0 & 0 & 6 & 13
\end{bmatrix}.
\end{align}
Note that, in this simple case study, we do not optimize the architecture or the convolutional code used in TCQ. The optimization of this quantizer structure, particularly taking into account the nature of the semantic information can be a promising research direction.

The quantized feature vectors are fed into a neural network where we have two fully connected hidden layers with ReLU activations, and the softmax activation is used at the output layer. For comparison purposes, we also quantize the features by a uniform scalar quantizer which performs an element by element quantization. 
As a benchmark, we compare the TCQ-based quantizer's accuracy with a full resolution case in which the features are directly fed into the neural network. The accuracy of the benchmark is $83.12\%$ where the elements of the feature vectors are represented by $32$ bits. To evaluate the performance of the TCQ quantizer relative to the benchmark, we normalize the classification accuracies by the accuracy of the benchmark and report the results in Table~\ref{tb:tcq_acc_tab_norm}.
In this example, the best classification accuracy for $2$-bit quantization is achieved with the Log-cosh cost while the MSE cost achieves higher accuracy than the other cost functions with $3$-bit and $4$-bit TCQ.  
\begin{table*}[ht]
\centering
\caption{Normalized classification accuracies of TCQ and uniform quantization where the accuracy of the benchmark is $83.12\%$, which is obtained by using a $32$-bit representation of subfeatures. The normalization is performed by dividing the classification accuracy of the TCQ based semantic communication case study by the benchmark accuracy.} \label{tb:tcq_acc_tab_norm}
\begin{tabular}{|c|c|c|c|c|c|c|}
\hline
\multirow{3}{*}{\begin{tabular}[c]{@{}c@{}}Number \\ of bits\end{tabular}} & \multirow{3}{*}{\begin{tabular}[c]{@{}c@{}}Compression \\ ratio\end{tabular}} & \multirow{3}{*}{\begin{tabular}[c]{@{}c@{}}Uniform scalar \\ quantization\end{tabular}} & \multicolumn{4}{c|}{TCQ}\\
\cline{4-7}  &  &  & 
\multicolumn{4}{c|}{Cost metrics} \\ 
\cline{4-7}  & & & MSE & \begin{tabular}[c]{@{}c@{}}Total variation\\ distance\end{tabular} & Log-cosh & \begin{tabular}[c]{@{}c@{}}Quantile\\ $\gamma = 0.5$\end{tabular} \\ \hline
2-bit & 0.06250 & 54.05\%  
& 67.06\% & 65.65\% & 67.12\%  & 63.31\% \\ \hline
3-bit & 0.09375 & 86.15\% & 91.87\% & 91.21\% & 91.86\%  & 90.80\% \\ \hline
4-bit & 0.12500 & 96.87\% & 97.58\% & 97.30\% & 97.45\%  & 96.59\% \\ \hline
\end{tabular}
\end{table*}

\subsubsection{Further Extensions for Effective Compression}
In many semantic signal processing applications, it is possible to further increase the compression rates. As an example, we consider the case study described in Section~\ref{subsec:caseStudy_video} where it is possible to track the individual objects by comparing their features as illustrated in Fig.~\ref{fig:deepsort_simple} using the similarity of the feature vectors for consecutive frames. These similarities can also be exploited to create a differential data storage and transmit system. That is, the first feature vector obtained for a specific individual is to be compressed as previously described. When a new feature vector arrives, the difference with the previous one can be computed, which represents the fresh information in the new feature vector.  One straightforward way to compress this differential information is to use TCQ with finer quantization levels suitable for the difference vectors, resulting in a higher resolution than sending all the feature vectors separately without utilizing the common information with the same strategy or communication costs. Furthermore, one may expect to have nearly \textit{sparse} difference vectors due to similarities, and consequently, it may be possible to further compress the information to be stored or transmitted via innovative compressed sensing approaches~\cite{baraniuk2007compressive}.

The described goal-oriented semantic signal processing framework differs from traditional sensor networks in terms of the scope of knowledge generated and processed in sensors/base stations, which may alter the quantization/compression schemes. Unlike traditional sensor networks, the class information of each feature vector is already generated in the sensors. With this additional information, the quantizers can be designed for specific classes resulting in a \textit{class-aware compression}. For instance, the transmit side can develop a performance indicator, e.g., using Kullback-Leibler distance and cross-entropy, to decide on the methods or metrics employed to compress the semantic information without losing the essential \textit{meaning}. Such an indicator can be used to evaluate the performance of the TCQ-based quantizers with different cost metrics; thus, the sensor can decide on the quantizer output to be sent. Moreover, the averages of feature vectors for each class can be determined and shared with the receiver offline to enable \textit{class-aware compression}. In most scenarios, the attribute set \eqref{eq:theta_ti} contains the class information and its associated features. Hence, instead of compressing and transmitting each feature vector, sensors may use the average feature information which is available at both the sensors and the base station. Similar to a differential data transmission setup, the fresh information in the feature vector can be extracted by taking the difference between the new feature vector and the average feature vector of the corresponding class. Furthermore, vector quantization or compressed sensing can be employed to compress the differential feature vector. An interesting line of research in the context of \textit{class-aware compression} is not only to use the class feature averages but also design a compressor for each class separately to exploit the shared knowledge of the sensors and the base station.

\subsubsection{Transmission over Noisy Channels}
As a further extension, joint source/channel coding resulting in more error-resilient systems can be employed~\cite{shannon1949communication} that are suitable for transmitting semantic information over noisy channels. In~\cite{goldsmith1998joint}, the authors consider the joint design of channel-optimized vector quantization (COVQ) and rate-compatible
punctured convolutional (RCPC) codes while~\cite{ho1996transmission} studies joint
source and channel coding using multi-carrier modulation.
The authors in~\cite{ruf1995rate} focus on image transmission where they analyze the rate-distortion behavior for joint source and channel coding with a wavelet-based sub-band source coding scheme and rate-compatible punctured convolutional (RCPC) code. 
Similar to these studies, a future line of research for goal-oriented semantic communication may be to jointly optimize the source and channel coding where the desired meaning in the attribute set is transmitted to a base station in a robust manner despite the presence of channel noise.

As a concrete example of semantic communication over noisy channels, consider an  $m$-bit quantization of the semantic information vector that is to be transmitted over a discrete memoryless channel (DMC).
The $b$-th element of the subfeature vector $\boldsymbol{ \theta} \in \mathbb{R}^d$ is quantized as $\boldsymbol{\hat \theta}$, where each element of $\boldsymbol{\hat \theta}$ can be represented by an $m$-bit binary sequence. 
As there are $2^m$ different $m$-bit sequences, let the sizes of both input and output alphabets of the DMC channel be $2^m$. Let us denote the $m$-bit input and output alphabets by $\mathcal{I}$ and $\mathcal{J}$, respectively, where $\mathcal{I} = \mathcal{J} = \{c_0, c_1, \cdots, c_{2^m-1}\}$. The channel transition probabilities are denoted by
\begin{equation}
    \mathbb{P}( Y = c_i|  X = c_j),
\end{equation}
which corresponds to the probability of receiving the $m$-bit $c_j$ sequence when the $m$-bit $c_i$ sequence is transmitted. In this case, the relevant cost at the receiver side becomes
\begin{equation} \label{eq:DMC_eq}
    L_{DMC} = \sum_{b = 1}^d \left( \sum_{k = 0}^{2^m-1} \mathbb{P}(Y_b = c_k| X_b = x_b) l(x_b,c_k) \right),
\end{equation}
where $\mathbb{P}(Y_b = c_k| X_b = x_b)$ is the transition probability for the $b$-th element of the subfeature vector representing the probability of receiving $m$-bit sequence $c_k \in \mathcal{J}$, when $x_b \in \mathcal{I}$ is transmitted, with the corresponding cost $l(x_b,c_k)$. Clearly, the overall cost function is assumed to be additive, and even though both $x_b$ and $c_k$ are $m$-bit binary sequences, the cost is calculated by considering their mapping into their real values, and the cost function can be one of the metrics described previously in this section. Note that $L_{DMC}$ is the average cost that takes into account the transition probabilities of the underlying channel model. 
As a further extension to the DMC channel model, one may also use different channel models, e.g., a Gaussian channel. In this case, instead of using the channel transition probabilities, the conditional probability density function defined by the channel should be used, and the summation over the variable $k$ in~\eqref{eq:DMC_eq} should be replaced with an integral.

Note that while it is not explored here, joint optimization of the vector quantization with physical-layer properties to have an efficient and robust semantic communications system in practice may be an interesting research direction. 

\section{Conclusions and Future Research Directions}
\label{sec:Conclusion}

In this paper, for a structured and universal representation and efficient processing of the semantic information in signals, we propose and demonstrate a formal semantic signal processing framework in which the semantic information is represented by a graph-based structure. In the proposed framework, following a preprocessing stage of raw signals, a semantic information extractor identifies and classifies components from a set of predefined application-specific classes. The states, actions, and semantic relations among the identified components are described by another set of predefined application-specific predicates. Furthermore, along with the identification and classification of the components, each node in the graph is associated with a hierarchical set of attributes that provide additional information in an organized way. The proposed semantic signal processing framework enables the use of internally or externally defined \textit{goals} that can vary with time. Using these goals, graphs can be grouped as those that are to be processed further and those that are of no interest. The further processing stages may include spatio-temporal tracking of the graph representations and a wide range of further signal processing operations on their more detailed set of attributes. At any point in the processing chain, the desired level of semantic information of those graphs which are of interest can be locally stored or shared with another processor through a communication protocol. Since a typical high bandwidth sensor data has sparse occurrences of interesting events, semantic signal processing of the sensor outputs results in remarkable compression rates. The wide range of applicability of the proposed goal-oriented semantic signal processing is illustrated over different examples and details are provided on how the proposed framework can be adapted for each use case.

The proposed semantic signal processing framework opens up multiple research directions in both theory and practice. First, available machine learning techniques should be assessed for their applicability in this framework for real-time, offline, and batch processing applications. For real-time applications where sensor data is semantically processed to observe and control the state of a system, semantic extraction should be completed within a certain deadline, depending on signal bandwidth and processing capabilities. Therefore, only those machine learning techniques that can extract semantic information with a low time complexity should be considered for this purpose. For applications that may allow offline processing such as semantic medical imaging, the semantic extraction and processing techniques must prioritize increasing the reliability of extracted meaning over the computational complexity. Based on the results of the assessment on both existing real-time and offline semantic extractors, desired features of new machine learning techniques should be identified for improved semantic extraction in the proposed framework. 

Along with the new ML techniques for semantic information extraction, new goal-oriented signal processing techniques should be developed to take advantage of the available semantic information on the identified signal components. The proposed graph-based structure for the semantic information not only captures the relationship among different components through a set of application-specific predicates but also provides a hierarchical set of attributes for each component. Therefore, signal processing techniques that first identify the components of interest and then process their attributes can provide improved time complexity over conventional signal processing techniques, by prioritizing to achieve the desired accuracy of processing only on the components of interest. These algorithms should be able to set the goals of semantic filtering and semantic post-processing based on the estimated states of the components of interest. Further research on goal generation and dissemination of goals should also be performed. Mapping of the global goals defined by either human operators in a sensor network or defined by manufacturers of semantic devices at a hardware level, to local goals that can be interpreted by the pertinent device languages should also be studied. 

Alternative hardware implementations of the proposed semantic signal processing framework should also be investigated. Since this framework can be implemented by using discrete subsystems or system-on-chip (SoC) architectures, a wide range of implementation alternatives exist. In the case of discrete subsystems, it is appropriate to combine the preprocessing and the semantic extraction stages in a sensor subsystem that generates data directly in the proposed graph-based structure with a hierarchical set of attributes. The aforementioned global goal definition by manufacturers can be employed in the semantic extraction stage at a hardware level, and the semantic filtering, semantic post-processing stages, and the storage unit can be built as a single subsystem. The transmission (or the storage) stage should be a versatile subsystem that can exchange relevant semantic data with a base station (or a storage unit). Detailed implementation of these subsystems should be carried out following an extensive analysis of their required specifications over multiple use cases. SoC implementation of the proposed framework will enable low-cost solutions for semantic signal processing and communication on a massive scale.  

The introduction of the proposed semantic framework enables improvements over the limits imposed by the classical information theory, which is only concerned with the compression and transmission of bits as opposed to the meaning in the underlying data. In order to understand and quantify the new fundamental limits, modeling of the semantic noise, concept of information, relevant lossless compression, channel capacity and rate-distortion formulations have to be developed and computed in a variety of settings. In addition, as the 
goal-oriented approach also directly affects the aforementioned limits, the nature and effects of goals should be incorporated into these theoretical investigations as well. For efficient storage and communication of the desired semantic information with the proposed graph-based language model, it is essential to employ appropriate compression techniques which reduce the memory/bandwidth requirements. To compress the biadjacency matrix representations of the graph-based structure, well-known methods in the literature, e.g., Huffman coding or universal codes, can be employed. For the compression of the dominant components in the corresponding attribute sets, the use of TCQ-based quantization for these bandwidth-consuming vectors with different cost metrics is investigated in this paper; and these studies should be further extended. After identifying useful performance metrics, joint optimization of vector quantization with the physical-layer properties of the channel should be investigated for an efficient and robust system for semantic communications.

Finally, over the challenging use cases for the next generation of intelligent distributed sensor networks and communication systems, the performance of the proposed semantic signal processing and communication methods should be compared with the available alternatives. The future of signal processing and communications requires a paradigm shift, and we believe that the proposed semantic framework may be a strong contender in developing the next generation of devices and systems whose performance surpass the limitations imposed by the classical information theory.

\section*{Acknowledgements}
We would like to thank Ercan E. Kuruoğlu, for his kind invitation to submit our work to the 30th anniversary special issue of the Digital Signal Processing Journal and for providing a thorough review with many constructive suggestions. 

\newpage
\bibliography{MK_bib, BT_bib, AA_Bib, MG_bib, sensors}

\end{document}